\documentclass{aa}  

\usepackage[utf8]{inputenc}
\usepackage{lscape}
\usepackage{natbib}

\usepackage{graphicx}	
\usepackage[none]{hyphenat}
\usepackage{txfonts}
\usepackage{hyperref}

\newcommand{\Ha} {H$\alpha$}
\newcommand{\ha} {H$\alpha$}
\newcommand{\Hi}{\textsc{Hi}}
\newcommand{\hi}{\textsc{Hi}}

\newcommand{\Deg}{${}^{\circ}$}

\newcommand{\Sun}{${}_{\odot}$}

\newcommand{\kms}{~km~s$^{-1}$}

\newcommand{\SNR}{signal-to-noise ratio}

\newcommand{\rc}{rotation curve}
\newcommand{\RC}{Rotation curve}
\newcommand{\rcs}{rotation curves}
\newcommand{\RCs}{Rotation curves}
\newcommand{\FOV}{field-of-view}

\begin{document}

     
\title{An \Ha~kinematic survey of the \textit{Herschel} Reference Survey -- I. Fabry-Perot observations with the 1.93m telescope at OHP\thanks{Based on observations collected at the Observatoire de Haute Provence (OHP) (France), operated by the CNRS.}.}    


     \author {G\'omez-L\'opez, J. A.\inst{\ref{inst1}}\thanks{Alternative \email{jagl.c06@gmail.com}}\and
 Amram, P.\inst{\ref{inst1}}\and
 Epinat, B.\inst{\ref{inst1}}\and
 Boselli, A.\inst{\ref{inst1}}\and
 Rosado, M.\inst{\ref{inst2}}\and
 Marcelin, M.\inst{\ref{inst1}}\and
   Boissier, S.\inst{\ref{inst1}}\and
    Gach, J.-L.\inst{\ref{inst1}}\and
 S\'anchez-Cruces, M.\inst{\ref{inst1}}\and
Sardaneta, M. M. \inst{\ref{inst1}}}

\institute{ Aix Marseille Univ, CNRS, CNES, LAM, Marseille, France. \email{jesus.gomez-lopez@lam.fr, philippe.amram@lam.fr, benoit.epinat@lam.fr, alessandro.boselli@lam.fr, michel.marcelin@lam.fr, monica.sanchez-cruces@lam.fr, minerva.munoz@lam.fr} \label{inst1}
\and 
Instituto de Astronom\'ia, Universidad Nacional Autonoma de M\'exico (UNAM), Apdo. Postal 70-264, 04510, Mexico City, Mexico. \email{margarit@astro.unam.mx} \label{inst2}
}
	


   \date{Received May 10, 2019; accepted August 22, 2019}



\sloppy

\abstract
{}
{We present new 2D high resolution Fabry-Perot spectroscopic observations of 152 star-forming galaxies which are part of the 
\textit{Herschel} Reference Survey (HRS), a complete $K$-band selected, volume-limited sample of nearby galaxies, spanning a wide range in stellar mass and morphological type.}{ Using improved data reduction techniques that provide adaptive binning based on Voronoi tessellation, using large \FOV\ observations, we derive high spectral resolution (R$>$10,000) \Ha\ datacubes from which we compute \Ha~maps and radial 2D velocity fields that are based on several thousand independent measurements. A robust method based on such fields allows us to accurately compute \rcs~and kinematical parameters, for which uncertainties are calculated using a method based on the power spectrum of the residual velocity fields.}{We check the consistency of the \rcs~by comparing our maximum rotational velocities to those derived from \Hi~data, and computing the $i$-band, NIR, stellar and 
baryonic Tully-Fisher relations. We use this set of kinematical data combined to those available at other frequencies to study 
for the first time the relation between the dynamical and the total baryonic mass (stars, atomic and molecular 
gas, metals and dust), and derive the baryonic and dynamical main sequence on a representative sample of the local universe.}
{}

\keywords{
Galaxies: fundamental parameters -- Galaxies: kinematics and dynamics -- Galaxies: spiral -- Galaxies: general -- Galaxies: statistics -- Galaxies: evolution. }

\maketitle

\section{Introduction}
\label{intro}
One of the main processes regulating galaxy evolution is star formation. The gas located along the disc of spiral galaxies
collapses inside molecular clouds to form stars following 
the Kennicutt-Schmidt law (\citealp{Schmidt:1959}; \citealp{Kennicutt:1989}). 
Besides enriching the Inter Stellar Medium ($ISM$) with chemical elements, massive 
evolved stars and supernovae inject a large amount of kinetic energy into the $ISM$ (\citealp{Van:2000}), 
driving the formation and evolution of new generations of stars. 
Finally, the star formation process affects also the Inter Galactic Medium 
($IGM$, \citealp{Boselli:panchromatic}).
\vspace{2.5mm}

We are still far from fully understanding the complexity of the star formation process. 
It is still unclear which mechanism triggers the collapse of the gas inside molecular clouds to form new stars; 
nowadays, this topic is still under debate. The kinematical properties in late-type galaxies seems 
to play a crucial role in triggering the star formation process, at large scales through the instability of the rotating disc
(\citealp{Toomre:1964}, \citealp{Larson:1992}; \citealp{Kennicutt:1998, Kennicutt:1998b}; \citealp{Tan:2000}; 
\citealp{Boissier:2003}) and the dynamical influence of the spiral arms (\citealp{Wyse:1986}; 
\citealp{Tan:2000}), while at small scales through turbulence (\citealp{Wang:1994}; 
\citealp{Corbelli:2003}; \citealp{Leroy:2008}; \citealp{Krumholz:2005}; \citealp{Krumholz:2012}; 
\citealp{Elmegreen:2015}).   
\vspace{2.5mm}

The study of the relations between the star formation rate ($SFR$), the gas column density ($\Sigma_{gas}$), 
and the kinematical properties of galaxies 
on strong statistical basis would require a well representative sample having multifrequency resolved 
images and 2D-spectroscopic observations. The distribution of the atomic and molecular gas can be derived using \hi~and CO
observations, while that of the star formation activity using direct tracers such as the Balmer H$\alpha$ emission line
or the UV emission, properly corrected for dust attenuation using for instance the far-infrared emission, 
or using SED fitting codes when multifrequency data are available.
\vspace{2.5mm}

The \textit{Herschel} Reference Survey (HRS, \citealp{Boselli:2010}) is a complete sample of 323 nearby 
galaxies defined with the purpose of studying the relation between the star formation process and the 
different components of the interstellar medium (ISM). This sample has been observed at all frequencies to
provide the community with the largest possible set of homogeneous data. This unique dataset includes
photometric data in the UV and in the optical bands (\citealp{Cortese:2012}; \citealp{Boselli:2011}, \citealp{Ferrarese:2012}), 
in the mid- and far-IR (\citealp{Ciesla:2012}; \citealp{Cortese:2014}; \citealp{Bendo:2012}; \citealp{Ciesla:2014}), 
and in the H$\alpha$ line\citep{Boselli:2015}. Spectroscopic data are also
avalibale in the visible (\citealp{Boselli:2013}, \citealp{Gavazzi:2004,Gavazzi:2018}) and in the radio at 21 cm (HI) and at 2.6 mm (CO)
\citep{Boselli:2014}.
\vspace{2.5mm}

Two-dimensional kinematical data are however still lacking. For this purpose, we are undertaking an \Ha~kinematic survey of the HRS star-forming galaxies using Fabry-Perot interferometry. This technique is perfectly adapted for this sample since it allows to gather, using typical integration times of $\sim$2 hours per galaxy and a 2m-class telescope, seeing-limited ($\sim$2 arcsec) datacubes of galaxies, within a large field-of-view (FoV$\sim$5.8$\times$5.8 arcmin$^2$), i.e. surrounding by their nearby environment, with high a spectral  resolution ($R>10,000$).  The large FoV allows to get high resolution \Ha\ spectra for hundreds to thousands independent spatial elements.
\vspace{2.5mm}
 
This paper presents new Fabry-Perot data for 152 HRS star-forming galaxies gathered during 9 observing runs at the Observatoire de Haute Provence (OHP). These data are used to derive 
the kinematical properties of the ionised gas at high spatial and spectral resolution. Combining this new 
set of data with other Fabry-Perot data available in the literature we study the relation between the dynamical 
and the baryonic mass, this last directly measured using multifrequency observations. 
We also derive, for the first time in the literature, the dynamical mass main sequence for a representative 
sample of the local universe.
\vspace{2.5mm}

The paper is organised as follows. Section \ref{sample} describes the HRS sample, section \ref{data} the 
observations and data reduction. In section \ref{dataanalysis} we derive the kinematical parameters. In 
section \ref{sanity} we compare several kinematical scaling relations derived for the HRS with those proposed in the literature. 
The summary and conclusions are given in Section 
\ref{conclusions}. All the data products, including comments on individual objects, are given in 
the different Appendices.
\vspace{2.5mm}

Consistently with our previous works, all the Fabry-Perot data will be made available on the HRS dedicated 
database HeDAM (\url{https://hedam.lam.fr/}), and on the Fabry-Perot 
database (\url{https://cesam.lam.fr/fabryperot}).

\section{The Sample}
\label{sample}

 \begin{figure}
\begin{center}
\includegraphics[width=8.3cm]{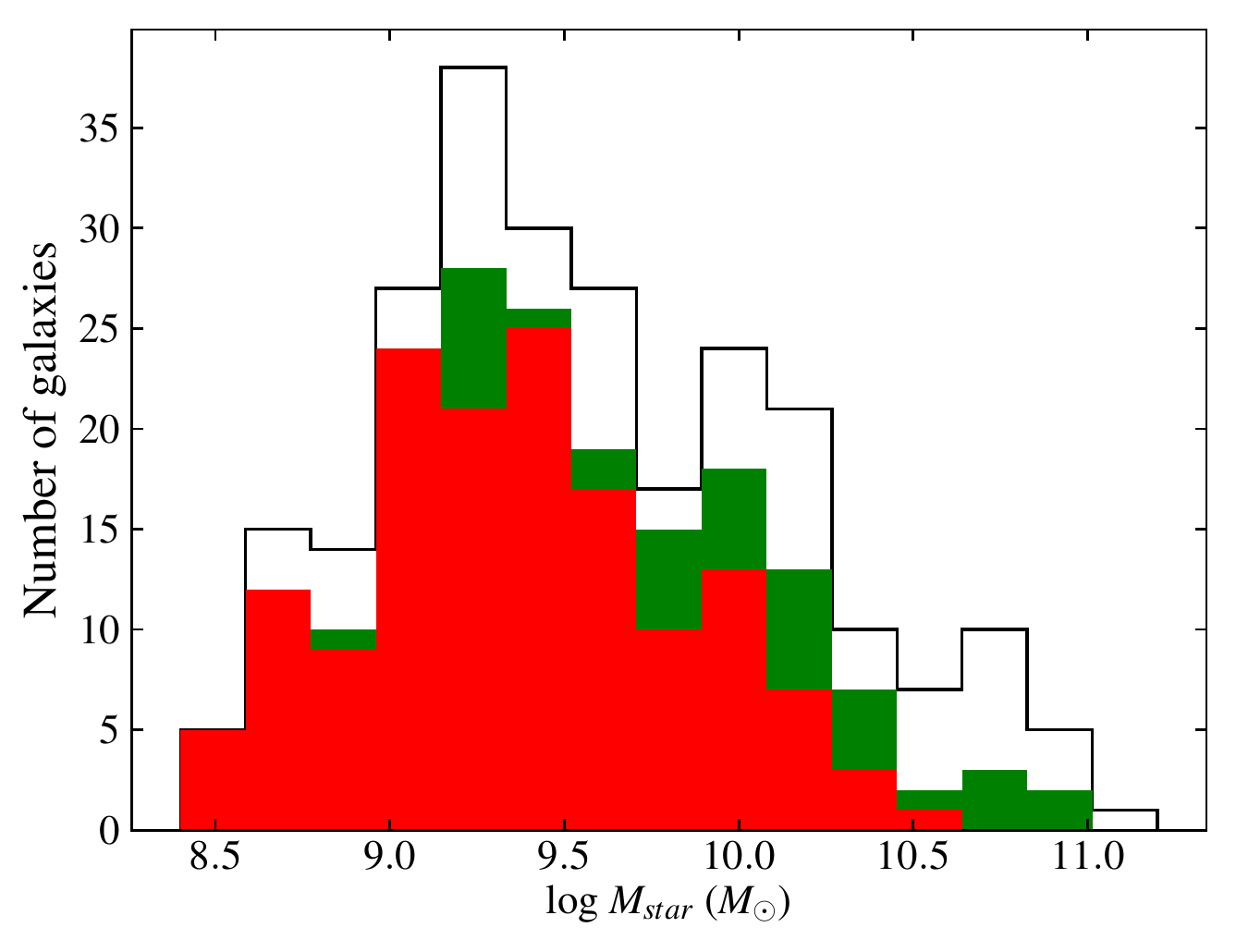}
\includegraphics[width=8.3cm]{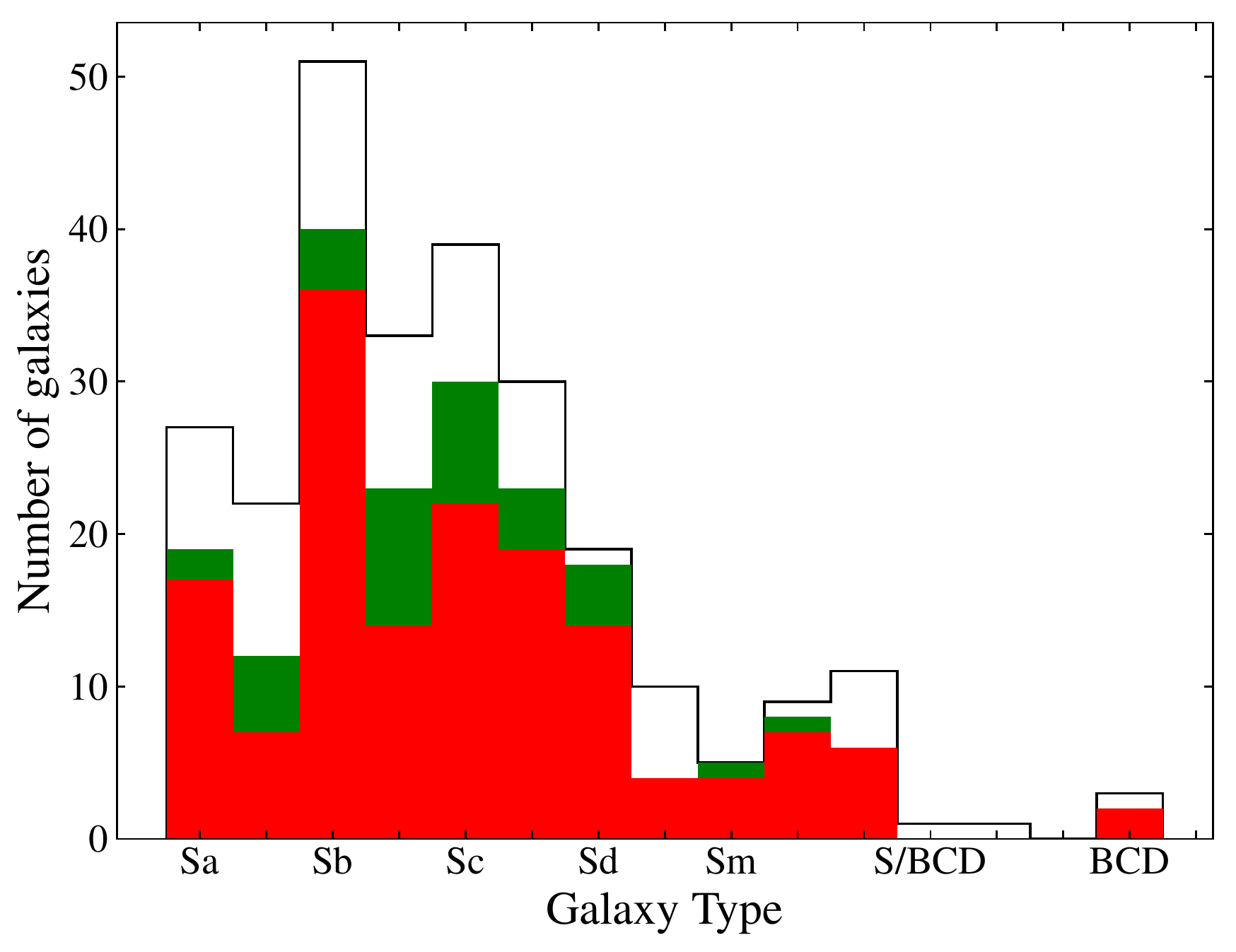}
\end{center}
\caption{ Top panel: Histogram of the stellar mass distribution of the HRS star-forming sample. Bottom panel:  
Histogram of the galaxy type distribution. Out of the 261 galaxies (white histograms), 152 have been observed at the 
OHP and are presented in this work (red), while 40 observations are available in the literature (green).} 
\label{fig:typemag}
\end{figure}
 
 The $\it{Herschel}$ Reference Survey (HRS) is  a $K$-band-selected (where the $K$-band is a proxy of stellar mass, \citealp{Gavazzi:1996}), 
 volume-limited (15 $<$ $D$ $<$ 25 Mpc) complete sample of 323 galaxies spanning a wide range in morphological type
 (from ellipticals to late-type spirals and irregulars) and stellar mass (10$^{8}$  $<$  $M_{star}$  $<$ 10$^{11}$ M\Sun).
 The sample, which has been extensively presented in \cite{Boselli:2010}, includes galaxies belonging to different environments, 
 from the Virgo cluster to isolated systems in the field. Out of the 323 galaxies of the sample, 261 objects are late-type systems, with an 
 ongoing star formation activity, as indicated by their strong Balmer line emission \citep{Boselli:2015}. These late-type systems are thus 
 adapted targets for a kinematical survey of the ionised gas using a Fabry-Perot interferometer. Thanks to its statistical completeness, the 
 HRS is becoming the ideal reference for local and high-redshift studies as well as for models and simulations. It is thus a tailored sample for
 tracing the kinematical scaling relation of a complete sample of nearby galaxies.
 \vspace{2.5mm}

In this work we present new Fabry-Perot data for 152 star forming galaxies observed at the OHP (see Table \ref{table_hrs}). 
Fabry-Perot are also available in the literature for other 40 galaxies from the following references: 
GHASP (\citealp{Garrido:2005}; \citealp{Epinat:2008a}), SINGS \citep{Daigle:2006sin}, Virgo cluster \citep{Chemin:2006}, 
and Loose Groups \citep{Marino:2013}. 
So far, Fabry-Perot data are thus available for 192 out of the 261 star-forming galaxies of the sample (73.6\% complete).  
Figure \ref{fig:typemag} compares the distribution in morphology and stellar masses (taken from \citealp{Cortese:2012}) 
of the HRS galaxies to the Fabry-Perot dataset presented in this paper and available from the literature. This figure shows that the available kinematical data constitue a representative subsample of the HRS, suitable for statistical studies, since they already include a large number of galaxies spanning a wide range in morphological type and stellar mass. \\

\section{Observations and Data Reduction}
\label{data}

Fabry-Perot observations scanning the \Ha~emission line of 152 late-type galaxies have been obtained along 91 nights at the OHP (9 runs from February 2016 to April 2018). The observations have been carried out in good photometric conditions and with a typical integration time of 2 hours per galaxy. The journal of the observations is given in Table \ref{tablelog}.  

\subsection{GHASP Instrumental Setup}
\label{GHASPsetup}

Fabry-Perot observations provide datacubes with dimensions $x$ (right ascension $\alpha$), $y$ (declination $\delta$) and $z$ (wavelength $\lambda$), containing \Ha~profiles for each pixel along the field-of-view (FoV). The instrument we use at OHP is GHASP, a focal reducer containing a scanning Fabry-Perot interferometer attached at the Cassegrain focus of the 1.93m telescope. The principles and characteristics of the GHASP instrument are extensively explained in \cite{Garrido:2002, Garrido:2003, Garrido:2004, Garrido:2005} and \cite{Epinat:2008a}. The focal reducer has an aperture ratio of $f/3.9$, with a FoV of $\simeq$5.9$\times$5.9 arcmin in a 512$\times$512 pixels window with a pixel size of $\simeq$0.69 arcsec/pix. The angular resolution is limited by the seeing, typically ranging between  $\simeq$1.5 and $\simeq$3 arcsec. The interference order of the Fabry-Perot interferometer we used is 798 at \Ha~rest wavelength, giving a free spectral range (FSR) of 376 km s$^{-1}$, reaching a spectral resolution of $R\simeq$10,000 (velocity sampling $\simeq$10 km.s$^{-1}$ for a resolution of $\simeq$30 km.s$^{-1}$). The pattern of interference rings varies along the FoV when the separation of the plates is changed by applying different voltage on the piezoelectric actuators. Because of that, the wavelength varies according to the position on the FoV, so we must calibrate the \Ha~profiles for each pixel by comparing them with reference profiles given by a monochromatic source such as a neon lamp (6598.95 \AA) in order to give the same wavelength origin (or phase) to all the pixels and associate a specific wavelength to each spectral channel. Two calibration cubes are taken, one before and one after each galaxy exposure, to account for flexion conditions and temperature variations. 
\vspace{2.5mm}

The detector, an Image Photon Counting System (IPCS), has the significant advantage of zero readout noise and enables to scan the interferometer quickly, taking 10s per scanning step, typically 32 (for 146 galaxies) or 48 channels (for 6 galaxies) to cover the whole FSR, so that we can repeat the FSR scanning many times (or cycles) and then add up the successive exposures, enabling to greatly reduce the effects of variation in atmospheric transmission and to correct any eventual telescope drift in a very efficient manner. 
\vspace{2.5mm}

To cover the \Ha~line in the range 6565\AA~to 6635\AA, we use seven different interference filters having a typical FWHM $\sim$15\AA, allowing to scan the isolated \Ha~line of ionised hydrogen (6562.78\AA) and rejecting other emission lines (e.g. [NII] contaminating contribution) as well as the sky background emission (mainly OH lines). If necessary, we tilt by a couple of degrees the corresponding filter to blue shift the central wavelength of the transmitivity peak the closest to the systemic velocity ($V_{sys}$) of the corresponding galaxy, as indicated in Table \ref{tablelog}.
\vspace{2.5mm}

\subsection{Comparison With Other Integral Field Spectrographs}
\label{Comparison with other integral field spectrographs}

The gain in FoV is significant with respect to other seeing-limited IFS instruments; e.g. CALIFA \citep{Sanchez:2012}, MaNGA \citep{Bundy:2015}, SAMI \citep{Bryant:2015} and MUSE \citep{Bacon:2017} have a FoV$\sim$1.30$\times$1.30, 0.20$\times$0.20 or 0.53$\times$0.53, 0.25$\times$0.25, and 1.0$\times$1.0 arcmin$^2$ respectively.  The large FoV of our instrument enables to cover the galaxies of the HRS up to their optical limit $r_{opt}$\footnote{The definition we adopt in this work for $r_{opt}$ is given in section 4.1.7.} , ranging from 0.16 to 2.55 arcmin,  with a mean/median value of 1.08/0.99 arcmin. In terms of effective radius $r_{eff}$,  $r_{opt}$ spans from 1.47 to 4.26 $r_{eff}$, with a mean spatial coverage of 2.56$\pm$0.59 $r_{eff}$. The mean/median radius reached in HRS sample is nevertheless smaller than the mean/median one in the GHASP sample \citep{Epinat:2008b, Epinat:2008a}, this is due to the galaxy environment. Galaxies in the GHASP sample are mainly located in an isolated environment while 25.8\% galaxies of our HRS sample are situated at a projected distance smaller that 3.4 Mpc from the Virgo center and are furthermore potentially affected by environmental effects (ram pressure stripping, tidal stripping, quenching, etc.). 
\vspace{2.5mm}

In order to quantify the spatial coverage of our observations, for each galaxy, we compute the number of independent seeing elements, hereafter refereed as $N_{beams}$. $N_{beams}$ is the total number of pixels covering a galaxy with a \SNR\ per bin$>$7 (see next paragraph) divided by the number of pixels per seeing element, for each observation. $N_{beams}$ are listed in Table \ref{kinparam}, they range from 66 to 21153 with a mean/median value of 3138/2066. For comparison, the half of the FoV\footnote{In order to make a rough approximation, we estimate for those instruments that, in average, half of the spaxels gives a spectrum with an acceptable \SNR.} of CALIFA, MaNGA and SAMI gives $N_{beams}$ equal to 234, 75 and 33, respectively.
\vspace{2.5mm}

Within a narrow spectral domain around \Ha, the gain in spectral resolution R is also significant with respect to other IFS instruments: CALIFA (R$\sim$850-1650), MaNGA (R$\sim$2000), SAMI (from R$\sim$1700 to R$\sim$4500), MUSE (R$\sim$3000). This allows to study the kinematics down to the scale of single HII regions ($\sigma \simeq$30 km.s$^{-1}$, \citealp{Boselli:2010}).   

\subsection{Data reduction}
\label{reduc}
The Fabry-Perot data reduction procedure\footnote{The Fabry-Perot data reduction pipeline is composed by the IDL based program \textit{computeeverything} and \textit{reducWizard} interface, both of them available on websites \url{https://projets.lam.fr/projects/computeeverything} and 
\url{https://projets.lam.fr/projects/fpreducwizard}, respectively.} basically follows the same used in \cite{Daigle:2006fp} and in \cite{Epinat:2008a}. The reduction procedure consists of: 
\vspace{2.5mm}

i) Obtaining the phase map. Such a map is computed from the rings of calibration cubes. The consistency of both calibrations is checked and the phase is calculated by averaging the two calibration exposures. The phase map gives the reference scanning channel as origin wavelength for the spectrum of each pixel. 

ii) Sorting and merging the data. This is done by removing strong variations in atmospheric transmission by comparing the recorded successive individual frames of the observations. Those frames having an inconsistent flux compared with the median flux are deleted and replaced by frames of adjacent cycles automatically.

iii) Correcting from any telescope drift and/or instrumental flexures along the different successive frames by using reference stars or bright HII regions. According to subsection \ref{GHASPsetup}, the wavelength varies with the position across the FoV, thus this correction from telescope drift implies a phase correction for each cycle (i.e. each $\simeq$5 or 8 minutes for 32 or 48 scanning channels per cycle respectively). 

iv) Obtaining wavelength calibrated cubes. The latter is done by calibrating the interference rings of the galaxy observation by using the phase map. The same wavelength origin is assigned for each pixel.

v) Performing a one-spectral-element Hanning smoothing of the spectrum which preserves the flux.

vi) Subtracting night skylines. The data cube is divided into sky-dominated and galaxy-dominated regions. Then a sky cube is built by fitting the sky dominated regions (polynomial fitting of $2^{nd}$ order in our case) and then making an interpolation of the sky spectrum on the galaxy-dominated regions. Such a sky cube is finally subtracted.

vii) Eventually, suppressing ghosts due to reflection at the interfaces air/glass of the interferometer when necessary.  

viii) Computing the astrometry to find the correct World Coordinate System for each galaxy dataset, making use of XDSS $R$-band images. This is done using $KOORDS$ routine of $KARMA$\footnote{$KARMA$ tools package is available on website \url{https://www.atnf.csiro.au/computing/software/karma/}.} \citep{Gooch:1996} by a systematic comparison between field stars that are present in both the broad-band XDSS images and our continuum images.

ix) Data processing of adaptive spatial binning based on Voronoi tesselation on the data. The latter preserves the spatial resolution of bright regions while the weak emission of diffuse gas areas is recovered. Due to the small ratio between the number of channels containing continuum and the channels containing the emission line, the criterion used for GHASP data relies on an estimate of the \SNR~ as the square root of the flux in the line, which is a simple estimation of the Poisson noise at the line flux. We aim to recover a \SNR $\geq$ 7 per spatial bin. Two kind of maps are generated: the binned maps with $N$ independent bins, where all the pixels belonging to each bin are affected to the same value and the bin-centroid maps, for which only one pixel per bin corresponding bin-centroid position, contains a value (the other pixels being set to NaN).  Those later maps are the ones that are used in practice, in order to provide the same statistical weight to each bin and not to each pixel.

x) Computation of the different momenta maps (flux, radial velocity, radial velocity dispersion, continuum). 

xi) FSR corrections on velocity fields, since some \Ha~profiles present FSR overlaps causing velocity jumps at some pixels. 

xii) The semi-automatic cleaning system of the velocity fields in order to delimitate the outskirts of the galaxy where there is no more \Ha~emission. The latter is done by a continuity velocity process on the velocity field, based on a cut-off value between contiguous velocity bins (typically $\sim$30 km s$^{-1}$ for a velocity field with an amplitude of $\sim$300 km s$^{-1}$). Regions with too low emission and very large bins are also discarded and thus masked. 

xiii) Correction of the velocity dispersion from the Line Spread Function (LSF) broadening. This is done by a quadratic difference between the velocity dispersion values of the observation maps and the mean dispersion due to the instrumental contribution.

\subsection{Flux Calibration and \Ha~Profiles}
\label{capro}
During the OHP runs, in order to maximize the observing time on the galaxies themselves, we do not observe flux calibration sources but we use instead the high quality \Ha~photometry, available for the whole HRS sample \citep{Boselli:2015}. Thus we perform an indirect calibration of the total \Ha~flux for the 152 datacubes in a similar way as in \cite{Epinat:2008b}, but using in our case the calibrated fluxes from \cite{Boselli:2015}. We correct their \Ha~+ [NII] fluxes for [NII] contamination (6584 and 6548 \AA) using the [NII]/\Ha~ratio derived from spectroscopic observations \citep{Boselli:2013}; for galaxies where there is no available [NII]/\Ha~ratio, we calculate it according to the $[NII]/H\alpha \;vs\; M_{star}$ relation in the $B$-band given in \cite{Boselli2009}. 
\vspace{2.5mm}

\begin{figure}
\begin{center}
\includegraphics[width=\columnwidth]{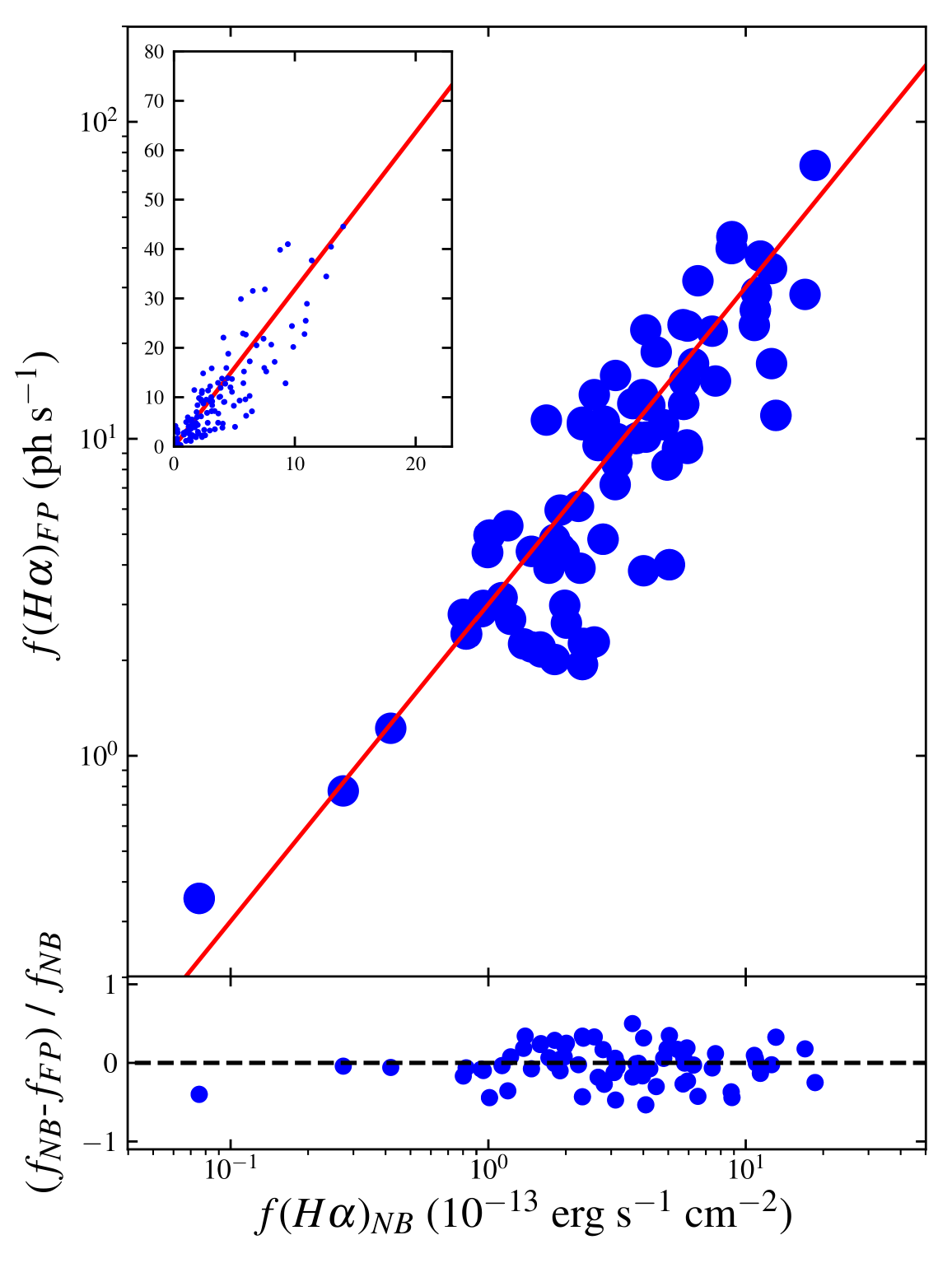}
\end{center}
\caption{Top panel (and top-left insert): \Ha~integrated fluxes measured by GHASP compared with \Ha~integrated fluxes from \cite{Boselli:2015}; the solid red line represents the OLS bisector linear regression on the data from which results our calibration; the insert at the top-left shows the same calibration plot but using a linear scale. Bottom panel: normalised residual values, where $f_{NB}$ are fluxes from \cite{Boselli:2015} and $f_{FP}$ are GHASP calibrated fluxes. } 
\label{fig:calib}
\end{figure}
 \vspace{2.5mm}

For each galaxy, the total \Ha~emission is computed by integrating the flux for each spatial element of the datacubes as described in \cite{Epinat:2008b}, taking into account:
 \vspace{2.5mm}
 
  i) the response of the instrument, 
  
  ii) the aperture of the instrument,
  
  iii) the exposure time (given in Table \ref{tablelog}),
  
  iv) using the velocity field to disentangle the FSR overlaps when necessary.
  
  v) the shift of the interference filter spectral range due to the tilt (if any) and the temperature,
  
  vi) the subtraction of the continuum contribution. 
  \vspace{2.5mm}
  
To minimize the foreground sky contamination, we only consider the spatial elements not masked in the final momenta maps (shown in Appendix \ref{maps}). The integrated \Ha~profiles per galaxy are shown in Appendix \ref{fig:profiless}. Figure \ref{fig:calib} shows the comparison between the Fabry-Perot fluxes and those from \cite{Boselli:2015}. The response of the GHASP camera being linear and since we are not comparing quantities having the same units, we use an Ordinary Least Squares (OLS) bisector linear regression, represented on the figure with a solid red line, to calibrate the Fabry-Perot H$\alpha$ fluxes. The fit is forced to pass through the origin for obvious physical reasons; the calibration coefficient is $1$ $ph\; s^{-1}=0.50 \pm 0.05\times 10^{-13}$ $erg$ $cm^{-2}$ $s^{-1}$. The integrated fluxes estimated for each galaxy using our calibration are given in Table \ref{kinparam}. 
 \vspace{2.5mm}

We measure the sensitivity of our Fabry-Perot dataset using two methods. The first one consists in computing the detection limits from the isophotes of VESTIGE \Ha~narrow band images \citep{Boselli:VESTIGE} for galaxies in common with the HRS, while the second method consists in calculating such limits directly on all our \Ha~monochromatic images using our estimated calibration. We find for each case a surface brightness detection limits of $\sim2.5 \pm 0.2 \times 10^{-17}$ erg s$^{-1}$ cm$^{-2}$ arcsec$^{-2}$ and $\sim 2.3 \pm 0.2 \times 10^{-17}$ erg s$^{-1}$ cm$^{-2}$ arcsec$^{-2}$  for a typical 2 hours exposure time, thus comparable to the typical sensitivity of narrow-band imaging. This sensitivity is close to that reached by \cite{Epinat:2008b} using similar integration times and predicted by Fig. 2 in \cite{Gach:2002} for a \SNR~ between 1 and 2.

\section{Data Analysis}
\label{dataanalysis}

\subsection{Kinematical Models and Rotation Curves}
\subsubsection{Kinematical Models}
\label{zhaocour}
\RCs~are computed from the velocity fields following the method described in \cite{Epinat:2008a}, which is a synthesis adapted to \Ha~data between the angular sector method, tilted-ring models and the fitting method used by \cite{BarnesSelwood:2003}. In 21cm-HI discs where warps are frequently observed at a distance from the center larger than the optical radius, the position angle ($PA$) and galaxy inclination ($i$) might vary as a function of the galacto-centric radius. Within the optical radius, the situation is different, as in \cite{Epinat:2008a}, we do not allow the $PA$ nor $i$ to vary because these variations are slight throughout the optical disc plane and due mainly to non-circular motions and velocity dispersion that behave as oscillations around a median value.
\vspace{2.5mm}

For each bin situated in the frame of the galactic plane of a disc, assuming that at first order the expansion velocity component and vertical motions are negligible, the velocity vector projected on the line-of-sight can be described as follows:
\\
\begin{equation}
V_{obs}$($r$)$\;$=$\;V_{sys}\;$+$\;V_{rot}$($r$)$ \; cos \, \theta \; sin \, \textit{i} 
\label{vobs}
\end{equation}
\\
where V$_{sys}$ is the systemic velocity of the galaxy, $i$ is the inclination, and $r$ and $\theta$ are the polar coordinates in the plane of the galaxy, both measured from the $PA$, $i$ and the kinematical center ($\alpha$, $\delta$) of the galaxy. V$_{rot}$($r$) is modeled assuming a four-parameters modified Zhao function \citep{Epinat:2008b} on the whole velocity field:
\\
\begin{equation}
V_{rot}(r)\;=\;v_t \frac{({r}/{r_t})^g} {1+(r/r_t)^a}
\label{vrot}
\end{equation}
\\
where $v_t$ is the effective "turnover" velocity, $r_t$ is the transition radius between the rising and flat part of the \rc~$g$ and $a$ are coefficients that parametrise the sharpness of the turnover. This leads  to a model of nine free parameters: four projection parameters ($PA$, $i$, galaxy center $\alpha$ and $\delta$) and five kinematical parameters ($V_{sys}$, $v_t$, $r_t$, $g$, $a$). The model parameters are obtained with a $\chi^{2}$ minimization based on the Levernberg-Marquardt method \citep{Press92numericalrecipes}, computing an iterative 3.5 $\sigma$ clipping on the observed bin-centroid velocity field. The number of unknowns will depend on how many initial guess parameters are fixed or let as free for Eqs. \ref{vobs} and \ref{vrot}: in our case, the galaxy center is carefully computed from the continuum image and always fixed; $PA$ and/or $i$ are taken from the literature and generally let as free or fixed in some cases (see subsection \ref{kinepar}). These initial guesses of projection parameters are used to fit the modified Zhao function to the velocity field in order to obtain the $V_{rot}$ starting parameters $v_t$, $r_t$, $g$ and $a$, which are then let always as free.
\vspace{2.5mm}

\subsubsection{Residual Velocity Fields}

 Since the kinematical models are calculated with a prior assumption that circular motions are dominant and those due to non-axisymetric perturbations are not part of a large scale pattern, the residual velocity field computed by subtracting the modeled velocity field to the original one reflects well the deviation from circular velocities. Since our model is simple, the resulting statistical uncertainties of kinematical parameters are too small. For that reason, the calculation of uncertainties is done by computing the power spectrum of the residual velocity fields and then considering several random phases, in order to compute residual fields that present the same kind of structure but placed differently. Through a Monte-Carlo method, we estimate robustly the standard deviations of the kinematical parameters over hundreds of simulated velocity fields (see \citealp{Epinat:2008b} for details). For consistency with \cite{Epinat:2008b} and with section 4.1.3, the residual velocities have been computed using the bin centroid maps.

\begin{figure}
\begin{center}
\includegraphics[width=\columnwidth]{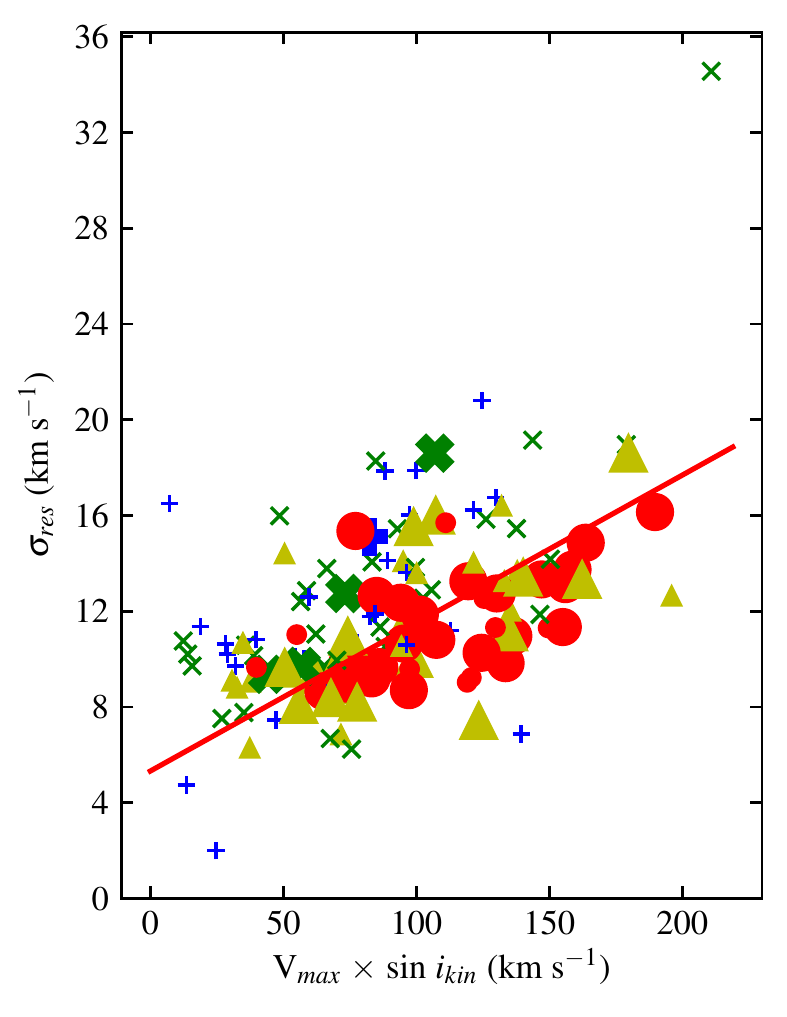}
\caption{Standard deviation of each residual velocity field as a function of the mean amplitude of the velocity field. The colors correspond to the different quality Flags: red dots are galaxies with Flag ``1", yellow triangles are galaxies with Flag ``2", green ``x" symbols are galaxies with Flag ``3" and blue ``+" symbols are galaxies with Flag ``4". The size of the symbols corresponds to the complementary  Flag``A" (big symbols) and ``B" (small symbols). The solid red line represents the bisector linear regression on the Flag ``A" data.}
\label{velrespaper}
\end{center}
\end{figure}

\begin{figure*}
\begin{center}
\includegraphics[width=13.9cm]{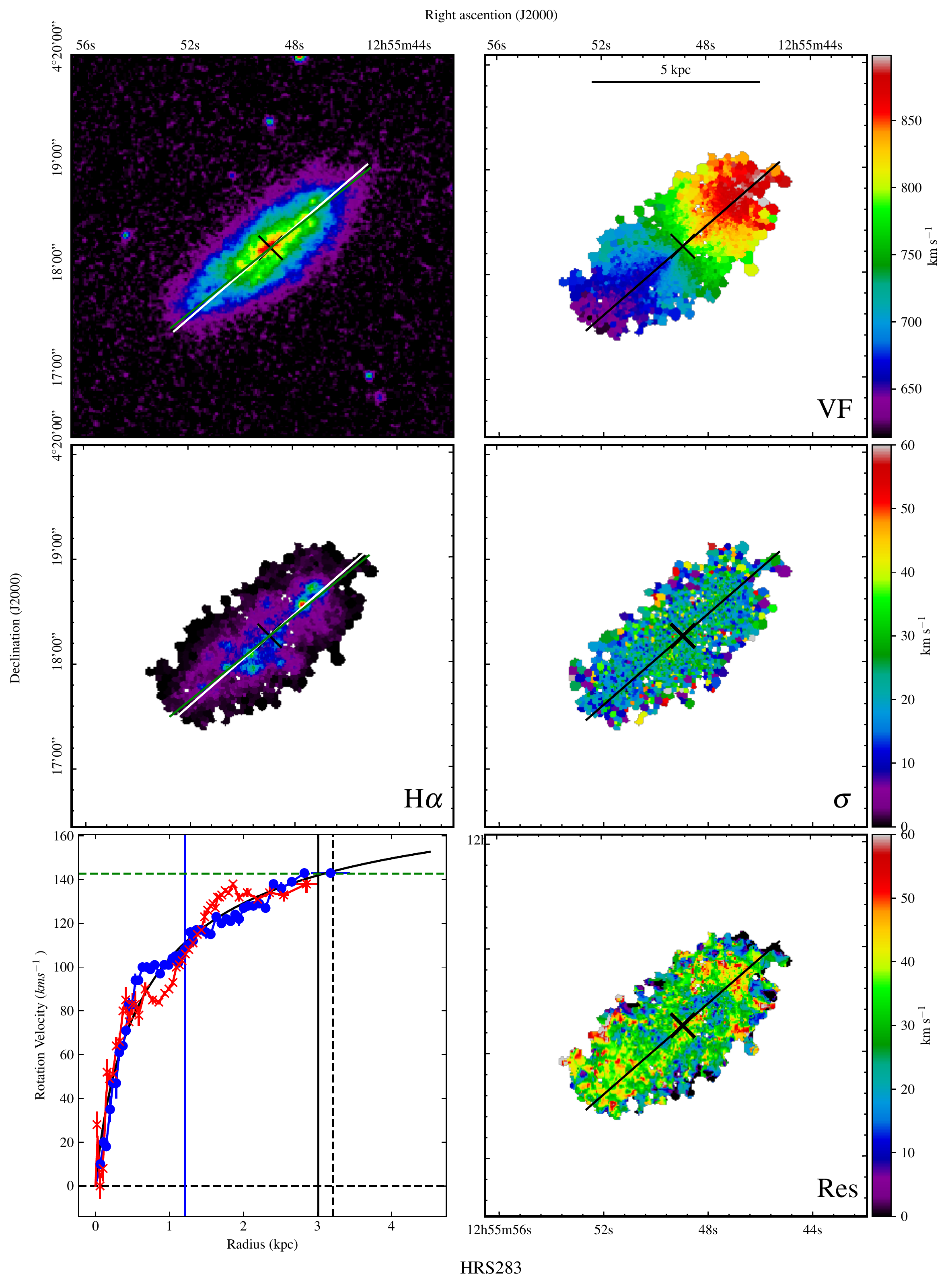}
\end{center}
\caption{Example of the products derived from Fabry-Perot observations. Top left: XDSS $R$-band image. Top right: \Ha~velocity field. Middle left: \Ha~monochromatic image. Middle right: \Ha~velocity dispersion field. Bottom left: \rc. Bottom right: residual velocity field.The black cross is the kinematical center. In the maps, the green line is the morphological major axis, while the black/white line is the estimated kinematical major axis, its length represents the $2 \times r_{opt}$. On the \rc, blue dots indicate the approaching side while red  crosses the receding side, the black solid vertical line represents the $r_{opt}$, and the black dashed line represents the $r_{25}$; the blue solid line the $r_{eff}$ and the green horizontal line the \Hi~$V_{max}$; the solid black curve is the Courteau function best fit to the \rc. The maps and plots for the all the galaxies of the HRS are online material, available at the CDS, on the HRS dedicated database HeDAM (\url{https://hedam.lam.fr/}), and on the Fabry-Perot database (\url{https://cesam.lam.fr/fabryperot }).}
\label{fig:hrs283}
\end{figure*}

\subsubsection{Rotation Curves}
Once the best set of parameters is derived using the method previously described, the \rc~is extracted from the velocity field. Since the velocity field is decomposed into rings, we consider the bin centroid velocity fields in order to avoid any artificial oversampling of rings. We use $n$ uncorrelated velocity bins per ring (or annulus) to provide each rotation velocity; $n=25$ optimizes the compromise between the higher \SNR~and the lower spatial resolution per ring \citep{Epinat:2008b}.  \RCs~are computed without correction from the non-circular motions along the disc. 

\subsubsection{Quality Flags on the Rotation Curves}
With the purpose of quantifying the quality of each \rc, we attribute a quality Flag using an automatic classification method extensively described in Appendix \ref{flags}, that goes from Flag ``1" (accurate estimation of the \rc) to Flag ``4" (poor estimation of the \rc), and attributing a supplementary Flag ``A" and Flag ``B", the latter corresponding to those peculiar cases leading to a non-realistic kinematical fitting such as the presence of a bar, asymmetries, high or low inclination, etc.   

\vspace{2.5mm}

For 11 cases over 152 (HRS 32, 35, 104, 136, 164, 184, 195, 249, 282, 291 and 300), characterised by a poor \SNR, the model does not match the rotation of the galaxy because of lack of \Ha~emission, thus no model is computed, neither residual velocity field nor \rc~can be plotted, and no kinematical parameters can be calculated.


\subsubsection{Quality Checks of the Residual Velocity Fields}

For each residual velocity field, the average value ($\overline{ Res}$) and the standard deviation of all the bins ($\sigma_{res}$) are computed (see Table \ref{kinparam}). Figure \ref{velrespaper} displays, for the whole dataset, the correlation between $\sigma_{res}$ (with a mean value $\sim$12.6 $\pm$ 2.7\kms, which is similar to the velocity of 13 \kms, computed in \citealp{Epinat:2008b}) and the mean amplitude of the velocity field (which is the maximum rotation velocity $V_{max}$\footnote{Defined in paragraph \ref{Maximum Rotation Velocity}}   multiplied by the $sin\;i_{kin}$). The bisector linear regression traces a trend which indicates that, in general, high velocity amplitude values are correlated with high $\sigma_{res}$ values. Nevertheless, many points for which $V_{max} \times sin\;i_{kin}$ $<$ $150$ km s$^{-1}$ are situated above the regression line, corresponding to galaxies with quality Flags ``3-B" and ``4-B", so that the distribution of such points with respect to the Y-axis seems to be related to the quality of the data and not to the line-of-sight velocity, thus resulting in a high $\sigma_{res}$ value (i.e. the outlier with $\sigma_{res}\sim$35 km s$^{-1}$ is the galaxy HRS 59, an almost edge-on object with a chaotic \rc~and flagged as ``3-B"); these specific cases are extensively described in Appendix \ref{notes}. The correlation is in agreement with \cite{Epinat:2008b} and \cite{Epinat:2008a}, and it does not depend on the galaxy type \citep{Epinat:2008b}. 

\subsubsection{Maximum Rotation Velocity}
\label{Maximum Rotation Velocity}

In order to compute our \Ha~$V_{max}$ without being perturbed by local variations in the \rcs~due for instance to crossing spiral arms, we fit the \rc~with analytic functions, between $r = 0$ and $r_{RC}$, the radius corresponding to the last ring of the \rc.  For this purpose, we select the Courteau profile \citep{Courteau:1997}. The original Courteau function describes the line-of-sight velocities from long-slit observations; because we are working with deprojected data, we used the following modified Courteau function:
\begin{equation}
v ( r ) =  v_c  \frac { \left(1+ \frac{r_t}{r}\right)^{\beta}}  {  \left(1+ \left(\frac{r_t}{r}\right)^{\gamma}\right)^{1/\gamma}}
\label{vcourteau}
\end{equation}
where the four free parameters are $v_c$, the asymptotic velocity; $r_t$ the transition radius between the rising and flat part of the \rc ; $\beta$, the drop-off or steady rise of the outer part of the \rc\ and $\gamma$, the sharpness of turnover beyond the $r_t$ radius. Because it does not affect much the result and decrease by one the degree of freedom, we used $\beta=0$ as suggested by \cite{Courteau:1997}. Furthermore, $V_{max}$ is defined by the maximum rotation velocity reached by the fitted modified Courteau profile within the optical radius.
The best fit is represented in Fig. \ref{fig:hrs283} and Figures in Appendix \ref{maps} with a black solid curve that goes through the velocity points in each \rc.
\\

It is important to mention that the choice of a given fitting function may introduce systematic biases in the estimation of $V_{max}$,  particularly when extrapolations are necessary where data do not extent further enough at large radii. For instance, modified Zhao and Courteau profiles have four free parameters.  A fair comparison of both functions should be done in reducing the degeneracy but in keeping the same number of free parameters, we furthermore set it at three; we fixed the parameters that have the weaker influence on the shape of the \rcs, namely $g=0$, for the modified Zhao function and $\beta=0$ for the modified Courteau function.  We conclude that this 3-parameters Zhao function is more sensitive to local oscillations that the 3-parameters Courteau one.  In other words, when the modified Zhao function is used to extrapolate the \rcs~outside the last observing radius, $V_{max}$ tends to follow the trend of the last data points rather than the general trend of the plateaus and, at the opposite, the modified Courteau function averages the mean trend of the whole outer \rcs.  In short, the modified Courteau function tends to extrapolate a flatter \rc\ and flatten local oscillation than the modified Zhao function and this have an impact on $V_{max}$. We choose the Courteau function because it provides a more conservative solution, as well as and for consistency with previous works on the GHASP sample.
\\

Alternatively to Zhao and Courteau profiles used in this work, other analytic functions are used in the literature to fit \rcs~like the empirical Polyex function (\citealp{Giovanelli:2002}; \citealp{Catinella:2005}; \citealp{Catinella:2006} and \citealp{Masters:2006}). Polyex profiles fit well a large variety of \rc\ shapes (94\% of the cases, including those declining at large radii, according to \citealp{Catinella:2005}).  A comparison between Polyex and the other fitting expressions is beyond the scope of this paper.


\subsubsection{Presentation of the Data}

We present in the Appendix \ref{maps} (online material at the CDS), for each galaxy (as in Fig. \ref{fig:hrs283}), six frames per Figure containing the XDSS image in the $R$-band, the \Ha~line-of-sight velocity field, the \Ha~monochromatic image (free of continuum, [NII] and night skylines contributions), the line-of-sight velocity dispersion field, the \rc~of the galaxy and the \Ha~residual velocity field. The white/black cross indicates the calculated kinematical center, while the solid green line traces the optical radius (called hereafter $r_{opt}$. For consistency with previous HRS works we choose $r_{opt}=r_{24}$, which is in fact $D_{24}/2$ in $r$-band, taken from \citealp{Cortese:2012}.  Nevertheless $r_{24}$ is not available for 5 galaxies, thus we used $r_{25}$ which is $D_{25}/2$ in $B$-band, taken from \citealp{Boselli:2010}) and convert it to $r_{24}$ following the relation $r_{24}=0.920\pm0.01$ $r_{25}$, obtained by comparing both radii for the whole HRS sample and fitting a bisector regression. The black/white solid line traces the kinematical major-axis deduced from our velocity field analysis; the $PA$ is the one calculated from our kinematical models (see subsection \ref{kinepar}).  In the \rc~plots (bottom left panel in Fig. \ref{fig:hrs283} and all Figures in Appendix \ref{maps}), both sides are superimposed in the same quadrant, using red crosses for the receding side and blue dots for the approaching side.  The solid vertical black line represents the $r_{opt}$; the black dashed line represents the $r_{25}$ in $B$-band, in order to compare the extent between $r_{24}$ and $r_{25}$. When available in \cite{Cortese:2012}, the effective radius $r_{eff}$ is plotted with a solid blue vertical line. If available in \cite{Boselli:2015}, the \Hi~$V_{max}$ is plotted as a horizontal green dashed line. We present the tables of the \rcs~in Appendix \ref{tables}. When no kinematical model can be derived, we plot instead the photometrical center and the $PA$ taken from \cite{Cortese:2012} without any residual velocity field nor \rc.


\subsection{Kinematical Projection Parameters}
\label{kinepar}

\begin{figure}
\begin{center}
\includegraphics[width=8.1cm]{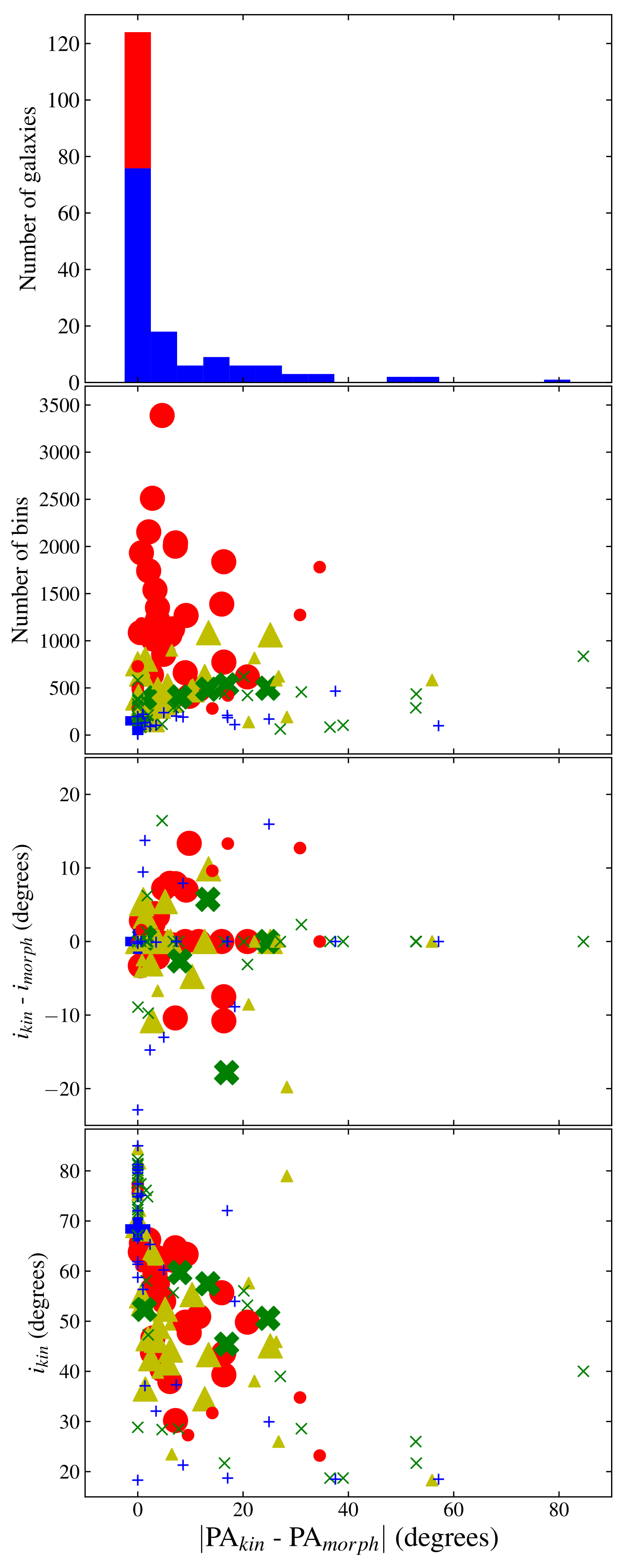}
\caption{Top panel: Histogram of the  the difference between the kinematical and morphological PA, where the red part of the histogram represent galaxies for which the kinematical $PA$ has been fixed to the morphological one. The X-axis in the last three panels represents the difference between the kinematical and morphological $PA$, as a function of: (top middle panel) the number of bins per velocity field, (bottom middle panel) the difference between the kinematical and morphological inclination and (bottom panel) the kinematical inclination. The conventions of color, symbol and marker size are as in Fig. \ref{velrespaper}. }
\label{papaper}
\end{center}
\end{figure}

As we saw in subsection \ref{zhaocour}, we need a set of initial guess parameters in order to compute our kinematical models. Such a set is composed either by parameters taken from high quality data available in the literature for the HRS (morphological measurements extracted from surface brightness photometry $PA_{morph}$, $i_{morph}$), systemic velocity from long-slit spectroscopy $V_{sys}$, or carefully calculated by us using morphological and kinematic parameters. The galaxy center $\alpha$ and $\delta$ is computed either as the nucleus in the continuum image or as the kinematical center of the velocity field. Table \ref{kinparam} shows the morphological parameters $PA_{morph}$ and ellipticity $\epsilon$ used to compute $i_{morph}$ for each galaxy, taken from \cite{Cortese:2012}. We computed $i_{morph}$ from the $\epsilon$ value following the procedure as in \cite{Masters:2010}:
\\
\begin{equation}
cos^2 \, \textit{$i_{morph}$} = \frac{ (1-\epsilon)^2 \; - \; q^2}{1 \;  - \;  q^2} 
\label{incmasters}
\end{equation}
\\
where $q$ depends on the galaxy type (\citealp{Haynes:1984}; \citealp{Masters:2010}).
 \vspace{2.5mm}

We use the initial guesses of projection parameters to derive a \rc~on which is fitted the modified Zhao function to obtain starting parameters $v_t$, $r_t$, $g$ and $a$. The different kinematical projection parameters computed by our models are accurately determined from different symmetry properties of the line-of-sight velocity field: typical accuracy reached by our method are sub-seeing for the coordinates of the galaxy center,  $\sim$0.5 arcsec along the major axis and $\sim$1 arcsec along the minor one; $\sim$3 km s$^{-1}$ for $V_{sys}$; $\sim$2$^{\circ}$ for $PA$ and 5$^{\circ}$-10$^{\circ}$ for kinematical inclination $i_{kin}$. We control those parameters using the method described in  \cite{Warner:1979}, \cite{Vanderkruit:1978} and \cite{Vanderkruit:1990}, which is based on the fact that residual velocity fields present characteristic patterns when one of several of those parameters is/are incorrectly determinated. Non-circular motions, mainly due to bar structures, spiral arms, HII complexes, are not considered in the kinematical models since we expect that, except in very low mass dwarf galaxies, such motions are low with respect to rotational ones in rotating systems. The modeled velocity is a purely rotating disk that optimally fits the observed velocity fields.  Non-circular motions are thus embedded in the residual velocity field, which is the difference between the observed velocity and the model. For the HRS, the mean velocity dispersion of the residual velocity field is 12.55$\pm$2.60 km s$^{-1}$; such motions may become more important in perturbed galaxies and systems in interaction, which rare in the HRS sample ($<$11\%, \citealp{Boselli:2010}). However, non-circular motions are taken into account in the determination of the uncertainties \citep{Epinat:2008b, Epinat:2008a}. 

\vspace{2.5mm}

Table \ref{kinparam} gives also the kinematical output parameters of the models and the best reduced $\chi^{2}_{red}$ giving the goodness of fit of the Levernberg-Marquardt method. In general, the $PA$ and $i$ values of the galaxy are left as free parameters when computing the best fitting. Nevertheless, for 42 galaxies (out of which 37 are Flag ``B" objects), we fix the kinematical $PA$ to the morphological value during the model computation because of the following reasons:

 \vspace{2.5mm}

 i) $i$  $\geq$ 70$^{\circ}$ (39 objects), because galaxies with high inclination have a low spatial coverage along their minor axis; and some of them have a spatial coverage actually too small (i.e. less than 150 velocity bins),
 
 ii) poor \SNR~(3 galaxies), because their kinematical model easily leads to non-realistic kinematical parameters.
\vspace{2.5mm}

 On the other hand, for 78 galaxies, we fix the kinematical $i$ to the morphological value during the fit process because:
 \vspace{2.5mm}

 i) $i$  $\geq$ 70$^{\circ}$ or $i$  $\leq$ 30$^{\circ}$ (63 objects), because an underestimation of the inclination leads to a overestimate on velocities and vice versa \citep{Tully:2000}.  
 
  ii) 15 galaxies showing velocity fields dominated by non-circular features.
 
 As already mentioned, some of galaxies relevant from those two later cases have in addition a spatial coverage or/and a \SNR~too low.
 
 \vspace{2.5mm}
The details about those special cases are given in the Appendix \ref{notes}, and the parameters $PA_{kin}$ or $i_{kin}$ are flagged with an asterisk in Table \ref{kinparam} when fixed during the model computation. 
\vspace{2.5mm} 

Figure \ref{papaper} shows in the top panel the histogram of the $\vert$$PA_{kin}$ - $PA_{morph}$$\vert$ distribution. For 42 galaxies over 140, the $PA_{kin}$ is fixed to the morphological one (red part of the histogram). Ignoring such galaxies, the peak of the histogram remains at $\vert$$PA_{kin}$ - $PA_{morph}$$\vert$ $= 0$ because the bin width centered at zero includes 44 galaxies with $\vert$$PA_{kin}$ - $PA_{morph}$$\vert$ $\leq$ 3$^{\circ}$; in fact, $\vert$$PA_{kin}$ - $PA_{morph}$$\vert$ $\leq$ $10$$^{\circ}$ for 105 galaxies over 140, and the median of the distribution is $2.47$$^{\circ}$  and the dispersion 7.85$^{\circ}$. Figure \ref{papaper} also displays the comparison of the difference between the kinematical $PA_{kin}$ and the morphological $PA_{morph}$ as a function of several parameters:
\vspace{2.5mm}

i) Number of bins per velocity field. The $\vert$$PA_{kin}$ - $PA_{morph}$$\vert$ values, as well as the quality flag, depend on the number of bins, which are related to the spatial coverage and \SNR~of the galaxy. Objects with a low number of bins are usually Flag ``B" for which $PA$ is not fixed, showing a $\vert$$PA_{kin}$ - $PA_{morph}$$\vert$ $>$ $25$$^{\circ}$, representing 11.4$\%$ of the whole sample. Cases for which $\vert$$PA_{kin}$ - $PA_{morph}$$\vert$ $>$ 50$^{\circ}$ belong to galaxies which are almost face-on with faint emission (HRS 68, HRS 154, HRS 255) or emission concentrated in the inner part (HRS 19, HRS 256); the details about these cases with Flag ``3-B" and ``4-B" are extensively described in Appendix \ref{notes}; nevertheless their velocity fields are good enough to allow the determination of $PA_{kin}$ values. 

ii) Difference between the kinematical and the morphological inclination ($i_{kin} - i_{morph}$). In this plot, if we do not take into account the data for which $i_{kin}$ has been fixed to the morphological one, we see that a possible under or over-estimation of the inclination does not depend on $\vert$$PA_{kin}$ - $PA_{morph}$$\vert$. Despite the fact that the median difference between the kinematical and morphological inclinations is almost equal to zero ($-0.2$$^{\circ}$), meaning that we probably do not introduce statistical biais to compute the inclination regardless the method used (morphological or kinematical), the dispersion of the $i_{kin} - i_{morph}$ distribution is quite large ($8.3$$^{\circ}$). 

iii) The kinematical inclination. $\vert$$PA_{kin}$ - $PA_{morph}$$\vert$ becomes larger as $i_{kin}$ decreases, showing good agreement with \cite{Epinat:2008b, Epinat:2008a}.
\vspace{2.5mm}

We also compare $\vert$$PA_{kin}$ - $PA_{morph}$$\vert$ as a function of the Galaxy Type and the Distance to the Virgo Cluster (A or B clouds), but we do not find any correlation with these two parameters and we do not show those plots here.
\vspace{2.5mm}

\begin{figure}
\begin{center}
\includegraphics[width=\columnwidth]{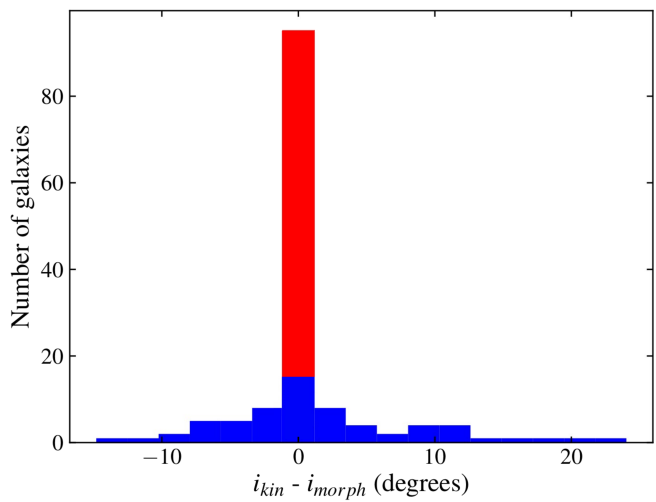}
\caption{Histogram of the difference between the kinematical and morphological inclination; the red part of the histogram represents the galaxies for which the kinematical inclination was fixed to the morphological one.}
\label{incpaper}
\end{center}
\end{figure}

Figure \ref{incpaper} displays the histogram of the variation between the morphological and the kinematical inclinations. For 78 galaxies over 140, $i_{kin}$ is fixed to the morphological value (red part of the histogram).  In order to test whether there was a correlation or not between the galaxy inclination (either $i_{kin}$ or $i_{morph}$) and the maximum rotational velocity $V_{max}$, we perform several OLS bisector regressions comparing both $i_{kin}$ and $i_{morph}$ values versus $V_{max}$ (not plotted here). We do not observe any significant relation, on the whole sample neither any of the subsamples as defined by their different quality flags.
\vspace{2.5mm}

Bars and Bulges are usually bright with respect to the disks, nevertheless, bulges contain a low content of warm gas while bars could be warm gas-poor or -rich.  In case of low gas content, Voronoi tessellations minimise the impact of the non-rotating bulge and of the bar on the determination of the disk parameters. In case of high gas content, Voronoi techniques do not spread out the impact of non-circular motions to outer regions. On the other hand, despite the accuracy of the $PA$ and $i$ computation, these two parameters may be inadequate in presence of multiple non-axisymmetric structures in the kinematics or/and in the morphology.  For instance, in the case of a strong bar, the determination of the $i$ and the $PA$ could vary with the radius within the bar, nevertheless we take this fact into account by measuring the $i$ and $PA$ outside the bar since our \rcs~are extended enough. Indeed, the bar length reaches the outskirt of the warm disk in only 2/152 galaxies. In addition, the estimation of $V_{max}$ is not significantly biased because of the presence of a bar; indeed, a peak followed by a decrease on the \rc\ is usually linked to the presence of a bar (i.e. galaxy HRS 60) and furthermore identified as such; otherwise the effect of the bar modify the inner slope of the \rc\ and possibly shift $V_{max}$ at a different radius (i.e. galaxy HRS 287), as described by \cite{Dicaire:2008}, \cite{Randriamampandry:2015} and \cite{Korsaga:2019}. Finally, we use a model to fit $V_{max}$ rather than fitting the raw \rc, precisely to average local non-circular motions. The position of the galaxy center is usually weakly affected by non-circular motions. In conclusion, the methods we use minimise the effects dues to non-circular motions.
\vspace{2.5mm}

\section{Discussion: Dynamical Masses}
\label{sanity}

Several models, simulations and multifrequency observations consistently indicate the mass as the principal driver of galaxy evolution (i.e. \citealp{Cowie:1996}, 
\citealp{Gavazzi:1996}, \citealp{Boselli:2001} for observations;  \citealp{Navarro:1996}, \citealp{Boissier:2003}, \citealp{Dutton:2012}, \citealp{Moster:2013}, 
\citealp{Dutton:2014} for models and simulations). The mass of galaxies is generally measured through the stellar mass which is 
derived from optical and near-IR imaging data. Nevertheless other components are present. These include dark matter, which is generally dominant at large radius, the mass of the different gas phases (atomic and molecular hydrogen, helium, ionised and hot gas, metals), and the mass of dust. 
The unique dataset available for the HRS allows us to directly measure most of the baryonic components: \Hi~and CO data 
are available for most of the star forming galaxies of the sample \citep{Boselli:2014}, the dust mass has been estimated from SED fitting using 
the far-IR data by \cite{Ciesla:2014}, and metallicities have been derived using integrated long slit spectroscopy by \cite{Hughes:2013}.
The kinematical data gathered in this work can be used to roughly estimate the total dynamical mass of a system using relation (\ref{dynleque}) 
\vspace{2.5mm}

In this section, we use the HRS to study the main scaling relations 
between the dynamical mass, the baryonic mass, and the star formation rate generally used to constrain models of galaxy formation and evolution.
The unique multifrequency coverage, which allows at the same time the accurate determination of the different baryonic components, of the star formation activity, 
and of the dynamical mass, combined with the sample definition (see sect. 2) and the statistics ($\sim$200 objects spanning a wide range 
in morphological type and mass), make the HRS the best sample available in the literature for this purpose.
\vspace{2.5mm}

To study the dynamical properties of the sample in the context of galaxy evolution, we proceed as follows:
\vspace{2.5mm}

i) Since we lack of Fabry-Perot data for 26.4\% of the HRS, we first test whether the \Hi~kinematical data can be used to derive the total dynamical mass of the missing galaxies without introducing any systematic bias.

ii) We check the statistical significance of our sample by comparing the Tully-Fisher relations to those derived in other representative samples generally 
used in the literature.  

iii) We then study the stellar and baryonic Tully-Fisher relations, the relation between the baryonic and dynamical mass, and finally we derive 
the main sequence for the first time using the dynamical mass \footnote{For a fair comparison, all references in the literature are scaled 
to $H_0$ $=$ $70$ km s$^{-1}$ Mpc$^{-1}$ used in this work, and stellar masses are corrected by 
a factor of $0.061\,dex$ (\citealp{Bell:2003}; \citealp{Gallazzi:2008}) consistently with the Chabrier IMF used in this work.}.

\subsection{$V_{max, HI}$ versus $V_{max, H\alpha}$}

\label{hi}
\vspace{2.5mm}
\begin{figure}
\begin{center}
\includegraphics[width=\columnwidth]{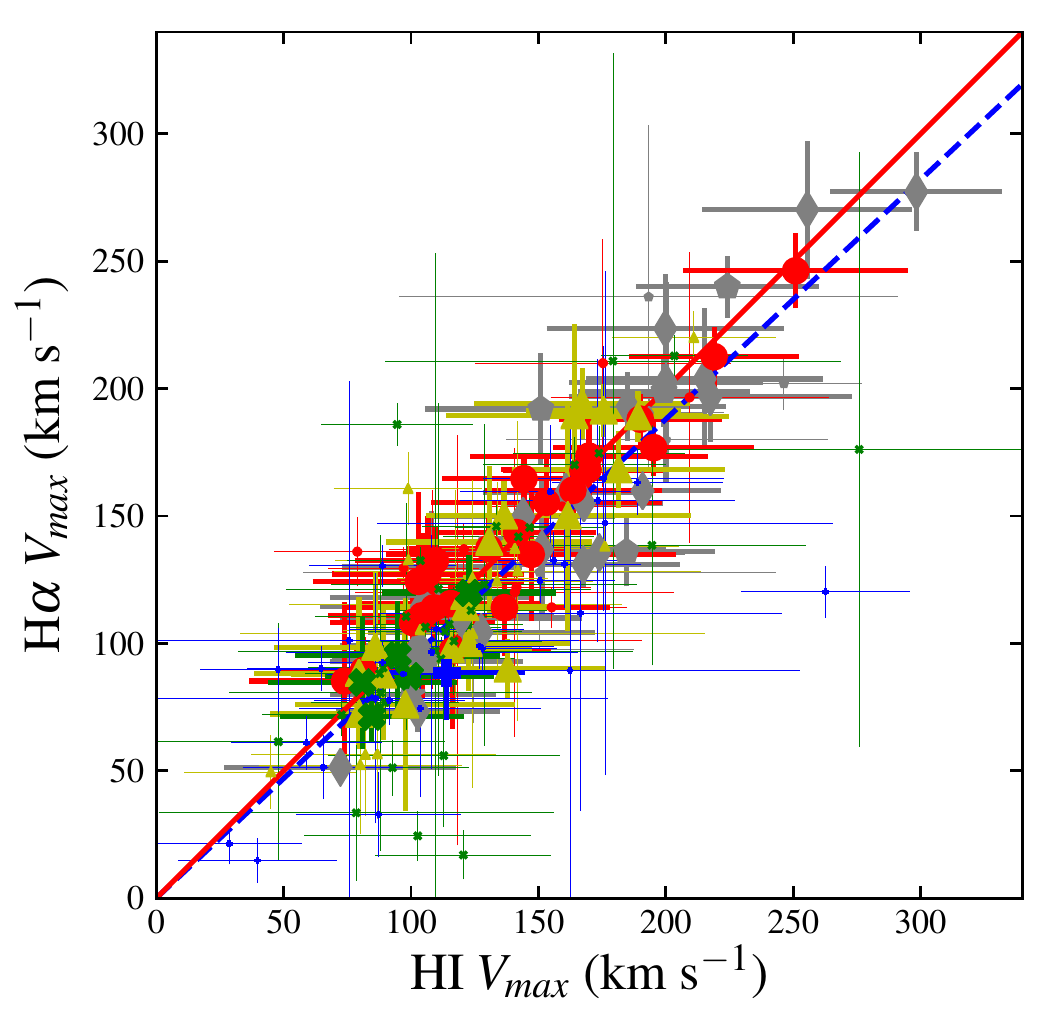}
\end{center}
\caption{Relation between $V_{max, HI}$ and $V_{max, H\alpha}$. The red solid and blue dashed lines represent the OSL bisector regression applied to a) the whole Fabry-Perot sample, and b) ignoring galaxies with Flag ``B''. Colors, symbols and markers are as in Fig. \ref{velrespaper}. Fabry-Perot data from the literature (grey symbols) are classified as: stars for galaxies belonging to the Loose Groups Survey, rhombuses for galaxies belonging to 
the Virgo Survey, and pentagons for galaxies belonging to the GHASP survey.}
\label{fig:vmaxhivmaxha}
\end{figure}

\cite{Boselli:2014} published a homogenised catalog of \Hi~data for the whole HRS. Their work provides line widths $W_{HI}$ measured at 50\% of the peak flux per galaxy. We compute $V_{max, HI}$ assuming that $W_{HI} = 2\,V_{max, HI}\,sin(i)$; where $V_{max, HI}$ is the maximal \Hi~rotation velocity. On the other hand,  we compute the maximal \ha\ rotation velocity $V_{max, H\alpha}$ from \rcs~derived from 2D velocity fields, thus intrinsically calculated in a more accurate way. 

\vspace{2.5mm}

Uncertainties on $V_{max, H\alpha}$ are calculated as the quadratic combination between the $i_{kin}$ error and the median dispersion of the \rc~rings beyond $r_t$, as in \cite{Epinat:2008a}. This is a much more realistic estimate of the uncertainty on the maximal rotational velocity than the one derived from the integrated \Hi~line profile $W_{HI}$, which is very low ($<$ 5 km s$^{-1}$). For that reason, in order to have comparable errors for both quantities, we derive $V_{max, HI}$ uncertainties considering the combination of the dispersion between $V_{max, HI}$ and $V_{max, H\alpha}$ values modulated by the inclination ($\sin{i_{kin}}$) and the uncertainties associated to inclination.
\vspace{2.5mm}

$V_{max, HI}$ and $V_{max, H\alpha}$ are computed comparing the kinematics of cold and warm gas phases respectively and using different methods; thus likening both quantities might present biases. A bias could come from the way $V_{max, HI}$ is computed using \Hi~emission lines; such measurements are done at certain percentage of the peak flux, here we measure it at 50\% of the peak flux but it could have been done at for instance 20\% and this might have an impact on $V_{max, HI}$ if the edges of \Hi~profiles are not sharp enough. In comparing $V_{max, HI}$ and $V_{max, H\alpha}$, we check whether systematic biases appear. Fig. \ref{fig:vmaxhivmaxha} shows the comparison between \Hi~and  \ha\ $V_{max}$ values using different symbols according to the quality of the \Ha~\rcs. The OSL bisector regression for the whole sample (Flag "A" + "B") gives  $V_{max, H\alpha}=0.95\pm0.02$ $V_{max, HI}$ with an intrinsic scatter of $0.82$, while for Flag "A"-only galaxies $V_{max, H\alpha}=0.9975\pm0.01$ $V_{max, HI}$  with an intrinsic scatter of $0.11$. The two sets of data give consistent results; however, for Flag "A" galaxies the slope of the relation is closer to one and the scatter is smaller than for the whole sample. We conclude that no systematic biases are present in the measurements and to explain the difference of slope between Flag "A"-only and Flag ("A" + "B") galaxies, we favour an explanation related to galaxy environments. Indeed, Flag "B" galaxies are mainly Virgo cluster objects which can be highly perturbed by their surrounding environment (\citealp{Boselli:2006, Boselli:2014B}). There are indeed clear indications that in cluster galaxies the star forming disc is radially truncated with respect to isolated objects (\citealp{Fossati:2013}, \citealp{Boselli:2015}) because of the oustide-in removal of the gas \citep{Boselli:2006}. For this reason, their H$\alpha$ and \Hi~\rcs~do not always reach the plateau, making the estimate of the maximal rotational velocity quite uncertain. The HI-deficiency parameter is generally used to measure the degree of perturbation in cluster objects: galaxies with $HI-Def > 0.4$ have a \Hi~gas content at least 2.5 times smaller than similar objects in the field \citep{Haynes:1984}. Most of the Flag "B" galaxies with a \rc~truncated at $r<0.6\,r_{opt}$ have indeed an HI-deficiency parameter $HI-Def > 0.4$ and are thus cluster perturbed galaxies.
Since we wish to trace the statistical properties of unperturbed systems, we on purpose avoid HI-deficient cluster galaxies which are known to have a lower baryonic mass than field objects (gas deficient objects, \citealp{Boselli:2006}, \citealp{Boselli:2014III}). For this reason, we will ignore galaxies with Flag ``B'' or $HI-Def > 0.4$ in the following analysis. The resulting sample includes 123 objects, out of which 80 (65\%) with \Ha~\rcs.  

\subsection{The Optical and NIR Tully-Fisher Relation}
\label{tully}

\begin{figure}
\begin{center}
\includegraphics[width=\columnwidth]{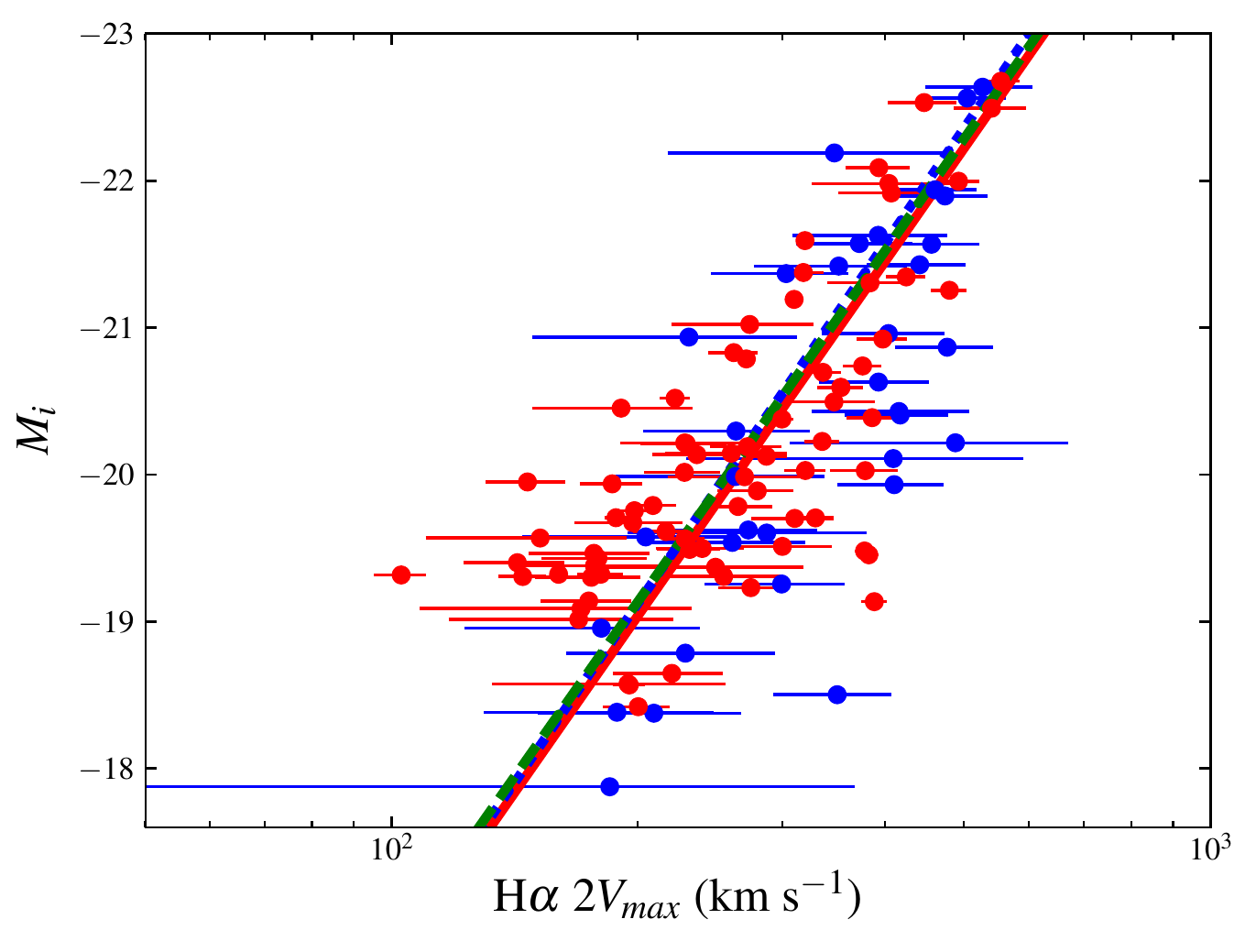}
\includegraphics[width=\columnwidth]{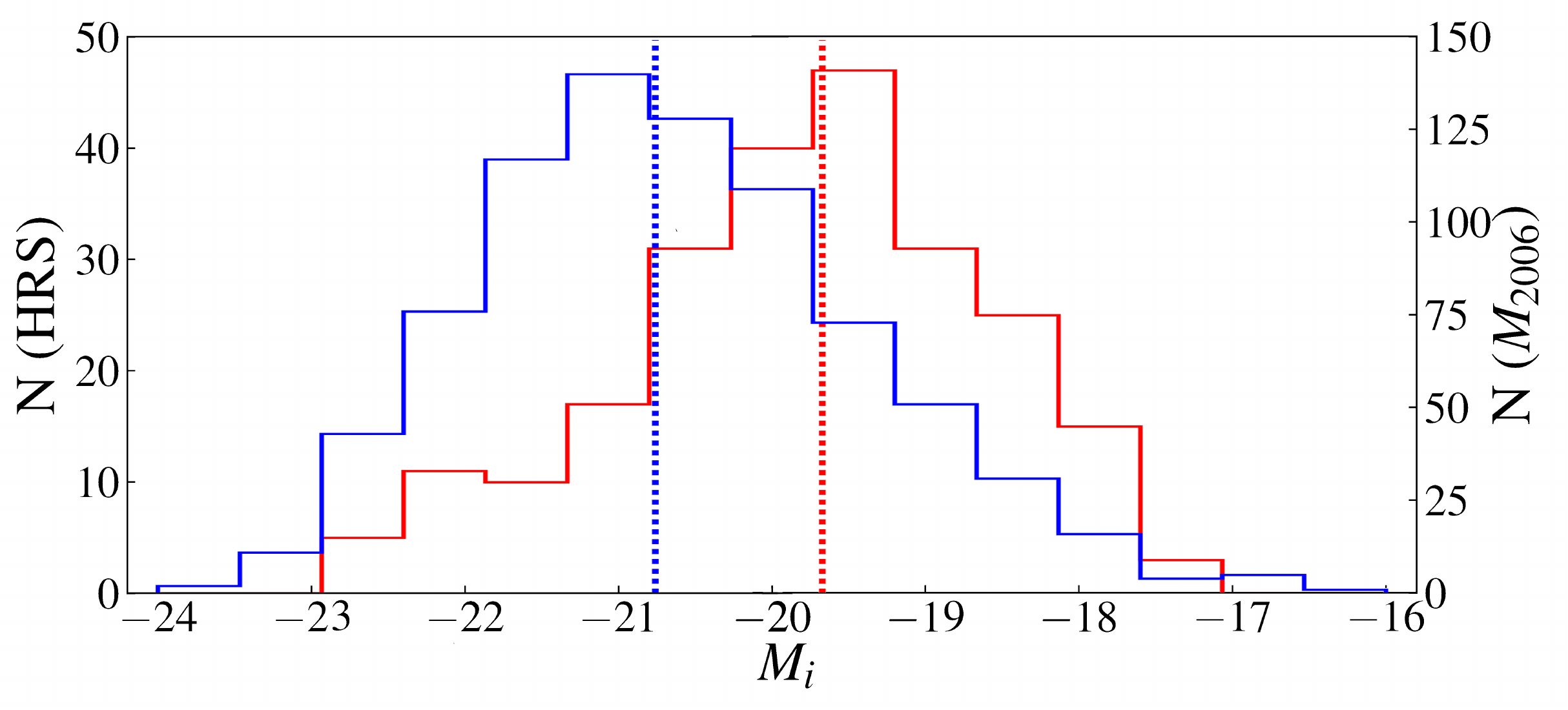}
\end{center}
\caption{Top panel: $i$-band Tully-Fisher relation. Red and blue dots indicate HRS galaxies with H$\alpha$ and \Hi~kinematical data. The solid red line indicates the OLS bisector regression to the HRS data. The dotted blue line represents the template $I$-band TF relation for the nearby galaxy sample of \cite{Masters:2006} but using an OSL bisector method (instead of the bivariate method used by those authors).  The dashed green line represents the median OSL bisector fit computed from the 100 subsamples of 135 galaxies matching our galaxy luminosity distribution, randomly selected from the \cite{Masters:2006} sample.
Bottom panel: $M_{i}$ distributions of the HRS (red) and the \cite{Masters:2006} samples $M_{2006}$ (blue); dotted lines indicate the median value of the corresponding distribution (HRS $=$ -19.67, $M_{2006}$ $=$ -20.76).} 
\label{fig:tfiband}
\end{figure}
\vspace{2.5mm}

\begin{figure}
\begin{center}
\includegraphics[width=\columnwidth]{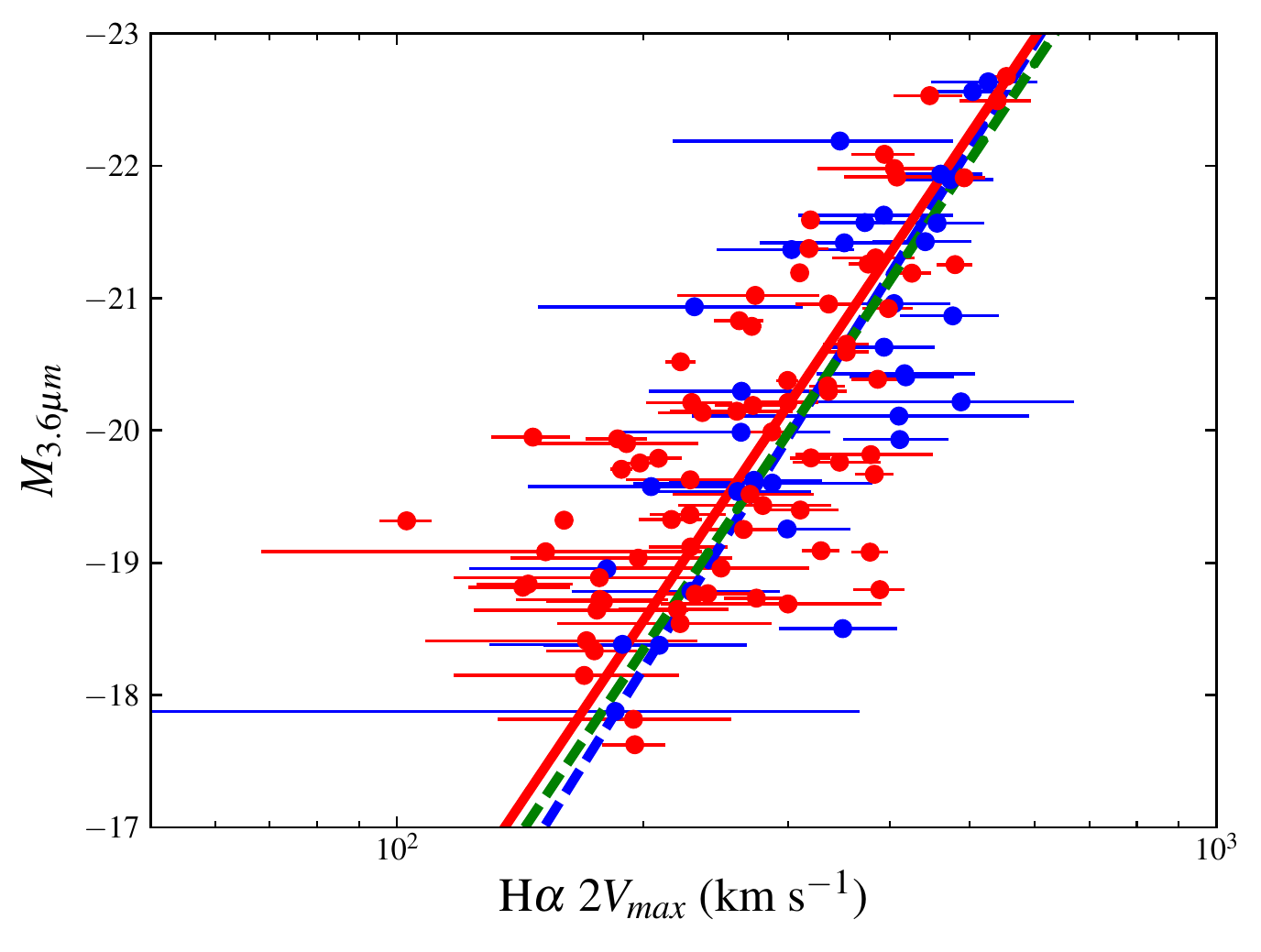}
\includegraphics[width=\columnwidth]{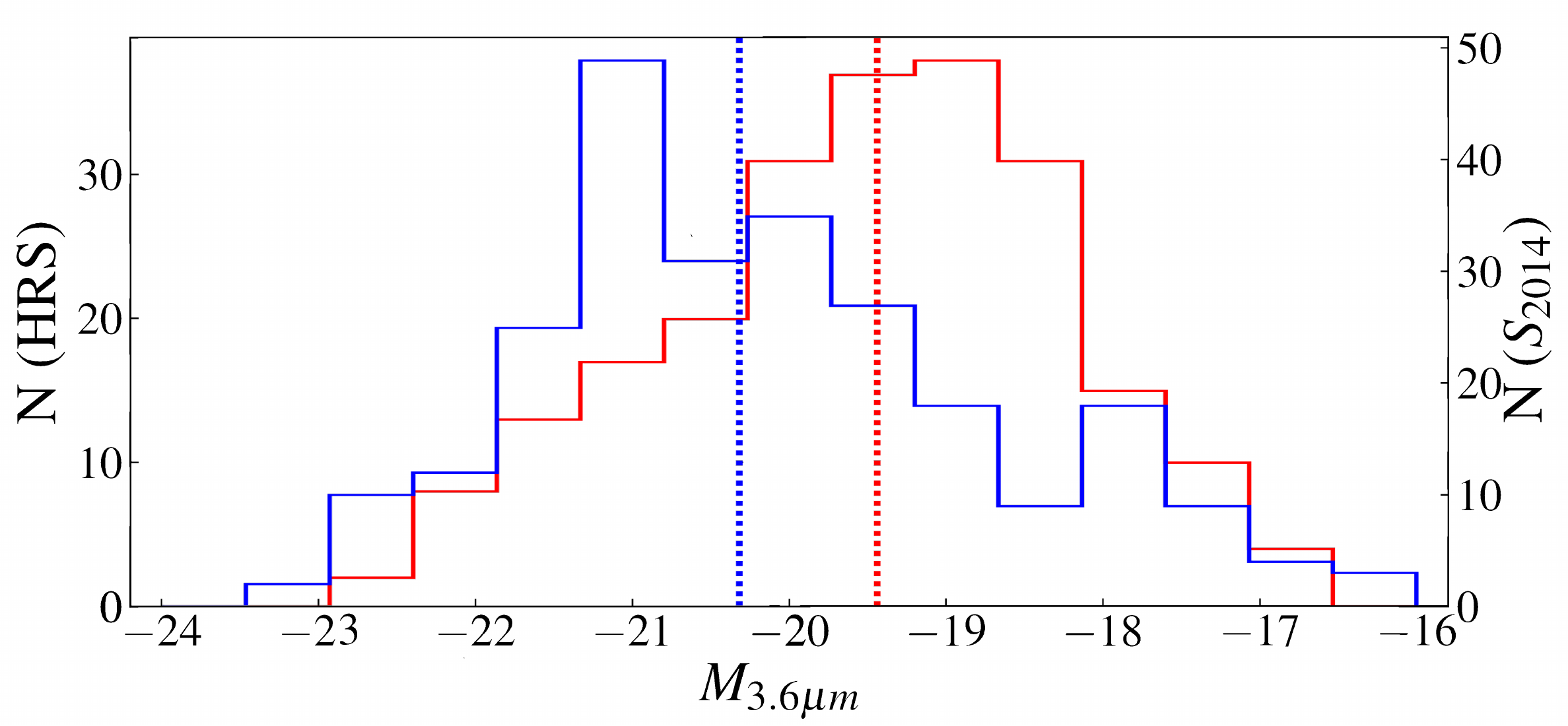}
\caption{NIR S4G-3.6$\mu$m band Tully-Fisher relation. The dashed blue line represents the Tully-Fisher relation determined by \cite{Sorce:2012} for nearby galaxies, while the solid red line the OLS bisector regression to our data. Colors and symbols as in Fig. \ref{fig:tfiband}. Bottom panel: $M_{3.6\mu m}$ distributions of the HRS (red) and the \cite{Sorce:2012} sample ($S_{2014}$, blue); dotted lines indicate the median value of the corresponding distribution (HRS is $=$ -19.44, 
$S_{2014}$ $=$ -20.32).} 
\label{fig:tfiracband}
\end{center}
\end{figure}
\vspace{2.5mm} 

\begin{figure}
\begin{center}
\includegraphics[width=\columnwidth]{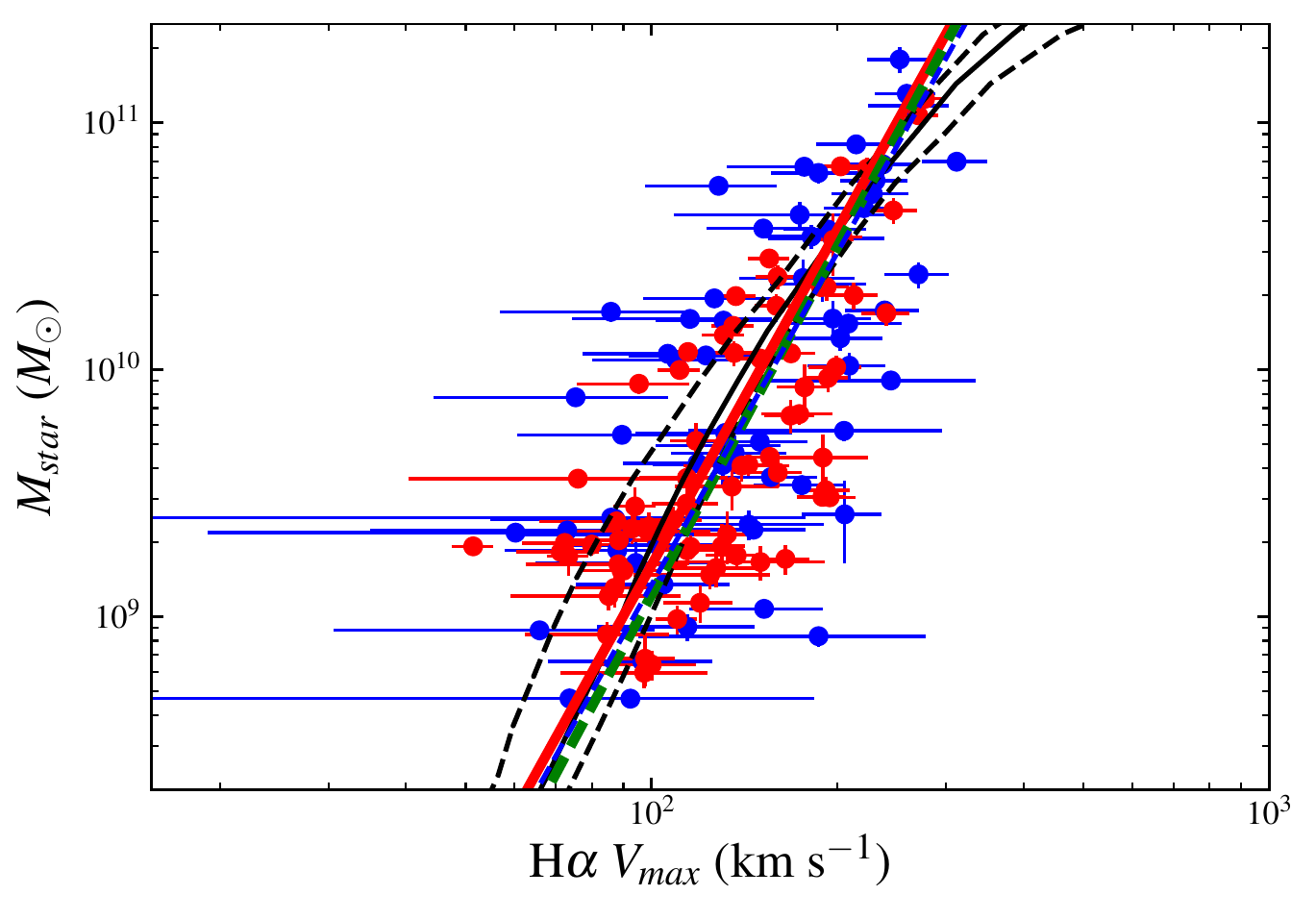}
\caption{Stellar Tully-Fisher relation. The dashed green line represents the relation determined by \cite{Bell:2001}, the dashed blue 
line that of \cite{McGaugh:2000}, the solid red line the OLS bisector regression to our sample. 
Colors and symbols are as Fig. \ref{fig:tfiband}.The solid black line
shows the predictions from the EAGLE cosmological hydrodynamical simulations for galaxies at $z\sim 0$ (\citealp{Schaye:2015}, with dashed lines 16\% and 84\% percentiles).} 
\label{fig:tfstel}
\end{center}
\end{figure}

\begin{table*}
\begin{center}
\caption{$i$-band Tully-Fisher relation.}
\label{itf_table}
\begin{tabular}{ccccc}
      \hline
  Authors & $\alpha \pm \sigma_{\alpha}$  & $\beta \pm \sigma_{\beta}$ & Fitting method& Number of galaxies\\
      \hline
This work 							& $-7.99 \pm 0.15$ 	& $-20.63 \pm 0.19$ & OSL bisector	& 123 \\ 
\cite{Giovanelli:1997b, Giovanelli:1997a}  	& $-7.67 \pm 0.33$ 	& $-20.83 \pm 0.32$ & bi-variate&782 \\     
\cite{Masters:2006}  					& $-7.85 \pm 0.10$ 	& $-20.68 \pm 0.05$ & bi-variate& 807 \\
\cite{Masters:2006} 					& $-8.14 \pm 0.28$ 	& $-20.73 \pm 0.05$ & OSL bisector	& 807 \\
\cite{Masters:2006}/HRS subsamples  	& $-7.91 \pm 0.18$ 	& $-20.71 \pm 0.08$ & OSL bisector	& 123     \\
 \hline
\end{tabular}
\end{center}
{\it{Note about Table \ref{itf_table}: Coefficients obtained for the $i$-band $TF$ relation $M_i  = (\beta \pm \sigma_{\beta}) + (\alpha \pm \sigma_{\alpha})\; (log2V_{max} - 2.5)$ in the literature and this work. The last line has been computed by averaging 100 subsamples randomly extracted from \cite{Masters:2006} and matching exactly our luminosity galaxy distribution.}}

\end{table*}

\begin{table*}
\begin{center}
\caption{NIR Tully-Fisher relation.}
\label{nir_tf_table}
\begin{tabular}{ccccc}
      \hline
   Authors & $\alpha \pm \sigma_{\alpha}$ & $\beta \pm \sigma_{\beta}$ & Fitting method& Number of galaxies\\
      \hline
    This work & $-9.23 \pm 0.26$ & $-20.19 \pm 0.18$ & OSL bisector & 123 \\ 
    \cite{Sorce:2012}  & $-9.77 \pm 0.19$ & $-20.14 \pm 0.09$ & Inverse fitting & 319 \\ 
    \cite{Sorce:2012} & $-9.89 \pm 0.21$ & $-20.17 \pm 0.11$  & OSL bisector & 319 \\ 
    \cite{Sorce:2012}/HRS subsample  & $-9.31 \pm 0.30$ & $-20.25 \pm 0.10$ & OSL bisector & 89  \\               
  \hline
\end{tabular}
\end{center}
{\it{Note about Table \ref{nir_tf_table}: Coefficients obtained for the NIR S4G-1 band $TF$ relation $M_{3.6\mu m}  = (\beta \pm \sigma_{\beta}) + (\alpha \pm \sigma_{\alpha})\; (log2V_{max} - 2.5)$ in the literature and this work.}}
\end{table*}

The Tully-Fisher relation ($TF$, \citealp{TullyFisher}), often used to constrain models and simulations of galaxy evolution, 
is a tight relation between the baryonic and the dynamical masses of late-type systems. This scaling relation has been derived in different photometric bands using different samples (e.g. \citealp{Giovanelli:1997b, Giovanelli:1997a}, \citealp{Steinmetz:1999}, \citealp{McGaugh:2000}, \citealp{Sakai:2000}, \citealp{Bell:2001}, 
\citealp{McGaugh:2005}, \citealp{Masters:2006}, \citealp{McGaugh:2012}, \citealp{Cortese:2014SAMI}, \citealp{Sorce:2012}, \citealp{Ponomareva:2017}, \citealp{Aquino:2018}), but comparisons with prior studies are really challenging.  
\vspace{2.5mm}

In the present study we use maximal rotation velocities derived from homogenous FP \Ha~\rcs~deduced for 2D velocity fields. Samples available in the literature often come from (i) heterogeneous sources, including \Ha\ long slit spectrography or \Hi~integrated line widths, (ii) they are different in absolute magnitude range, (iii) are based on slightly different photometric bands and (iv) have been studied using different fitting methods.  For example, \cite{Giovanelli:1997b, Giovanelli:1997a} used rotational velocities derived either from 21 cm spectra or optical emission line long–slit spectra; $\sim$60\% of the galaxies in use in \cite{Masters:2006} have their rotational velocities measured using HI, the other $\sim$40\% using \Ha\ long-slit data.  \cite{Giovanelli:1997b, Giovanelli:1997a} and \cite{Masters:2006} used bi-variate fittings in $I$- and $i$-bands respectively while  \cite{Sorce:2012} used inverse fitting methods and photometry in the 3.6$\mu$m-band. 
\vspace{2.5mm}

A direct comparison galaxy per galaxy with several previous works is difficult because often galaxy names are not given or the number of galaxies in common is quite reduced. For instance, \cite{Aquino:2018} using CALIFA data, and \cite{Cortese:2014SAMI}, using SAMI data, do not specify the name of the galaxies they used to compute the kinematics of gas. We have only three galaxies in common (NGC 3370, 4535 and 4536) with \citealp{Ponomareva:2017} who used resolved \Hi~data, for which $V_{max}$ is respectively 152$\pm$4, 195$\pm$6 and 161$\pm$10 \kms, while we reach $V_{max}$ values of  160$\pm$9, 201$\pm$15 and 160$\pm$7 \kms, thus compatible with our results. On the other hand, \rcs~deduced from IFUs typically only extend up to smaller radius $r_{opt}$ (e.g. $r_{opt}\sim r_{eff}$, \citealp{Cortese:2014SAMI});
furthermore, $V_{max}$ measured from IFU surveys are often measured at radii where the maximum of the \rc~is not reached yet. While, thanks to the large FoV of our instrument, we know the actual size of the warm disk because it is not limited by the FoV.
\vspace{2.5mm}

\cite{Aquino:2018} and \cite{Ponomareva:2017} samples are smaller than our HRS one (respectively 42 and 32 galaxies with gas kinematics) while \cite{Cortese:2014SAMI} sample, with its 193 galaxies, is comparable to ours. But even more important than the sample size is the galaxy luminosity distribution. To compare different galaxy samples, an important issue is thus to compare the galaxy luminosity distributions of the samples. It is indeed known that the slope of the TF relation is known to slightly change with luminosity \citep[e.g.][]{Schaye:2015}. We check hereafter how the statistically limited HRS compares with works which are generally based on several hundreds of galaxies.
\vspace{2.5mm}

We directly compare the HRS data with two other samples, one in the optical \citep{Masters:2006} and the other in the IR \citep{Sorce:2012}, see Figures \ref{fig:tfiband} and \ref{fig:tfiracband}, Tables \ref{itf_table} and \ref{nir_tf_table}. We use the \cite{Masters:2006} sample because it has been qualified by these authors as a "template calibrator sample", by the way this sample is partially extracted from the \cite{Giovanelli:1997b, Giovanelli:1997a} sample. It consists of 807 galaxies  while the present HRS consists of 123 objects.  In order to fairly compare the results we compute the TF relationship on the whole \cite{Masters:2006} sample using the OSL bisector.   The result we obtain is, as expected, different from the one those authors found using the bivariate method. 
\vspace{2.5mm}

We hereafter quantify how the difference in the TF relationships are related to the statistical significance of the sample and mostly, to their galaxy luminosity distribution using the same fitting method, the OSL bisector.  To quantify those effects, we randomly pick up one hundred subsamples of 123 galaxies from the \cite{Masters:2006} sample with exactly the same luminosity distribution, magnitude bin per magnitude bin, as the HRS. For each of those one hundred subsamples we refit the data and compute the TF slopes and intercepts.  Averaging these 100 iterations and computing their r.m.s., we obtained a slope and zero point values closer to the ones obtained with the HRS than the ones for the whole \cite{Masters:2006} sample.  We see that the values obtained using the HRS sample are within the 1-$\sigma$ uncertainty for both parameters. Both whole sample and subsamples agree quite well, which means that the impact of the sample is not important in this magnitude range.
\vspace{2.5mm}

Using always the OSL bisector method, we make a similar check in the 3.6 $\mu$m-band using the sample of \cite{Sorce:2012} containing 319 galaxies. The \cite{Sorce:2012} sample is larger than the HRS but not large enough to randomly simulate the same galaxy magnitude distribution as the HRS. We furthermore build subsamples of 89 galaxies matching approximatively our magnitude distribution. Despite the impact of sample selection seems larger for the \cite{Sorce:2012} sample then for the \cite{Masters:2006} sample, the results are nevertheless similar to those obtained for the $i$-band. 
\vspace{2.5mm}

We can thus conclude from these analyses that, despite the difference in luminosity distribution and our smallest sample,  the HRS represents well other larger samples generally used as reference in the literature.

\subsection{Stellar and Baryonic Tully-Fisher Relations}
\subsubsection{The Stellar Tully-Fisher Relation}
\label{steltf}
Figure \ref{fig:tfstel} shows the stellar TF relation for the HRS. Stellar masses (from \citealp{Cortese:2012}) have been derived using the $i$- and $g$-band SDSS photometry following the prescription of \cite{Zibetti:2009} (Chabrier IMF). The OSL bisector regression and the best fit published in the literature are given in Table \ref{star_tf_table}. As for the $i$- and 3.6$\mu$m-bands, the stellar TF relation derived for the HRS is in good agreement
with compute the  generally taken as reference in the literature \citep{Bell:2001, McGaugh:2000}. It well matches also that derived on a much smaller sample with Fabry-Perot data by \cite{Torres:2011}. 
\vspace{2.5mm}

We also compare our data with the predictions of the EAGLE cosmological hydrodynamical simulations (\citealp{Schaye:2015}, \citealp{Ferrero:2017}) on $z\sim0.1$ late-type galaxies (Fig. \ref{fig:tfstel}). The agreement between the slope ($4.51\pm0.68$) and intrinsic scatter ($\sim0.16$ dex) of the HRS stellar $TF$ relationship and the EAGLE $TF$ predictions (mean slope $\sim4.84$ and intrinsic scatter $\sim0.14$ dex) is very good.

\begin{table*}
\begin{center}
\caption{Stellar Tully-Fisher relation.}
\label{star_tf_table}
\begin{tabular}{cccc}
      \hline
   Authors & $\alpha \pm \sigma_{\alpha}$ & $\beta \pm \sigma_{\beta}$ & Number of galaxies. \\
      \hline
    This work & $4.51 \pm 0.68$ & $0.19 \pm 0.64$ &123 \\ 
        \cite{McGaugh:2000} & $4.47 \pm 0.54$ & $0.19 \pm 1.23$  & $\sim$550 \\     
    \cite{Bell:2001} & $4.68 \pm 0.40$ & $-0.27 \pm 0.88$ & 79  \\     
    \cite{Torres:2011} & $4.48 \pm 0.38$ & $0.21 \pm 0.83$  & 40 \\
  \cite{Schaye:2015}$^a$ & $\sim 4.84$  &   & EAGLE simulation \\ 
        \cite{Ferrero:2017} & $4.10 \pm 0.05$ & $0.28 \pm 0.02$ & 867 simulated galaxies\\

  \hline
\end{tabular}
\end{center}
Coefficients obtained for the stellar $TF$ relation $log\;M_{star}(M_{\odot})   = (\beta \pm \sigma_{\beta}) + (\alpha \pm \sigma_{\alpha})\; (logV_{max})$ 
in the literature and this work, assuming a Chabrier IMF.\\
\scriptsize{$^a$ Mean slope and intrinsic scatter calculated using 8 quantiles situated within $8.0 > log \; M_{star}\,(M_{\odot})> 11.5 $.}\\
\end{table*}

\vspace{2.5mm} 
\subsubsection{The Baryonic Tully-Fisher Relation}

\begin{figure}
\begin{center}
\includegraphics[width=\columnwidth]{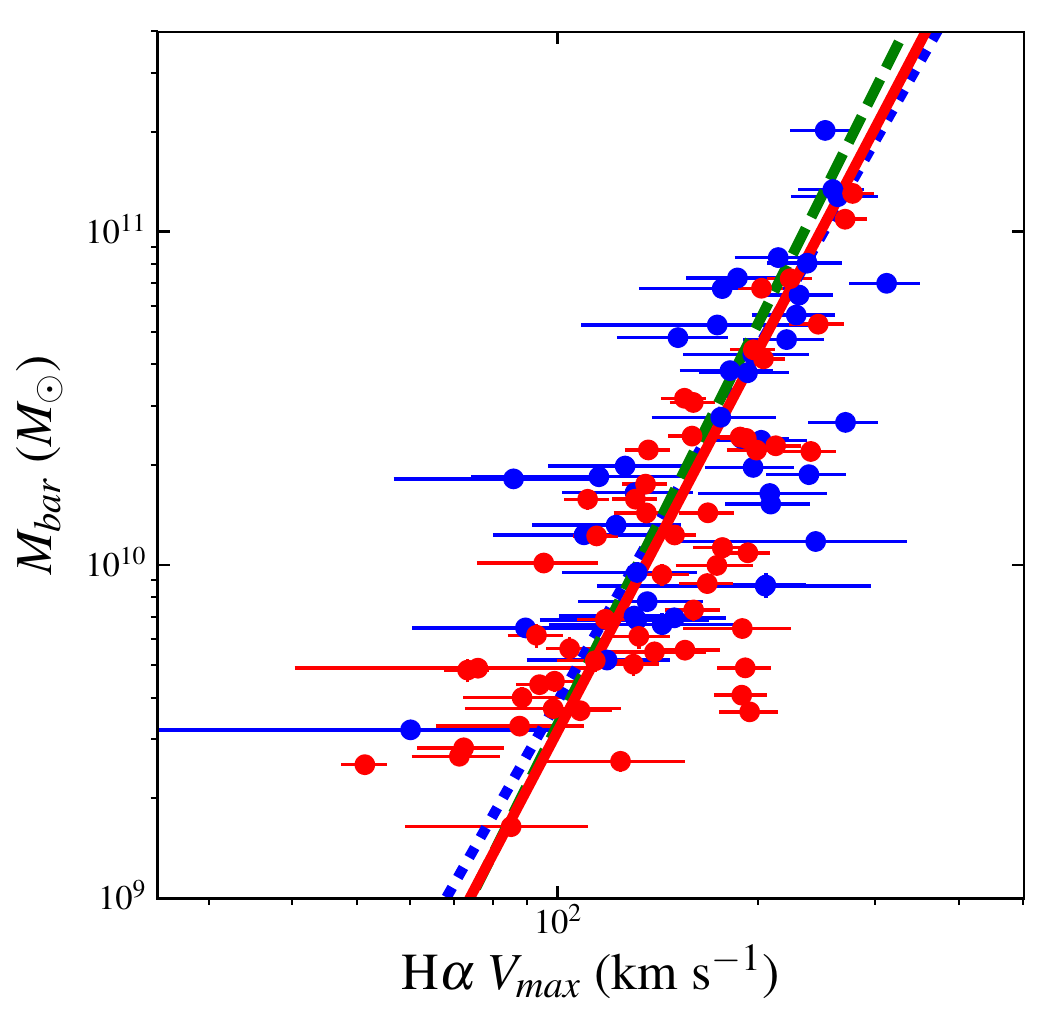}
\caption{Baryonic Tully-Fisher relation. The dashed green line represents the relation determined by \cite{McGaugh:2000}, the dotted  
blue line that of \cite{Bell:2001}, the solid red line the OLS bisector regression to our sample. 
Colors and symbols are as Fig. \ref{fig:tfiband}.} 
\label{fig:tfbar}
\end{center}
\end{figure}

We can also derive the baryonic TF relation (Fig. \ref{fig:tfbar}), where the baryonic mass is defined as:

\begin{equation}
{M_{bar} = M_{star} +  M_{gas} + M_{dust}},
\end{equation}
\\
and the gas mass is defined as follows:

\begin{equation}
M_{gas} =  M_{HI} + M_{H_2} + M_{He} + M_z ,
\end{equation}
\\
where $M_{HI}$ is the atomic gas mass, $M_{H_2}$ the molecular gas mass, $M_{He}$ the helium mass and $M_z$ the metal mass. The hydrogen mass $M_H = M_{HI} + M_{H_2}$ is directly measurable (\citealp{Boselli:2014}, where $M_{H_2}$ is measured using a luminosity dependent $X_{CO}$ conversion factor, \citealp{Boselli:2002}). $M_{He}$ is derived from the primordial nucleosynthesis helium fraction, $Y \equiv M_{He}/M_{gas} \simeq $0.28 \citep{Pagel:2009}, using the relation:

\begin{equation}
M_{gas} = \frac{1}{1-Y-Z} \times M_{H},
\end{equation}
\\
where $Z=M_z/M_{gas}$ is the metal fraction which can be derived from metallicity measurements using the relation \citep{McGaugh:2000}:
\begin{equation}
 log\,(Z/Z_{\odot}) = log\,(O/H) - log\,(O/H)_{\odot},
\end{equation}
 taking $Z_{\odot}=1.34\times10^{-2}$ and $12+log\,(O/H)_{\odot} = 8.69$ from \cite{Asplund:2009}. 
\vspace{2.5mm} 

As previously mentioned, all baryonic components are directly measurable for the HRS, with the exception of the ionised and hot gas masses which we consider as negligible. Table \ref{proportions} lists the average contribution of each baryonic component to the total baryonic mass.  The mean mass and percentage of each baryonic component has been computed using the whole HRS. We note that the gas (plus dust) content represents more than the half of the stellar content $(M_{gas = HI+H_2+He}+M_z+M_{dust})/M_{star}\simeq$56\% and the metal content represents 2\% of the total gas (plus dust) content ($M_z/(M_{gas = HI+H_2+He}+M_z+M_{dust})\simeq$0.02). 
\vspace{2.5mm}

\begin{table}
\begin{center}
\caption{Mean contribution of the baryonic components.}
\label{proportions}
\begin{tabular}{ccc}
      \hline
  Baryonic comp & Average $\%$ & Average mass ($M_{\odot}$)  \\
      \hline
        $M_{star}$ & $65 \pm 15\%$ & $5.24 \pm1.27  \times10^{9}$  \\ 
        $M_{HI}$ & $19 \pm 6\%$ & $1.53 \pm 0.45\times10^{9}$  \\ 
        $M_{He}$ & $11.4 \pm 3.3\%$ & $9.19 \pm 2.67\times10^{8} $  \\ 
         $M_{H_2}$ & $5.2 \pm 2.8\%$ & $4.45 \pm 2.25 \times10^{8}$  \\
         $M_{z}$ & $0.87 \pm 0.35\%$ & $7.03 \pm 2.83 \times10^{7}$  \\  
        $M_{dust}$ & $0.16 \pm 0.07\%$ & $1.32 \pm 0.58 \times10^{7}$  \\  
  \hline
\end{tabular}
\end{center}
Average contribution of each baryonic component to the total baryonic mass in percentages and solar masses.
\end{table}

The OLS bisector regression to the data is compared to the values published in the literature in Table \ref{barylit}.
In these published works, the baryonic mass is generally derived summing the contribution of the stellar mass to that of the \Hi~gas mass, and assuming a constant $M(HI)/M(H_2)$ contribution for the molecular gas component. These works also lack of high quality 2D velocity fields.  Despite these major differences, these works give results consistent with ours.
\vspace{2.5mm}


\begin{table*}
\begin{center}
\caption{Baryonic Tully-Fisher relation.}
\label{barylit}
\begin{tabular}{ccccc}
      \hline
   Authors & $\alpha \pm \sigma_{\alpha}$ & $\beta \pm \sigma_{\beta}$ & Nb galaxies &  $M_{bar}$ \\
      \hline
    This work & $3.79 \pm 0.39$ & $1.91 \pm 0.63$ & 123 & $M_{star} + 1.4(M_{HI}+ M_{H_2})+ M_{z} + M_{dust}$  \\ 
   \cite{McGaugh:2000} & $3.98 \pm 0.12$  &  $1.42 \pm 0.25$ & $\sim$550& $M_{star} + 1.4\,M_{HI}$ \\ 
   \cite{Bell:2001} & $3.53 \pm 0.20$  &  $2.56$ & 79 &  $M_{star} + M_{HI}$ \\ 
    \cite{McGaugh:2005} & $3.19 \pm 0.14$ & $ 3.06$ &74& $M_{star} + 1.4\,M_{HI}$ \\ 
   \cite{Torres:2011} & $3.64 \pm 0.37$ & $1.95 \pm 0.61$ &40 & $M_{star} + (\sim1.4\,M_{HI})$ \\ 
  \cite{Zaritsky:2014} & $3.50 \pm 0.20$ & $2.96$ &1468 & $M_{star} + M_{HI}$\\
  \cite{McGaugh:2015} & $4.04 \pm 0.09$  &  $1.44 \pm 0.18$ &26&  $M_{star} + 1.33\,(M_{HI}+ M_{H_2})$\\
  \cite{Lelli:2016} &   $3.71 \pm 0.08$ &  $2.10 \pm 0.18$  &118& $M_{star} + 1.33\,M_{HI}$ \\
  \cite{Papastergis:2016} &  $3.75 \pm 0.11$ & &97& $M_{star} + M_{HI}$ \\
  \cite{Ponomareva:2018} &  $2.99 \pm 0.20$  & $2.71 \pm 0.56$ & 32 &  $M_{star} + M_{HI}+ 1.4\,M_{H_2}$\\
  \hline
\end{tabular}
\end{center}
Coefficients obtained for the baryonic $TF$ relation $log\;M_{bar}(M_{\odot}) = (\beta \pm \sigma_{\beta}) + (\alpha \pm \sigma_{\alpha})\; (logV_{max})$ in the literature and this work.
\end{table*}

We can also compare our results to the predictions of models and simulations. According to \cite{Dutton:2014}, \cite{Moster:2013} and  \cite{Zu:2015}, N-body 
simulations predict for the baryonic $TF$ relation an intrinsic scatter of $\sim 0.15$, consistent with the dispersion of our sample ($0.16$). \\

\subsection{Baryonic versus Dynamical Masses}
\label{masota}

\begin{figure}
\begin{center}
\includegraphics[width=8.0cm]{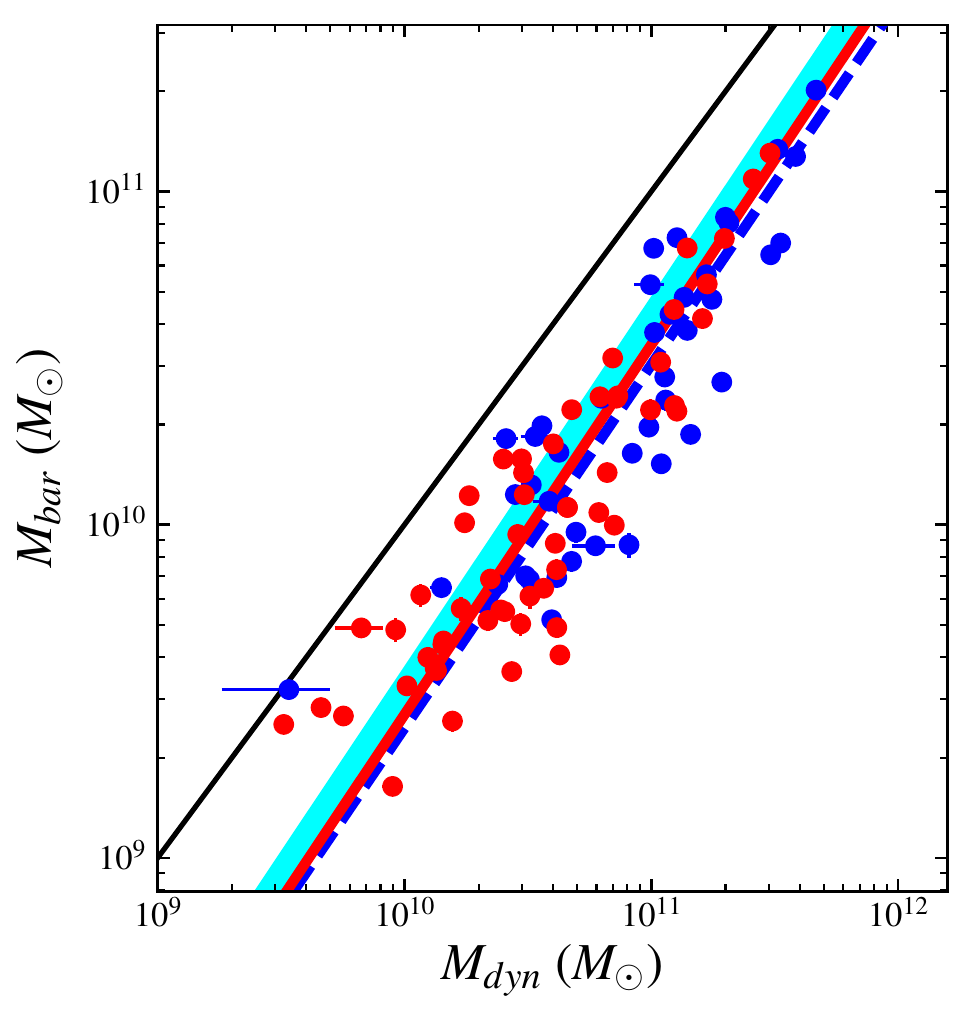}
\end{center}
\caption{Baryonic versus dynamical mass. Colors and symbols are as in Fig. \ref{fig:tfiband}. The dashed blue line represents the best fit of \cite{Torres:2011}, while the solid red line the OLS bisector regression fit to our sample. The solid black line shows the 1-1 relation, and the cyan shaded area the mass interval $0.6$ $\leq$ $\alpha$ $\leq$ $1.0$. $M_{bar}$ are computed within $r_{opt}$, while $M_{dyn}$ from $V_{max}$ computed by Courteau's profile extrapolation within $r_{opt}$.} 
\label{fig:mazorra}
\end{figure}

\begin{figure}
\begin{center}
\includegraphics[width=\columnwidth]{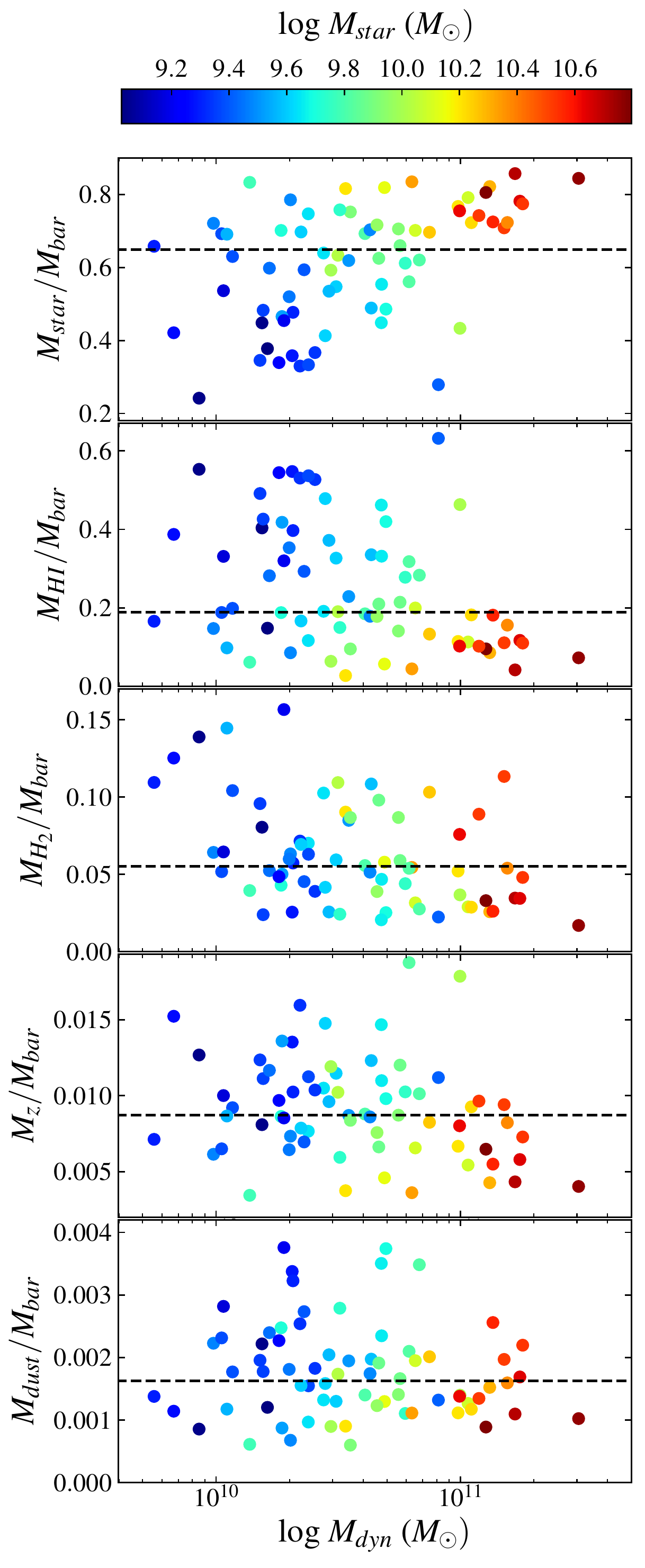}
\caption{Variation of the different baryonic components to the total baryonic mass of the total dynamical mass, from top to bottom: stellar (first panel), \Hi~(second panel), H$_2$ (third panel), metal elements (fourth panel), and dust (fifth panel). The black dashed line shows the mean contribution of the corresponding baryonic component. The colors represent the stellar masses in logarithmic scale.}
\label{fig:masotas}
\end{center}
\end{figure}

\begin{table}
\begin{center}
\caption{Baryonic versus Dynamical Masses}
\label{bardyn}
\begin{tabular}{ccc}
      \hline
   Authors & $\alpha \pm \sigma_{\alpha}$ & $\beta \pm \sigma_{\beta}$  \\
      \hline
    This work & $1.11 \pm 0.12$ & $-1.67 \pm 0.59$  \\ 
    \cite{Torres:2011} & $1.09 \pm 0.04$ & $-1.34 \pm 0.42$   \\     
  \hline
\end{tabular}
\end{center}
Coefficients obtained for relation $M_{bar}(M_{\odot})  = (\beta \pm \sigma_{\beta}) + (\alpha \pm \sigma_{\alpha})\; log\;M_{dyn}$ by \cite{Torres:2011} 
and this work.
\end{table}
\vspace{2.5mm}

Dynamical masses are computed following the prescription of \cite{Lequeux:1983}, which is a variation of the Virial theorem that attempt to account for a variety of galaxy mass distribution, from a purely flat to a purely spherical model : 
\begin{equation}
M_{dyn}(M_{\odot}) = \alpha\;r_{opt}\;V^{2}_{max}\;/\;G,
\label{dynleque}
\end{equation}
where $G$ is the gravitational constant and $\alpha$ is a coefficient which depends on the model flatness of mass distribution ($0.6 \leq \alpha \leq 1.0$). Figure \ref{fig:mazorra} shows the relationship between baryonic and dynamical masses. Since we are taking into account neither luminosity profiles nor models of mass distribution, we chose for the comparison the spherical model $\alpha = 1.0$. In order to quantify the impact of this extreme assumption, we plot a cyan shaded area representing the interval $0.6$ $\leq$ $\alpha $ $\leq$ $1.0$. The values of the OLS bisector fit are given in Table \ref{bardyn}. The two variables are strongly correlated.  The fraction of baryons is lower in low-mass-gas-rich galaxies 
($\sim$20-25\% of the total dynamical mass) than in high-mass galaxies with low SFR ($\sim$35-45\% of the total dynamical mass), consistently with what seen in the previous $TF$ relations (Figs. \ref{fig:tfiband}, \ref{fig:tfiracband}, \ref{fig:tfstel} and \ref{fig:tfbar}). Figure \ref{fig:masotas} shows how the contribution of each baryonic component changes as a function of the dynamical mass and of the stellar mass (color symbols) within the HRS. Using a simple linear regression (not plotted for clarity), we estimate the trend of the y-axis (respectively, from the top to the bottom: the fraction of stars $M_{star}/M_{bar}$, neutral gas $M_{\hi}/M_{bar}$, molecular gas $M_{H2}/M_{bar}$, metals $M_{z}/M_{bar}$ and dust $M_{dust}/M_{bar}$) as a function of the dynamical mass ($\log$ M$_{dyn}$).  A positive correlation is only observed between the fraction  $M_{star}/M_{bar}$ and the dynamical mass, with a statistical coefficient of determination $R^2\sim0.42$.  Weak negative correlations are found in the fourth next cases. Both slope and $R^2$ coefficient become increasingly weaker from the second top panel to the bottom one. The bottom one, i.e. $M_{dust}/M_{bar}$, is in fact compatible with a zero slope ($R^2\sim0.03)$.  The gaseous components as well as the metal and dust components, are relatively more abondant, relatively to the baryonic component, in low mass galaxies (log$\,M_{dyn}$ $<$ 10.5$M_{\odot}$), while the stellar component does in massive objects. The correlation and the dispersion around this correlation between the stellar and the dynamical masses, already observed in Fig. \ref{fig:mazorra}, is again underlined by these plots. This is indeed expected given the rapid evolution in the past of massive galaxies, which already transformed most of the gas into stars, with respect to a fairly constant star formation activity in low mass systems possible thanks to their large gas reservoirs \citep{Sandage:1986,Gavazzi:1996,Gavazzi:2002,Boselli:2001,Boissier:2003}. 

\vspace{2.5mm}

Dynamical masses are computed within $r_{opt}$ and furthermore do not provide the total masses since they only account for the visible part of galaxies.  In addition, dynamical masses do not take into account possible variation in the bulge-to-disc ratio nor in the actual disk or spherical shape of the baryonic matter distributions. Despite those rough approximations,
the observed dispersion in the relation ($\sim$0.12 dex) is comparable to that predicted by semi-analytic models of galaxy formation in a $\Lambda$CDM context ($\sim$0.15 dex,  \citealp{Dutton:2012}, \citealp{Dutton:2014}).
\vspace{2.5mm}

\subsection{Baryonic and Dynamical Mass Main Sequences}
\label{masssfr}

The tight relationship between $SFR$  and $M_{star}$ is generally called main sequence
(i.e. \citealp{Guzman:1997}; \citealp{Brinchmann:2000}; \citealp{Bauer:2005}; \citealp{Bell:2005}; \citealp{Spapovich:2006}; 
\citealp{Reddy:2006}; \citealp{Noeske:2007}; \citealp{Salim:2007}; \citealp{Elbaz:2007}; \citealp{Daddi:2007}; \citealp{Pannella:2009}; \citealp{Rodighiero:2010}; 
\citealp{Peng:2010}; \citealp{Karim:2011}; \citealp{Whitaker:2012}; \citealp{Speagle:2014}; \citealp{Torrey:2014}; \citealp{Gavazzi:2015}; \citealp{Renzini:2015}; 
\citealp{Sparre:2015}; \citealp{Cano:2016}; \citealp{Hsieh:2017}; \citealp{Medling:2018}). The slope of this obvious scaling relation 
(bigger galaxies have more of everything), its bending at high luminosities, its scatter, and its variation as a function of redshift and environment, are often used 
to trace the evolution of galaxies with cosmic time. Observational evidence indicates mass as a principal driver of galaxy evolution (e.g. \citealp{Cowie:1996}; \citealp{Gavazzi:1996}; \citealp{Boselli:2001}). Consistently with these obervational results, hydrodynamic cosmological simulations suggest that
the evolution of galaxies is mainly gouverned by the dark matter halo in which they reside (\citealp{Governato:2012}; \citealp{Brooks:2014}; \citealp{Christensen:2014}). It would thus be 
interesting to derive the main sequence using dynamical masses rather than with stellar masses.

\vspace{2.5mm}

\begin{figure}
\begin{center}
\includegraphics[width=8.5cm]{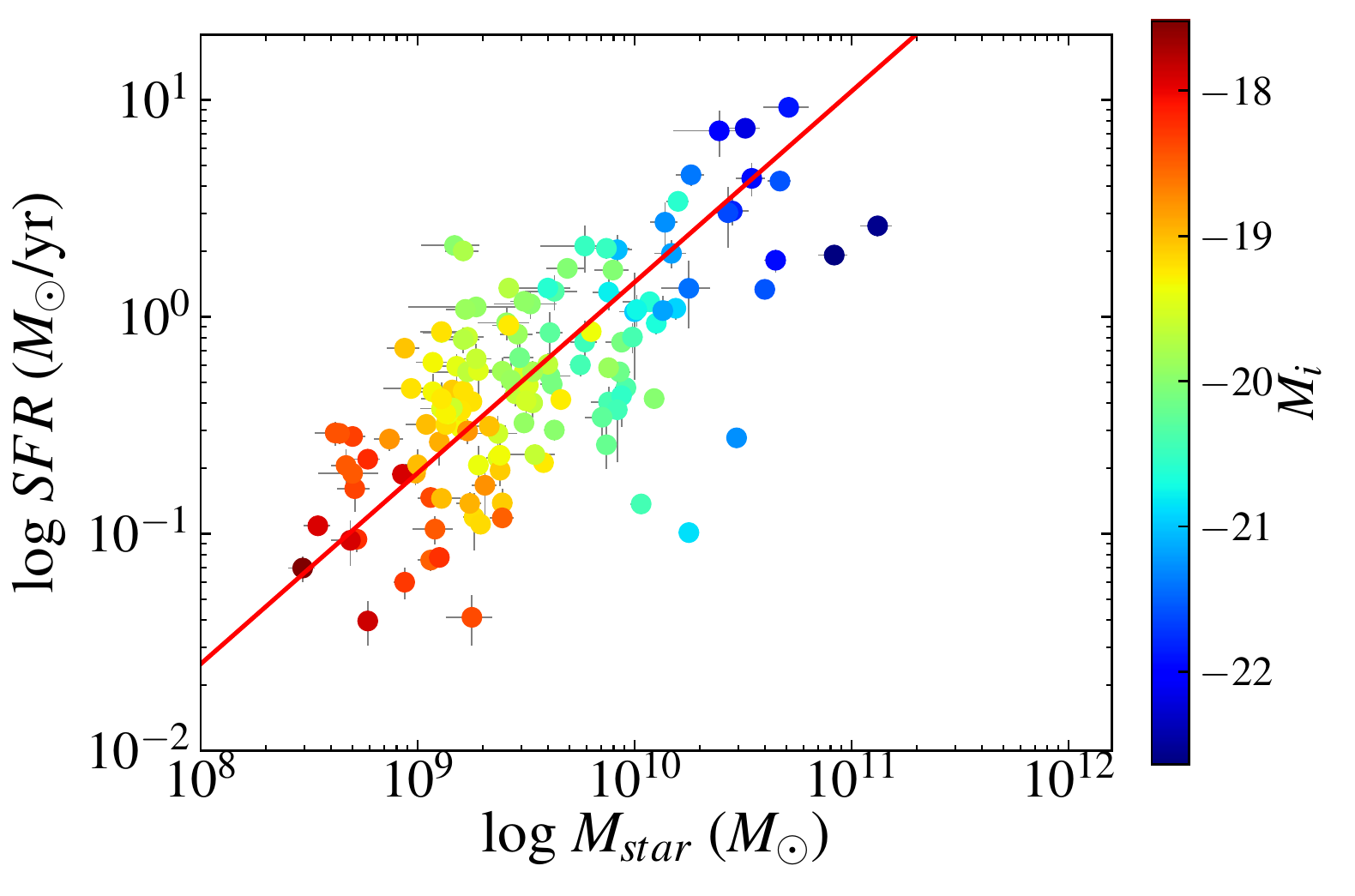}
\includegraphics[width=8.5cm]{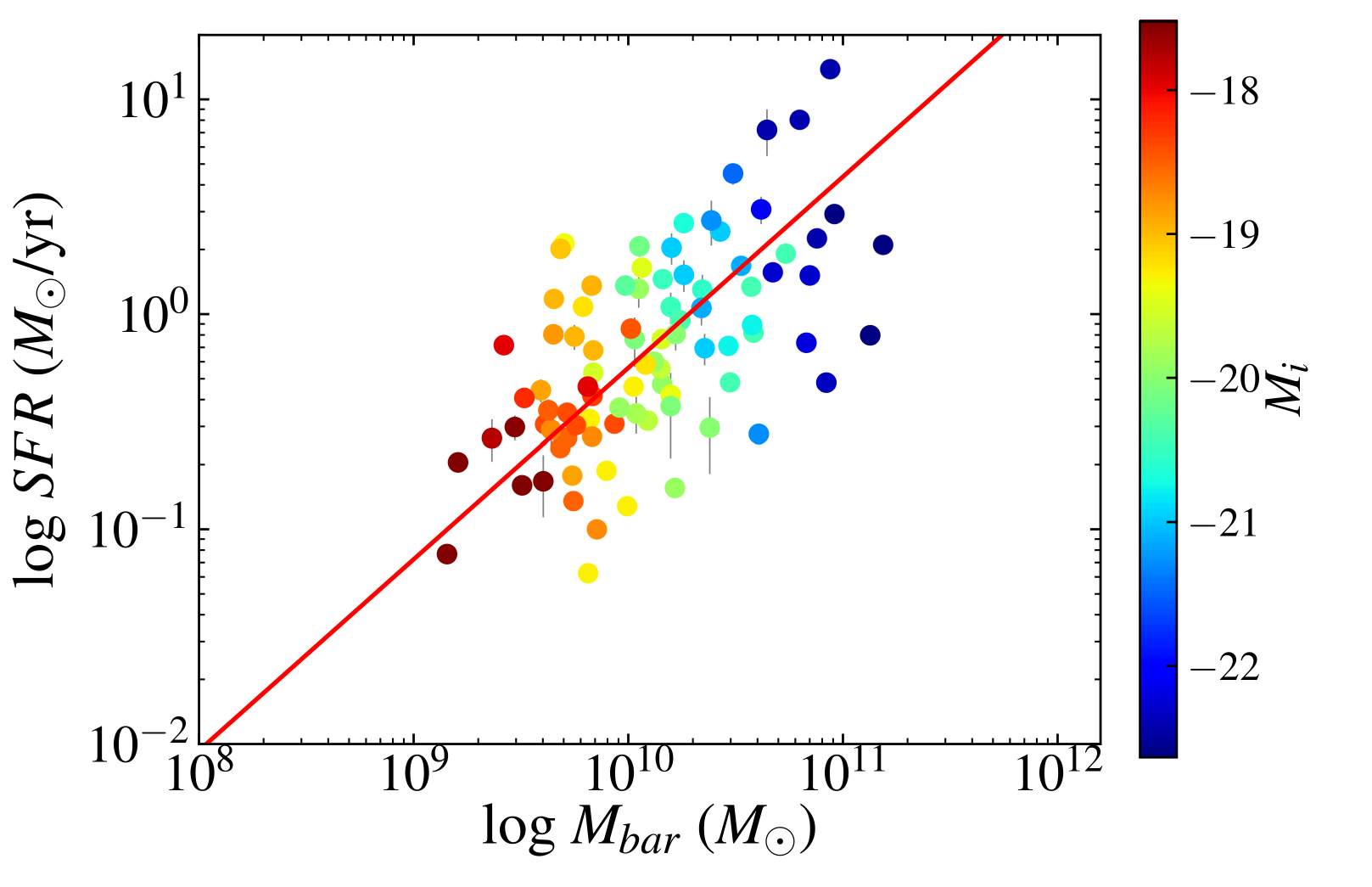}
\includegraphics[width=8.5cm]{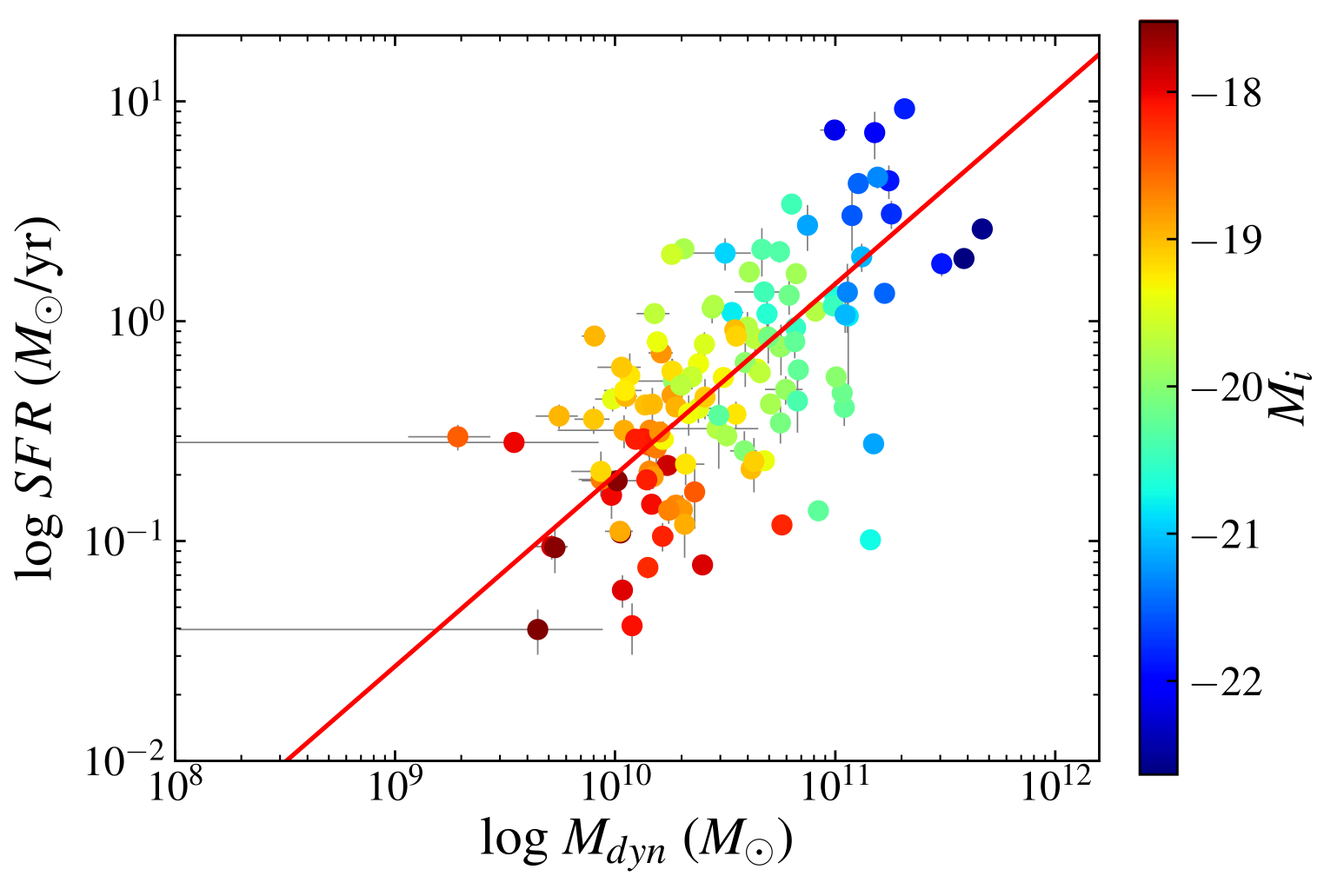}
\caption{Relationship between the SFR and $M_{star}$ (top panel), $M_{bar}$ (middle panel), and $M_{dyn}$ (bottom panel). The OLS bisector regressions are represented with the solid red lines. The colors represent the $i-band$ magnitudes in logarithmic scale.} 
\label{fig:mainseq}
\end{center}
\end{figure}

The stellar main sequence relation has been derived and analysed for the HRS
by \cite{Ciesla:2014}, \cite{Boselli:2015} and \cite{Boselli:2016A}. Thanks to the unique dataset in our hands, we can compare for the first time
in the literature the main sequence relations derived using $M_{star}$, $M_{bar}$, and $M_{dyn}$ (Fig. \ref{fig:mainseq}).
\vspace{2.5mm}

 \begin{table}
\begin{center}
\begin{tabular}{cccc}
      \hline
Galaxy mass &  $\alpha \pm \sigma_{\alpha}$ & $\beta \pm \sigma_{\beta}$ & Scatter   \\
      \hline
Stellar mass & $0.88 \pm 0.10$& $ -8.64 \pm 0.67$ & 0.29\\ 
 Baryonic mass & $0.89 \pm 0.16$ & $-9.05 \pm 0.77$ & 0.30\\ 
 Dynamical mass & $0.87 \pm 0.19$  &  $-9.45 \pm 0.56$ & 0.29\\
  \hline
\end{tabular}
\caption{Stellar, baryonic and dynamical main sequence.}
\label{mainseq2}
\end{center}
Coefficients obtained for the different main sequence relations $log\,SFR (M_{\odot} yr^{-1}) = (\beta \pm \sigma_{\beta}) + 
(\alpha \pm \sigma_{\alpha})\; log\;M_{galaxy}(M_{\odot})$ for the HRS star-forming sample. $SFR$ are derived using a Chabrier IMF.
\end{table}
Consistently with \cite{Boselli:2015}, we use the whole HRS late-type sample but ignoring the HI-deficient galaxies. The $SFR$ is studied considering the mean value derived using the three tracers $M_{dyn}$, $M_{bar}$ and $M_{star}$.  The number of galaxies differs for the three cases according to the data available for each tracer. OLS bisector fits to the data have been computed. Fig. \ref{fig:mainseq} and Table \ref{mainseq2} show that the similar main sequence relations are present using the three different galaxy mass estimators. The slope and the scatter of the three relations are comparable. They obviously differ in their zero points given that $M_{dyn}$ $>$ $M_{bar}$ $>$ $M_{star}$. The best fit values can be taken for comparison with the predictions of models and simulations. We recall that these relations are valid within the mass range 3$\times$10$^8$$\lesssim$$M_{star}$$\lesssim$10$^{11}$ M$_{\odot}$, or equivalently 3$\times$10$^9$$\lesssim$$M_{dyn}$$\lesssim$4$\times$10$^{11}$ M$_{\odot}$.  The correlation between the magnitudes (M$_i$), the three different mass tracers and the SFR is also pointed out on this figure with the expected scatters.

\section{Conclusions}
\label{conclusions}

We present new high resolution Fabry-Perot observations gathered at the OHP of 152 star-forming galaxies belonging to the \textit{Herschel} Reference 
Survey. Combined with those available in the literature (40 objects), previously collected with the same facility, an homogeneous set of kinematical data 
is now available for 192 objects (73.6\%~ of the sample).
By now, this is the first work presenting \Ha~high resolution spectroscopic data of the HRS, with a typical 
spatial sampling of $\sim$2 arcsec and a spectral resolution of R$\sim$10,000.
Using improved data reduction pipelines, we compute the \Ha~momenta, optimizing the spatial resolution given a \SNR~through an adaptive binning method based on Voronoi tessellations. We also derive accurate kinematical models and parameters,  
residual velocity maps, and rotation curves. 
\vspace{2.5mm}

i) We derive the $i$- and the 3.6 $\mu$m-band Tully-Fisher relations and compare them to those obtained for larger samples in the literature.
Despite the difference in the kinematic data (Fabry-Perot \Ha~\rcs~in this work vs. \Hi~line width profiles or long slit optical spectra
in the literature), in the dynamical range of the sample, and in the statistics, the different TF relations are very consistent, suggesting 
that the HRS can be taken as a representative sample for these scaling relations in the local universe.

ii) Thanks to its unique multifrequency coverage, we use this dataset to derive the baryonic TF relation and the relation between the 
baryonic mass and the dynamical mass of galaxies. The baryonic mass is for the first time measured using direct estimates
of the stellar, atomic, molecular, dust and metal mass of galaxies.
The intrinsic scatter in the baryonic TF relation ($\sim$0.16 dex) is in agreement with the predictions of cosmological simulations. The baryonic
component is dominated by the stellar mass in massive objects and by the total gas mass (\Hi~and H$_2$) in low mass systems. The dust content 
is just a very small fraction ($\sim$0.2\%) of the total baryonic mass, while the contribution of metals is fairly constant at $\sim$0.9\%.

iv) We computed the baryonic and dynamical main sequence, finding relations with a similar slope and intrinsic scatter 
than the stellar mass main sequence. 
\vspace{2.5mm}   

The observations of the whole star-foming galaxy sample will be completed 
in forthcoming runs. Combined with those available at other frequencies, these 2D-spectroscopic data 
make the HRS a unique dataset for studying on strong statistical basis the role of gas kinematics on the 
process of star formation, a major step towards the understanding of the mechanisms that drive galaxy evolution.
 
\begin{acknowledgements} 
Based on observations taken at the Observatoire de Haute Provence (OHP, France), operated by the French CNRS. The authors warmly thank Olivier Boissin from LAM and the OHP team for its technical assistance before and during the observations, namely the night team: Jean Balcaen, Stéphane Favard, Jean-Pierre Troncin, Didier Gravallon and the day team led by François Moreau as well as Dr. Auguste Le Van Suu, the Head of Observatoire de Haute Provence-Institut Pythéas.
The authors thank the Mexican grants CONACYT-253085 and DGAPA-UNAM PAPIIT-IN109919 which have extensively supported this work.
This work was supported by the Programme National Cosmology et Galaxies (PNCG) of CNRS/INSU with INP and IN2P3, co-funded by CEA and CNES.
This research has made use of data from the HRS project. HRS is a Herschel Key Programme utilising guaranteed time from the SPIRE instrument team, ESAC scientists and a mission scientist. This research has also made use of the NASA/IPAC Extragalactic Database (NED) which is operated by the Jet Propulsion Laboratory, California Institute of
Technology, under contract with the National Aeronautics and Space
Administration. The authors have also made an extensive use of the HyperLeda Data base (\url{http://leda.univ-lyon1.fr}). The Digitized Sky Surveys were produced at the Space Telescope Science Institute under U.S. Government grant NAG W-2166. The images of these surveys are based on photographic data obtained using the Oschin Schmidt Telescope on Palomar Mountain and the UK Schmidt Telescope. The plates were processed into the present compressed digital form with the permission of these institutions. JAGL thanks the Consejo Nacional de Ciencia y Tecnologia (CONACYT) of Mexico for the scholarship awarded during the PhD studies at Aix-Marseille University.  The authors are very grateful to the referee Dr. Octavio Valenzuela for the very constructive comments and suggestions.
\end{acknowledgements} 

\bibliographystyle{aa}
\bibliography{biblio} 




\appendix


\clearpage
\section{Quality Flags of the Rotation Curves}
\label{flags}

As described in the previous subsection \ref{zhaocour}, the maximum rotational velocity ($V_{max}$) per galaxy has been obtained by fitting a modified Courteau function to the rotation curve. We considered as $V_{max}$ the maximum velocity value reached by the Courteau profile between $r = 0$ and the radius corresponding the last crown of the \rc. The estimated $V_{max}$ per galaxy are shown in Table \ref{kinparam}, except for those galaxies for which no kinematical model was reached and therefore no rotation curve was obtained. 
\vspace{2.5mm}

\begin{figure}
\begin{center}
\includegraphics[width=\columnwidth]{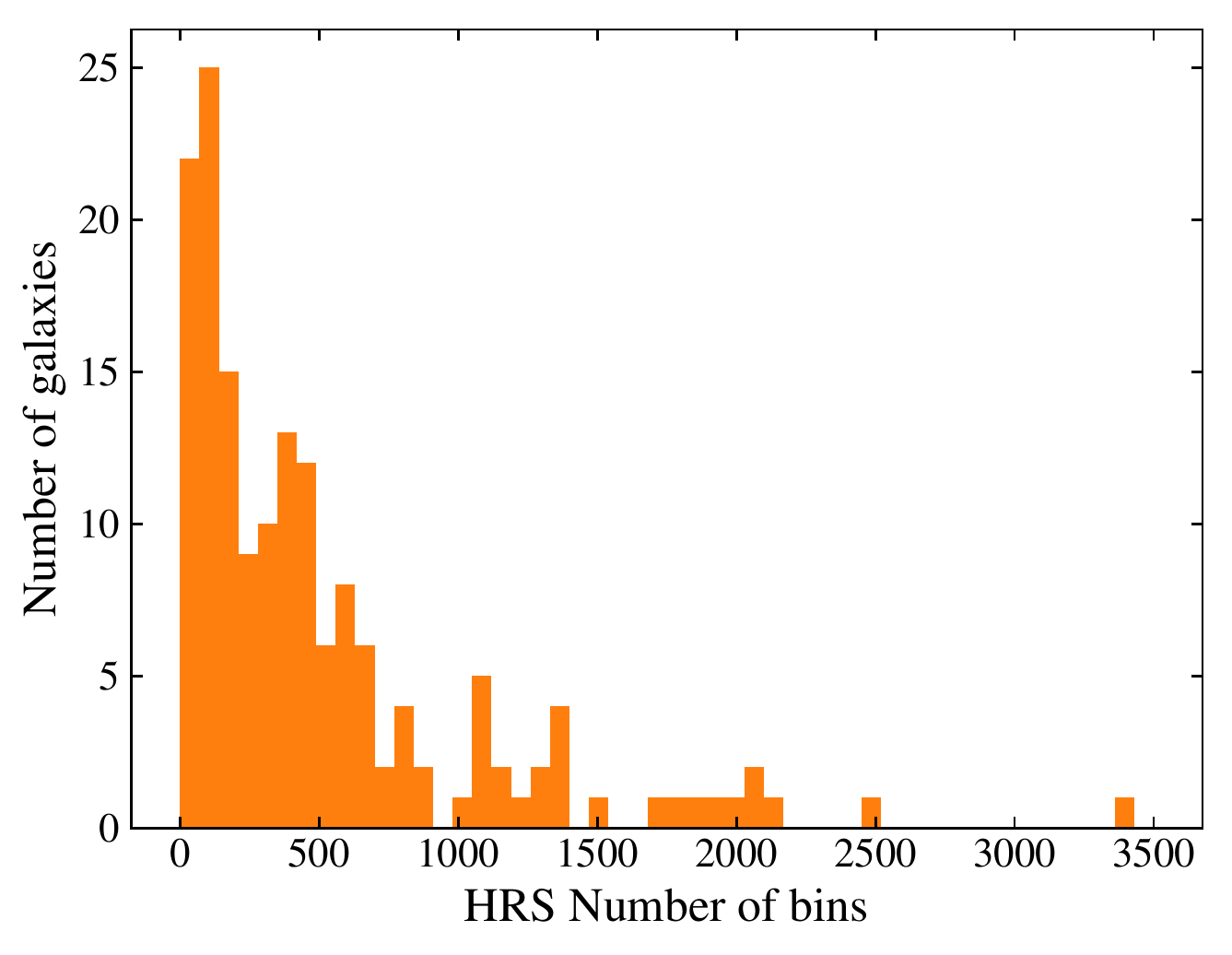}
\end{center}
\caption{ Histogram of the number of spatial bins used to compute the rotation curves. Galaxies with high spatial coverage and/or high S/N ratio have the highest number of bins.} 
\label{fig:histbins}
\end{figure}

In order to quantify the quality of each \rc~and to attribute a quality flag, we used an automatic classification method based on the next parameters: 
\vspace{2.5mm}

i) The ratio between $r_{RC}$ and $r_{opt}$. The extension of the rotation curve with respect to the optical radius is directly related to the S/N ratio of the data and the spatial coverage/distribution of \Ha~emission, resulting in an accurate or poor estimation of kinematics and thus influencing the quality of the rotation curve. 

ii) The number of bins computed by the adaptive spatial binning when obtaining the different momenta (see Fig. \ref{fig:histbins}). High number of bins means good S/N and/or high spatial coverage of the galaxy, and vice versa, resulting in an accurate or poor estimation of kinematics and thus influencing the quality of the rotation curve.

iii) The mean velocity error of the \rc. 

iv) The ratio between the radius of the receding and approaching side of the \rc. If one side is shorter than the other, its contribution to the \RC~will be smaller and the quality of the \rc~will be affected.

 v) We fitted a Spline function to both the receding and the approaching side in order to calculate the root mean square difference (RMSD) between both sides of the \RCs. The value of the RMSD is directly related to the symmetry of the velocity field from which we estimate the \rc~and thus influencing its quality.
 \vspace{2.5mm}

\begin{figure}
\begin{center}
\includegraphics[width=\columnwidth]{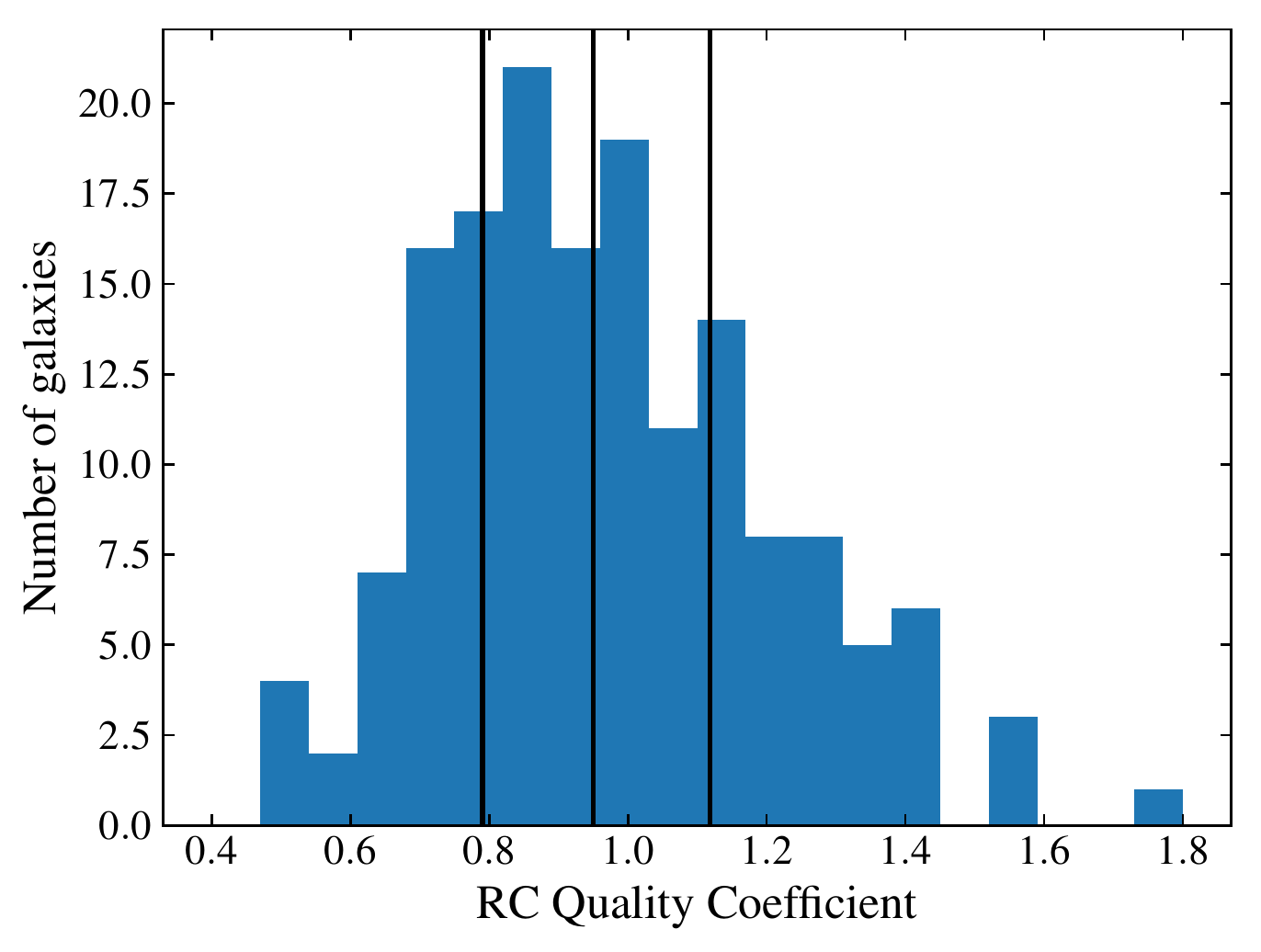}
\end{center}
\caption{ Histogram of the quality coefficient of the rotation curves. This distribution was divided in 4 quantiles with the same number of galaxies per quantil (35), in order to obtain four different Flag quality values indicated by the three vertical lines starting from the right to the left: Flag ``1", Flag ``2", Flag ``3", Flag ``4".} 
\label{fig:COURTEAUHIST}
\end{figure}

 All these five quantities were normalized in order to be summed and to obtain a quality classification coefficient to flag the rotation curves (see Fig. \ref{fig:COURTEAUHIST}). These flags are given in the Table \ref{kinparam}:
\vspace{2.5mm}

Flag 1) Accurate estimation of the \rc~and $V_{max}$, 

Flag 2) Reliable estimation of the \rc~and $V_{max}$,

 Flag 3)  Fairly reliable estimation of the \rc~and $V_{max}$ probably reached,
 
 Flag 4) Poor estimation of the \rc~and $V_{max}$ probably not reached. 
 \vspace{2.5mm}
 
In addition to these 4 flags, we defined a supplementary quality classification flags called Flag ``A" and Flag ``B". The Flag ``B" corresponds to those peculiar cases leading to a non-realistic kinematical fitting:
\vspace{2.5mm}

a) Galaxies for which $i$ $\geq$ 70$^{\circ}$, since the absorption effects due to high inclination lead to velocity underestimations along the disc, and a little variation in $i$ produces a very high variation in velocity determinations.

b) Objects for which $i$ $\leq$ 25$^{\circ}$, because a little variation in $i$ produces a very high variation in velocity determinations.

c) The difference between the morphological PA and the kinematical PA is bigger than 20$^{\circ}$. A low S/N ratio or small spatial coverage lead to poor kinematical fittings and therefore a poor estimation of the PA. On the other hand, a velocity field supported by high velocity dispersion values related to big asymmetries, strong bars, strong defined spiral arms, among other non-circular motions, will influence the determination of the kinematical PA. Finally, morphological PAs have systematically higher uncertainties than kinematical ones, specially in galaxies with low inclination as showed in \cite{Epinat:2008a}.
 
d) the $r_{RC}/r_{opt}$ ratio is lower than 0.6. 
  
e) The number of bins of the adaptive spatial binning is less than 200. Poor S/N or small spatial coverage lead easily to non-realistic kinematical fittings.
  
 f) the ratio between the \Ha~$V_{max}$ of our Fabry-Perot dataset and the $V_{max}$ of HI data \citep{Boselli:2015} is bigger than 1.5 or lower than 0.5. Since the \rc~is derived from the velocity field, poor S/N or small \Ha~emission spatial coverage, in addition to asymmetries in the velocity field, could easily lead to an under/over estimation of the V$_{max}$ of the galaxy.
   
 g) the normalized RMSD between both sides of the RC is lower than 0.6. 
\vspace{2.5mm}

\vspace{2.5mm}

\vspace{2.5mm}

\clearpage
\section{Notes on Individual Galaxies}
\label{notes}



HRS 1. Because this galaxy is not far from edge-on, the kinematical inclination was difficult to determine and we adopted the morphological value when computing the \rc. Otherwise the agreement between the kinematical and morphological PA of the major axis is very good. The \rc~is fairly regular and symmetric but does not reach the $r_{24}$ radius in $r$-band. Both models (Courteau and Zhao) suggest that it goes on rising beyond that radius.

HRS 2. There is a fairly good agreement for the PA of the major axis (4$^{\circ}$ difference only). Our \rc~almost reaches the $r_{24}$ radius in $r$-band for both sides. Courteau's model gives the best result since it extrapolates our \rc~towards the \Hi~$V_{max}$ whereas Zhao's model suggests a declining curve beyond the $r_{24}$ radius.

HRS 5. Because this galaxy is almost edge-on, the kinematical inclination was difficult to determine and we adopted the morphological value when computing our \Ha~\rc. Otherwise the agreement between the kinematical and morphological PA of the major axis is perfect. The rotation of the redshifted side appears more chaotic because of some faint \Ha~emission out of the disc which does not seem significant and could be ignored (N.B. It is not seen on the \Ha~+ [NII]  image from \citealp{Boselli:2015}). Our curve exhibits a chaotic behavior of the redshifted side but the blueshifted side is fairly regular and both sides reach the effective radius in $r$-band where the velocity is close to the \Hi~$V_{max}$ for both.

HRS 6. For this edge-on galaxy we adopted the morphological values for the inclination, as well as for the PA of the major axis, when computing our \Ha~\rc. The rotation of the blueshifted side is more chaotic and the outermost part was not taken into account when tracing the \rc. Indeed, the residual velocity field shows exceedingly large values in the outer parts of that side. The \Ha~+ [NII]  image from \cite{Boselli:2015} confirms that there is almost no emission of ionized gas beyond 30 arcsec from the nucleus on that side of the galaxy. Surprisingly, in spite of this deficiency, our \Ha~rotation velocity reaches much higher values than the \Hi~$V_{max}$ for both sides, redshifted as well as blueshifted. This is all the most surprising that our \rc~barely reaches the effective radius in $r$-band and both models (but Zhao more than Courteau) suggest that the curve goes on rising beyond that radius. 

HRS 9. The \Ha~emission is mainly seen on the southern and western part of the galaxy, where the \Ha~+ [NII]  image from \cite{Boselli:2015} also shows bright spots. The velocity field is dominated by an almost constant value around 1410 km s$^{-1}$ which could be due to a contamination of the observed velocities by an OH night-sky line (\cite{Osterbrock:1996}). As a result, the residual velocity field exhibits strong values on both sides (north and south) of the galaxy and we did not take into account the corresponding velocities when tracing our \Ha~\rc. Also, this contamination of the velocity field could explain the difference observed between kinematical and morphological inclination (9$^{\circ}$) as well as for the PA of the major axis (18$^{\circ}$). The resulting \rc~is quite chaotic for both sides and barely reaches the effective radius in $r$-band for the redshifted side. The velocity we find at this radius is close to the \Hi~$V_{max}$.    

HRS 10. We adopted the morphological value of the inclination for computing our \Ha~\rc. Otherwise there is an excellent agreement between the kinematical and morphological PA of the major axis (2$^{\circ}$ difference). The velocity field is regular and the resulting \rc~is pretty good and symmetric, reaching a plateau slightly below the \Hi~$V_{max}$. The curve almost reaches the $r_{25}$ radius in B-band (quasi coincident here with the $r_{24}$ radius in $r$-band).

HRS 11. There is an excellent agreement between the kinematical and morphological inclination (2$^{\circ}$ difference) as well as for the PA of the major axis (1$^{\circ}$ difference). The velocity field is regular, except for some high velocity points on the eastern edge of the disc (probably spurious since the \Ha~+ [NII]  image from \cite{Boselli:2015} shows no significant emission at that place). Our \Ha~\rc~is fairly regular and symmetric, it almost reaches the $r_{25}$ radius in B-band for the redshifted side.  Zhao's model suggests that the curve goes on rising towards a maximum reached at the $r_{24}$ radius in $r$-band, whereas Courteau's model suggests a continuous rising beyond that radius. However, the velocity reached by our curve at the effective radius is already much higher than the \Hi~$V_{max}$.

HRS 12. The \Ha~emission is limited to the inner disc of this galaxy and the outer parts of our velocity field are clearly not significant as suggested by the \Ha~+ [NII]  image from \cite{Boselli:2015} (showing an abrupt fall of emission on the edge of the inner disc). This is confirmed by the residual velocity field where extreme values are seen all around the disc. This peculiar concentration of the ionized gas in the center of the galaxy probably explains the discrepancy observed between the kinematical and morphological inclination (9$^{\circ}$ difference) but, more specifically, that observed between the kinematical and morphological PA of the major axis (21$^{\circ}$ difference).
Anyway the resulting \rc~is fairly symmetric (when ignoring the outer parts of the velocity field) and steeply rising up to the effective radius in $r$-band where it reaches the \Hi~$V_{max}$. Note however that Zhao's model exhibits a strange behavior while Courteau's model suggests that the curve reaches a plateau. 

HRS 13. The \Ha~+ [NII]  image from \cite{Boselli:2015} suggests that the southwestern extension of the disc, near the strong spiral arm, is spurious (which is confirmed by the extreme values of the residual velocity field) but the easternmost points, on the major axis, seem real. There is a fairly good agreement between the kinematical and morphological inclination (4$^{\circ}$ difference), also for the PA of the major axis (7$^{\circ}$ difference). The resulting \rc~exhibits a slight asymmetry between redshifted and blueshifted side but the average is satisfying and reaches a plateau close to the \Hi~$V_{max}$ just after the effective radius in $r$-band. 

HRS 17. We find an excellent agreement between the kinematical and morphological inclination (3$^{\circ}$ difference), also for the PA of the major axis (2$^{\circ}$ difference) when computing our \Ha~\rc. The resulting curve exhibits a perfect plateau (at the \Hi~$V_{max}$) extending between the effective radius and the $r_{24}$ radius in $r$-band. 

HRS 18. Although the kinematical and morphological inclinations are in perfect agreement, we find a difference of 17$^{\circ}$ between the kinematical and morphological PA of the major axis. In fact the agreement is good for the inner part of the disc but the PA changes for the outermost isophotes of this almost face-on galaxy. Anyway the velocity field exhibits a clear gradient and the resulting \Ha~\rc~is fairly symmetric but does not go very far beyond the effective radius in $r$-band. Also, the curve is continuously climbing and bends slightly after the effective radius in $r$-band, with the outermost points remaining below the \Hi~$V_{max}$.  

HRS 19. This galaxy has a nice spiral structure designed by many HII regions but it is almost face-on and exhibits a very faint velocity gradient, so that its \Ha~\rc~is hard to draw. We adopted the morphological value of the inclination when computing the \rc~and found a huge difference (53$^{\circ}$) between the kinematical and morphological PA of the major axis, likely to be explained by the shape of the outermost isophotes of the disc. The resulting \rc~is quite chaotic and reaches a radius between the effective radius and the $r_{24}$ radius in $r$-band. Also, the maximum of this curve is only half of the \Hi~$V_{max}$.

HRS 21. For this edge-on galaxy, we adopted the morphological values for the inclination as well as for the PA of the major axis, when computing the \Ha~\rc. There is a marked velocity gradient and, athough it looks like a solid body rotation, the \rc~probably reaches a plateau (as suggested by Courteau's model) beyond the effective radius in $r$-band since it reaches the \Hi~maximum there.

HRS 23. Our continuum image is contaminated by many ghost images of field stars as well as by a ghost image of the galaxy itself (about 20 arcsec below the true image of HRS023) caused by a reflection of light between the Fabry-Perot interferometer and the interference filter. The extension of the \Ha~emission on the southwestern side of the disc is apparently linked with a ghost image of a field star, otherwise our monochromatic image seems clean when comparing with the \Ha~+ [NII]  image from \cite{Boselli:2015}. The velocity field is rather irregular as confirmed by the residual velocities, displaying extreme values on both sides. Nevertheless the \Ha~\rc~is acceptable and continuously rising up to the $r_{25}$ radius (at least for the blue side) where it reaches velocities 50\% above the \Hi~$V_{max}$. Moreover, the agreement between the kinematical and morphological inclination is perfect, as well as for the PA of the major axis.

HRS 25. This galaxy exhibits a nice velocity field and a fairly regular and symmetric \Ha~\rc, reaching a plateau at the same value as the \Hi~$V_{max}$ and almost extended up to the $r_{24}$ radius in $r$-band (also not far from the $r_{25}$ radius in B-band). We find an excellent agreement between the kinematical and morphological inclination (2$^{\circ}$ difference), as well as for the PA of the major axis (less than 1$^{\circ}$ difference). Both models (Zhao and Courteau) agree that our \rc~reaches a plateau beyond the optical radius, with a rotation velocity slightly above the \Hi~$V_{max}$.

HRS 26. For this almost edge-on galaxy, we adopted the morphological values for the inclination as well as for the PA of the major axis, when computing the \Ha~\rc. Although the \Ha~emission is limited to the inner disc of this galaxy, the \rc~reaches the effective radius in $r$-band and climbs up to the \Hi~$V_{max}$.

HRS 27. Despite its faint inclination (29$^{\circ}$ according to the morphology but 21$^{\circ}$ only according to the kinematics) this small galaxy exhibits a well marked velocity gradient. The kinematical PA of the major axis is slightly different from the morphological one. The resulting \Ha~\rc~almost reaches the $r_{24}$ radius in $r$-band (at least for the blueshifted side) and seems to climb up to a plateau at the level of the \Hi~maximum velocity. 

HRS 28. There is a good agreement between the kinematical and morphological inclination, also for the PA of the major axis, and the resulting \Ha~\rc~(ignoring the northeastern extension) is fairly satisfying despite some kinks on the blueshifted side. This curve reaches the $r_{25}$ radius for both sides and seems to approach a maximum value there, above the \Hi~$V_{max}$. 

HRS 32. This face-on galaxy, although it is of Sb type, has almost none \Ha~emission, as confirmed by the \Ha~+ [NII]  image from \cite{Boselli:2015}. Our \Ha~image and velocity field are mainly noise and we could not draw any \Ha~\rc~for this galaxy. 

HRS 35. As for HRS 32, this face-on galaxy has almost none \Ha~emission. Our \Ha~image and velocity field are mainly noise and we could not draw any \Ha~\rc~for this galaxy. 

HRS 37. Although there is 11$^{\circ}$ difference between the kinematical and morphological inclination and 16$^{\circ}$ difference between the kinematical and morphological PA of the major axis, our \Ha~velocity field produces a very nice and symmetric \Ha~\rc. It just reaches the $r_{25}$ radius in B-band and probably stays flat afterwards but significantly below the \Hi~$V_{max}$ (150 km s$^{-1}$ instead of 200 km s$^{-1}$).

HRS 38. There is a good agreement between the kinematical and morphological inclination and a perfect agreement between the kinematical and morphological PA of the major axis. Our \Ha~\rc~is fairly satisfying and reaches the $r_{24}$ radius in $r$-band for both sides. The maximum velocity reached there is the same as the \Hi~$V_{max}$ and the shape of the curve suggests that it is not far from reaching a plateau. 

HRS 39. For this almost edge-on galaxy, we adopted the morphological values for the inclination as well as for the PA of the major axis, when computing the \Ha~\rc. Although our \Ha~\rc~barely reaches the effective radius in $r$-band, it is not far from the \Hi~$V_{max}$ there. This is consistent with Courteau's model suggesting that it reaches a plateau just after, whereas Zhao's model suggests a continuous rising beyond.

HRS 40. The agreement between the kinematical and morphological inclination is good and it is perfect for the PA of the major axis. Our \Ha~\rc~shows a solid body rotation up to the effective radius in $r$-band. The redshifted side suggests however that the curve reaches a plateau slightly after this radius, with a velocity matching the \Hi~$V_{max}$. This is consistent with Courteau's model whereas Zhao's model suggests a decreasing curve in the outer parts.

HRS 44. This galaxy is almost face-on, hence a faint velocity gradient explaining the bad quality of the \Ha~\rc. We adopted the morphological value of the inclination for computing the \Ha~\rc~and find a 38$^{\circ}$ difference between the kinematical and morphological PA of the major axis. Our \Ha~\rc~is very chaotic for both sides (especially the blueshifted side) but almost reaches the $r_{24}$ radius in $r$-band, with velocities greater than the \Hi~$V_{max}$ (100 km s$^{-1}$ instead of 75 km s$^{-1}$). However, the outermost parts of our \Ha~map and velocity field are very noisy and must be considered with care.

HRS 47. Our \Ha~map and velocity field cover fairly well the optical extent of this galaxy. There is a good agreement between the kinematical and morphological inclination, also for the PA of the major axis. Despite some dispersion, our \Ha~\rc~is fairly regular and symmetric; it reaches the $r_{24}$ radius in $r$-band where the velocity is close to the \Hi~$V_{max}$ (150 km s$^{-1}$). Both models (Courteau and Zhao) suggest however that the curve must go on rising beyond.

HRS 48. The spiral pattern of this galaxy is well marked and our \Ha~map and velocity field follow this structure. The velocity gradient is well marked, despite a faint inclination of the disc. We adopted the morphological value of the inclination for computing our \Ha~\rc~and find a 37$^{\circ}$ difference between the kinematical and morphological PA of the major axis. Our \rc~is quite chaotic (especially the redshifted side) but seems to reach a plateau before the effective radius in $r$-band, with a velocity close to the \Hi~$V_{max}$ (around 120 km s$^{-1}$).

HRS 50. This galaxy has a nice velocity field and \Ha~\rc. We find a good agreement between kinematical and morphological inclination, as well as for the PA of the major axis. The \rc~is perfectly symmetric and almost reaches the $r_{24}$ radius in $r$-band where the velocity is close to the \Hi~$V_{max}$. Both models (Courteau and Zhao) suggest that it reaches a plateau just after. 

HRS 52. The \Ha~+ [NII]  image of this galaxy from \cite{Boselli:2015} shows many small HII regions dispersed in the disc that we miss here. Indeed our \Ha~map and velocity field only show the central part of the galaxy and none of these small regions. We find a fairly good agreement between the kinematical and morphological value of the PA of the major axis. Our \rc~shows a steep rise and bends clearly before reaching the effective radius in $r$-band. The velocity almost reaches the \Hi~$V_{max}$ there and both models suggest that the \rc~reaches its maximum there. However, Courteau's model suggests a plateau whereas Zhao's model suggests a decreasing curve beyond the effective radius.

HRS 56. The resulting curve is fairly symmetric and reaches a plateau between the effective radius and the $r_{24}$ radius in $r$-band, with a velocity close to the \Hi~$V_{max}$ (around 190 km s$^{-1}$). Here again Courteau's model suggests a plateau whereas Zhao's model suggests a decreasing curve beyond the effective radius. 

HRS 59. This galaxy is almost edge-on and the blueshifted side (west) on our \Ha~map seems brighter and more extended than the redshifted side (east), maybe because of dust absorption. We adopted the morphological values for the inclination as well as for the PA of the major axis, when computing our \Ha~\rc. Despite some dispersion, our curve is fairly symmetric up to the effective radius where, after a steep rising, it reaches the \Hi~$V_{max}$. Beyond that radius, the curve is  mainly traced by the blueshifted side which is quite chaotic. Anyway, the curve seems to follow a plateau at about 200 km s$^{-1}$, between the effective radius and the $r_{24}$ radius in $r$-band (coincident here with the $r_{25}$ radius in B-band).

HRS 60 (KPG 209B). The \Ha~emission is very faint on the redshifted side (south) of this barred galaxy, as can be seen also on the \Ha~+ [NII]  image from \cite{Boselli:2015}. As a result, our velocity field and \Ha~\rc~mainly rely on the blueshifted side (north). The agreement between kinematical and morphological inclination is not very good (13$^{\circ}$ difference) but it is better for the PA of the major axis (5$^{\circ}$ difference). Our \rc~is very chaotic, perhaps also because of the strong bar in the center. The curve barely reaches the effective radius in $r$-band where its velocity seems to stop rising although it is much smaller than the \Hi~$V_{max}$ (about 125 km s$^{-1}$ compared with more than 300 km s$^{-1}$). 

HRS 61. This galaxy is almost edge-on and the agreement between kinematical and morphological inclination is not very good (9$^{\circ}$ difference) but it is perfect for the PA of the major axis. The outermost extension to the north of our \Ha~map and velocity field is doubtful (there is no hint of any counterpart there on the \Ha~+ [NII]  image from \citealp{Boselli:2015}). It gives large values on the residual velocity field and it was not taken into account when tracing our \Ha~\rc. Our curve is asymmetric, with a redshifted side rising steeper than the blueshifted side.  The curve almost reaches the $r_{25}$ radius in B-band where its velocity is slightly above the \Hi~$V_{max}$. Both models (Courteau and Zhao) suggest that the \rc~goes on rising beyond the $r_{25}$ radius. 

HRS 62. In spite of a slight disagreement between kinematical and morphological inclination (7$^{\circ}$ difference) as well as for the PA of the major axis (9$^{\circ}$ difference) our \Ha~\rc~is fairly symmetric and regular. The curve reaches the $r_{25}$ radius in B-band (coincident with the $r_{24}$ radius in $r$-band for this galaxy) and its velocity there is slightly above the \Hi~$V_{max}$. The curve extends beyond the $r_{25}$ on the blueshifted side and both models (Courteau and Zhao) suggest that the plateau is not reached.

HRS 63. There is a good agreement between kinematical and morphological inclination, as well as for the PA of the major axis. Our \Ha~\rc~is fairly symmetric and regular, almost reaching the $r_{24}$ radius in $r$-band for both sides. The curve reaches its maximum there, with a velocity close to the \Hi~$V_{max}$ (around 160 km s$^{-1}$) although Courteau's model suggests that it goes on rising beyond that point.  

HRS 64. This galaxy is almost edge-on and we adopted the morphological values for the inclination as well as for the PA of the major axis, when computing our \Ha~\rc. In spite of some dispersion, our curve is fairly symmetric and almost reaches the $r_{24}$ radius in $r$-band for the blueshifted side. The maximum reached is not far from the \Hi~$V_{max}$ and both models (Courteau and Zhao) suggest that our curve reaches a plateau at the $r_{24}$ radius (although slightly declining beyond that point for Zhao). 

HRS 65. The \Ha~emission is rather patchy and very faint outside of the spiral pattern of the galaxy. Anyway, we find a fairly good agreement between kinematical and morphological inclination, as well as for the PA of the major axis. Our \Ha~\rc~is quite chaotic and asymmetric but reaches the $r_{25}$ radius in B-band (almost coincident here with the $r_{24}$ radius in $r$-band). The velocity there is slightly below the \Hi~$V_{max}$ but both models (Courteau and Zhao) suggest that the curve, although clearly bending, is still rising beyond that radius.

HRS 66. We find a good agreement between kinematical and morphological inclination, as well as for the PA of the major axis. Our \Ha~\rc~is fairly regular and symmetric, it almost reaches the $r_{24}$ radius in $r$-band for both sides. The maximum velocity is slightly above the \Hi~$V_{max}$ and both models (but Courteau better than Zhao) suggest that the curve is about to reach a plateau after that radius. 

HRS 67. The faint extension of our \Ha~map and velocity field, on the southern side of the galaxy,  seems spurious (no counterpart can be seen on the \Ha~+ [NII]  image from \citealp{Boselli:2015}). We adopted the morphological value of the inclination for computing our \Ha~\rc~and find a 25$^{\circ}$ difference between the kinematical and morphological PA of the major axis. Anyway our \Ha~\rc, despite some dispersion, is fairly regular and symmetric. The curve barely reaches the effective radius in $r$-band, where the velocity is found to be close to the \Hi~$V_{max}$. Both models (Courteau and Zhao) suggest that the curve goes on rising beyond the $r_{24}$ radius in $r$-band but Courteau's model favors a plateau after that radius.

HRS 68. This galaxy is almost face-on and the \Ha~emission is concentrated in the center of the disc, as confirmed by the \Ha~+ [NII]  image from \cite{Boselli:2015}. As a result, the dispersion is very high in the outer parts of the velocity field, as can be seen also on the residual velocity field where extreme values are reached on both sides (blueshifted and redshifted). Furthermore, we adopted the morphological value of the inclination for computing our \Ha~\rc~and find a huge 57$^{\circ}$ difference between the kinematical and morphological PA of the major axis. The resulting curve is quite chaotic and asymmetric, barely reaching a radius of 5 arcsec (about half the $r_{25}$ radius in B-band). Also, the velocity remains below 100 km s$^{-1}$ whereas the \Hi~$V_{max}$ is around 160 km s$^{-1}$.  

HRS 70. We find an acceptable difference (12$^{\circ}$) between the kinematical and morphological PA of the major axis. Despite this difference and some dispersion of the velocities in the very center, our \rc~is fairly regular and symmetric. It reaches the $r_{24}$ radius in $r$-band for both sides, with a rotation velocity close to the \Hi~$V_{max}$. Both models (but Courteau more than Zhao) suggest that the curve is about to reach a plateau beyond that radius. 

HRS 72. Our \Ha~velocity field clearly suggests that this galaxy is not as inclined as suggested by its morphology (59$^{\circ}$ instead of 79$^{\circ}$). Also, the kinematical PA of the major axis is found to be 28$^{\circ}$ different from that suggested by the outermost isophotes of the disc. The resulting \Ha~\rc~is fairly regular and symmetric, revealing a pure solid body rotation up to the $r_{24}$ radius in $r$-band. The rotation velocity at this radius is significantly higher than the \Hi~$V_{max}$ (about 160 km s$^{-1}$ against 110 km s$^{-1}$). 

HRS 74. Although this galaxy has a rather faint inclination, our \Ha~velocity field exhibits a marked velocity gradient. We find an acceptable difference (13$^{\circ}$) between the kinematical and morphological PA of the major axis. Despite some lack of \Ha~emission in the very center (confirmed by the \Ha~+ [NII]  image from \citealp{Boselli:2015}) the \rc~can be traced with confidence and clearly shows that a maximum is reached for both sides at the effective radius in $r$-band. This maximum is significantly higher than the \Hi~$V_{max}$ (about 190 km s$^{-1}$ instead of 165 km s$^{-1}$). Zhao's model suggests that the curve goes on declining beyond, whereas Courteau's model favors a plateau. 

HRS 76. The lack of \Ha~emission on the northern side (redshifted) of this galaxy is confirmed by   the \Ha~+ [NII]  image from \cite{Boselli:2015}. The high inclination of the disc led us to adopt the morphological values for the inclination and for the PA of the major axis when computing our \Ha~\rc. Despite a rather strong dispersion, the \rc~is acceptable for the blueshifted side and seems to reach a plateau beyond the effective radius (but the redshifted side is difficult to exploit). Our maximum rotation velocity is close to the \Hi~$V_{max}$.

HRS077. Our \Ha~image and velocity field covers at best the optical disc of this galaxy. The agreement between morphological and kinematical values is excellent for the inclination as well as for the PA of the major axis. Our \Ha~\rc~is perfectly symmetric and, after a very steep rise, the velocity remains constant beyond the effective radius and close to the \Hi~$V_{max}$. The plateau extends almost out to the $r_{24}$ radius in $r$-band (close here to the $r_{25}$ radius in B-band). 

HRS 78. The agreement between morphological and kinematical values is rather good for the inclination as well as for the PA of the major axis. The \Ha~emission is brighter on the northern side (blueshifted) of this galaxy, as can be seen also on the \Ha~+ [NII]  image from \cite{Boselli:2015}. As a result, the \rc~is better traced by the blueshifted side whereas the redshifted side is quite chaotic (also the residual velocity field shows some extreme values on the redshifted side). Our curve almost reaches the $r_{24}$ radius in $r$-band and both models (Courteau and Zhao) suggest that it goes on rising for a while. Indeed the maximum rotation velocity reached by our curve is still much below the \Hi~$V_{max}$ (almost half).

HRS 79. This galaxy is almost edge-on and we adopted the morphological values for the inclination and for the PA of the major axis when computing our \Ha~\rc. The points surrounding the main body of the galaxy on our different maps are but noise and are not taken into account for tracing our \Ha~\rc. The \Ha~emission is more extended on the blueshifted side and the \rc~is also more extended on that side, going beyond the $r_{25}$ radius in B-band whereas it just outpasses the effective radius on the redshifted side. The maximum velocity rotation of our curve is clearly higher than the \Hi~$V_{max}$ (130 km s$^{-1}$ instead of 100 km s$^{-1}$) and both models (but Zhao better than Courteau) suggest that the curve is not far from reaching a plateau beyond the $r_{25}$ radius.

HRS 80. The \Ha~emission of this galaxy is very patchy and irregular, with only a few HII regions on the redshifted side (south). Furthermore its disc has a faint inclination and the velocity gradient is faint, leading us to adopt the morphological values for the inclination and for the PA of the major axis when computing our \Ha~\rc. Despite some dispersion, our \rc~is fairly symmetric and both sides exhibit a strange behavior in the central part, with a clear bump within the 10 first arsec. Then the curve is mainly traced by the blueshifted side, rising up steeply before bending slightly.  It extends halfway between the effective radius and the $r_{24}$ radius in $r$-band where its velocity is close to the \Hi~$V_{max}$. Zhao's model suggests that the \rc~goes on rising whereas Courteau's model favors a plateau beyond the effective radius.

HRS 82 (KUG 1201+163). The \Ha~emission is concentrated in some bright spots in the very center of the disc of this galaxy. We find an excellent agreement between the kinematical and morphological PA of the major axis. The central part of our \rc~rises steeply (steeper for the redshifted side) with a solid body rotation, almost out to the effective radius. Then the redshifted side goes on rising  (up to the \Hi~$V_{max}$ of 80 km s$^{-1}$) whereas the blueshifted side reaches a plateau at 60 km s$^{-1}$. Courteau's model suggests that the \rc~reaches a plateau at 75 km s$^{-1}$ beyond the $r_{24}$ radius in $r$-band whereas Zhao's model suggests that the curve reaches a maximum there before declining beyond. 

HRS 83. The \Ha~emission is limited to the inner part of the disc of this galaxy. Because of the strong inclination, we adopted the morphological value of the inclination for computing our \Ha~\rc~and find a perfect agreement between the kinematical and morphological PA of the major axis. Our \rc~almost reaches the effective radius in $r$-band and exhibits a solid body rotation for both sides. Our maximum rotation velocity is close to the \Hi~$V_{max}$ and it is probable that we reach a plateau there (as suggested by Courteau's model) whereas Zhao's model suggests that it goes on rising beyond.

HRS 84. The \Ha~emission is much brighter on the southern part of the disc of this galaxy, as can be seen also on the \Ha~+ [NII]  image from \cite{Boselli:2015}. As a result, our \Ha~\rc~is mainly traced by the blueshifted side (south). Also, there is some lack of emission in the very center of the disc, so that we have no points to trace the rising part of the \rc~which is very steep here. The agreement between morphological and kinematical values is very good for the inclination as well as for the PA of the major axis. Our \rc~almost reaches the $r_{25}$ radius in B-band (here slightly shorter than the $r_{24}$ radius in $r$-band). We find rotation velocities higher than the \Hi~$V_{max}$ beyond the effective radius and both models (Courteau and Zhao) suggest that it goes on rising beyond our last measured points.

HRS 86 (KPG 322A). The \Ha~emission covers fairly well the optical disc of this galaxy, although the \Ha~+ [NII]  image from \cite{Boselli:2015} suggests that we miss some faint HII regions in the southern part of the disc. The agreement between morphological and kinematical values is not too bad for the inclination (10$^{\circ}$ difference) as well as for the PA of the major axis (7$^{\circ}$ difference). Our \Ha~\rc~is fairly regular and symmetric, with a solid body rotation shape up to the $r_{24}$ radius in $r$-band where it reaches a velocity close to the \Hi~$V_{max}$. Despite some dispersion, the outermost points of both sides suggest that the curve reaches its maximum there and the blueshifted side exhibits a plateau extending beyond the $r_{25}$ radius in B-band.  

HRS 88 (KPG 322B). Despite some holes, the \Ha~emission covers fairly well the optical disc of this galaxy.  The agreement between morphological and kinematical values is not too bad for the inclination (13$^{\circ}$ difference) but rather bad for the PA of the major axis (31$^{\circ}$ difference). Our \Ha~\rc~is quite chaotic in its central rising part, especially for the redshifted side (the residual velocity field shows extreme values on that side, in the inner part of the disc, and more generally  all around the nucleus). The \rc~is more quiet and symmetric beyond the effective radius and reaches the $r_{24}$ radius in $r$-band for both sides. Altough we find a rotation velocity significantly higher than the \Hi~$V_{max}$, even before reaching the effective radius, both models (Courteau and Zhao) suggest that our curve goes on rising much beyond our last measured points.

HRS 92. This galaxy is almost edge-on and we adopted the morphological values for the inclination and for the PA of the major axis when computing our \Ha~\rc. The signal to noise is rather faint, especially on the redshifted side (northwest) and our \rc~is quite chaotic, barely reaching the effective radius. The curve exhibits a solid body rotation reaching a maximum velocity around 120 km s$^{-1}$. Courteau's model suggests a plateau slightly below the \Hi~$V_{max}$ (150 km s$^{-1}$) beyond the effective radius whereas Zhao's model favors a velocity peak at the effective radius.   

HRS 95. The very thin stripe of \Ha~emission of this galaxy, as can be seen also on the \Ha~+ [NII]  image from \cite{Boselli:2015}, led to think that its disc is more inclined than suggested by the morphological value (56$^{\circ}$).  Anyway, we adopted that value when tracing our \Ha~\rc. We find a rather good agreement between morphological and kinematical values for the PA of the major axis (7$^{\circ}$ difference).  Our \rc~exhibits a solid body rotation for both sides, with a steep rise up to the effective radius in $r$-band. Then the curve bends clearly and reaches a maximum for both sides (around 85 km s$^{-1}$ for the redshifted side and 110 km s$^{-1}$ for the blueshifted side, slightly below the 115 km s$^{-1}$ of the \Hi~$V_{max}$). Courteau's model suggests that our curve reaches a plateau there, whereas Zhao's model favors a velocity peak at the effective radius. 

HRS 98. This galaxy is almost edge-on and we adopted the morphological values for the inclination and for the PA of the major axis when computing our \Ha~\rc. The curve is fairly regular and symmetric but barely reaches the effective radius in $r$-band. Courteau's model suggests that the curve clearly bends after the effective radius, reaching a plateau close to the \Hi~$V_{max}$ at the optical radius, whereas Zhao's model suggests that it goes on rising much beyond.   

HRS 104. This Sb type galaxy has no significant \Ha~emission, as confirmed by the \Ha~+ [NII]  image from \cite{Boselli:2015}. No \Ha~velocity field nor \rc~could be extracted from our data. Note also that the isolated spot at about 30 arcsec north from the nucleus is probably but noise (maybe a ghost image).

HRS 106. The \Ha~emission covers fairly well the optical disc of this galaxy. Because of its faint inclination, we adopted the morphological value for the inclination when tracing our \Ha~\rc. We find a rather good agreement between morphological and kinematical values for the PA of the major axis (8$^{\circ}$ difference). Despite some chaotic behavior in the center (first 5 arcsec), our \rc~is fairly regular and symmetric. The curve reaches the effective radius for both sides and the blueshifted side (south) extends out to the $r_{24}$ radius in $r$-band. The rotation velocity reached there is close to the \Hi~$V_{max}$. Anyway both models (Courteau as well as Zhao) suggest that the curve goes on rising beyond that radius.

HRS 107. This galaxy is almost edge-on and we adopted the morphological values for the inclination and for the PA of the major axis when computing our \Ha~\rc. The velocity gradient of the velocity field is well marked but the signal to noise is rather bad, so that our \rc~suffers from dispersion. Our curve does not even reach the effective radius in $r$-band and the maximum velocity reached is between 100 km s$^{-1}$ and 110 km s$^{-1}$, depending on the side, below the \Hi~$V_{max}$ (close to 130 km s$^{-1}$). 

HRS 112. There is some diffuse \Ha~emission in the disc of this galaxy but its surface brightness is very faint, as can be seen also on the \Ha~+ [NII]  image from \cite{Boselli:2015}. As a result, the signal to noise ratio of our \Ha~map is also very faint and, despite the relatively high inclination of this galaxy, no velocity gradient can be guessed on our velocity field. We adopted the morphological values for the inclination and for the PA of the major axis when computing our \Ha~\rc, but no realistic curve could be obtained.

HRS 117. This galaxy is almost edge-on and we adopted the morphological values for the inclination and for the PA of the major axis when computing our \Ha~\rc. The \Ha~emission is limited to a few spots in the central part of the disc of this galaxy, mainly on the southwestern side, as can be seen also on the \Ha~+ [NII]  image from \cite{Boselli:2015}. As a result, our \rc~is only traced by the blueshifted side (southwest) but nevertheless almost reaches the effective radius in $r$-band. The rotation velocity reached at this radius is close to the \Hi~$V_{max}$ and both models (Courteau and Zhao) suggest that the curve goes on rising beyond.

HRS 118. Apart a few bright spots in the center, the \Ha~emission of this galaxy is rather faint in this galaxy, as can be seen also on the \Ha~+ [NII]  image from \cite{Boselli:2015}. Nevertheless it covers fairly well the optical disc of the galaxy. We adopted the morphological value for the inclination and find an acceptable agreement between the morphological and kinematical values of the PA of the major axis (26$^{\circ}$ difference) when tracing our \Ha~\rc. The resulting curve is fairly regular and symmetric, almost reaching the $r_{24}$ radius in $r$-band for both sides. The rotation velocity reached there is close to the \Hi~$V_{max}$, however some patches of \Ha~emission northward of the disc enable to extend our \Ha~\rc~much beyond the $r_{25}$ radius in B-band on the redshifted side, suggesting that the \rc~goes on rising beyond the optical disc. Note that Courteau's model fails to fit our data correctly, contrary to Zhao's model.

HRS 121. For this galaxy, when tracing our \Ha~\rc, we adopted the morphological values for the inclination as well as for the PA of the major axis. The resulting curve is fairly regular and symmetric, extending midway between the effective radius and the $r_{24}$ radius in $r$-band (almost coincident here with the $r_{25}$ radius in B-band). Both sides of the curve (although blueshifted better than redshifted) almost reach the \Hi~$V_{max}$. Zhao's model suggests that the maximum of our \Ha~curve is almost reached by our outermost blue point whereas Courteau's model favors a continuous rising, even beyond the optical radius.

HRS 132. The \Ha~emission is mainly produced by some bright spots lying along the major axis. This galaxy is almost face-on. The \rc~is quite chaotic for both sides, more especially the redshifted side (northeast) for which there is a counter rotation in the first 5 arcsec, with negative values, before the curve rises steeply. The residual velocity field exhibits high values there, but also extreme values all around the disc. Anyway, both sides of the curve extend beyond the effective radius in $r$-band, with maximum values of the velocity rotation close to the \Hi~$V_{max}$.

HRS 133. The  comparison of our \Ha~map with the \Ha~+ [NII]  image from \cite{Boselli:2015} show that we miss the outermost emission regions on the western side (redshifted) of the disc. This galaxy is almost edge-on and we adopted the morphological values for the inclination as well as for the PA of the major axis when tracing our \rc. Both sides of the curve are very chaotic in the central part (first 15 arcsec) then the trend is that of a solid body rotation. The curve barely reaches the effective radius in $r$-band, where its rotation velocity is clearly below the \Hi~$V_{max}$. Indeed both models (Courteau and Zhao) suggest that our \Ha~curve goes on rising and reaches the \Hi~$V_{max}$ at the $r_{24}$ radius in $r$-band.  

HRS 134. Although this edge-on galaxy is classified as Sc type, its \Ha~emission is limited to two bright spots on each side of the nucleus, as confirmed by the \Ha~+ [NII]  image from \cite{Boselli:2015}. We adopted the morphological values for the inclination as well as for the PA of the major axis when tracing our \Ha~\rc. However, because of the lack of velocity points, no significant curve could be obtained.

HRS 136. This Sa type galaxy has no significant \Ha~emission, as confirmed by the \Ha~+ [NII]  image from \cite{Boselli:2015}. The spots on our maps are probably more noise than signal and no \Ha~\rc~can be obtained for this galaxy.

HRS 139. When tracing the \Ha~\rc~of this galaxy, we find a fairly good agreement between the morphological and kinematical values of the PA of the major axis (15$^{\circ}$ difference). The resulting \rc~is fairly regular and symmetric, with both sides extending midway between the effective radius and the $r_{24}$ radius in $r$-band. The rotation velocity reaches the \Hi~$V_{max}$ at the effective radius and the curve goes on rising beyond although with a smaller slope. 

HRS 141. The \Ha~emission of this galaxy is rather faint and patchy but the velocity field is nice. We find a good agreement between the morphological and kinematical values of the inclination (6$^{\circ}$ difference) and a perfect agreement between the morphological and kinematical values of the PA of the major axis (less than 2$^{\circ}$ difference). The resulting \Ha~\rc~is fairly regular and symmetric and reaches the effective radius in $r$-band for both sides. The rotation velocity reached there is slightly above the \Hi~$V_{max}$ and both models (Courteau and Zhao) suggest that it goes on rising beyond, although with a smaller slope.

HRS 142. There is a strong \Ha~emission in the inner disc of this Sa type galaxy but it is very faint beyond the effective radius, as can be seen also on the \Ha~+ [NII]  image from \cite{Boselli:2015}. Despite the relatively high inclination of the disc, the gradient of the \Ha~velocity is very faint, suggesting that the behavior of the ionized gas do not reflect that of the main body of the galaxy. Also, the outer parts of our velocity field have a very low signal to noise ratio and could be contaminated by a nightskyline. We adopted the morphological value for the inclination when tracing our \rc. The first 15 arcsec of the curve show a rising part fairly symmetric for both sides, almost out to the effective radius in $r$-band. Beyond that point the curve becomes quite chaotic but with a trend to decrease for both sides, casting doubts on the quality of our data in the outer parts of the disc of this galaxy. However, the fact that the rotation velocity reached at the effective radius is much below the \Hi~$V_{max}$ (20 km s$^{-1}$ against 120 km s$^{-1}$) favors our hypothesis of a peculiar behavior of the ionized gas in the very center of this galaxy. Could it be the remnant of a merger?
 
HRS 143. This galaxy is almost edge-on and we adopted the morphological values for the inclination and for the PA of the major axis when computing our \Ha~\rc. The resulting \rc~is fairly regular and symmetric, with both sides extending midway between the effective radius and the $r_{24}$ radius in $r$-band. Courteau's model suggests that the curve goes on rising beyond and reaches the \Hi~$V_{max}$ between the $r_{24}$ radius in $r$-band and the  $r_{25}$ radius in B-band, whereas Zhao's model favors a maximum reached at the $r_{24}$ radius, with a rotation velocity slightly below the \Hi~$V_{max}$. 

HRS 145. We adopted the morphological value for the inclination of the disc of this galaxy and find a good agreement between the morphological and kinematical values of the PA of the major axis (4$^{\circ}$ difference) when computing our \Ha~\rc. The resulting curve is somewhat chaotic in the center (first 7 arcsec) where the \Ha~emission is less bright, otherwise it is fairly regular and symmetric, with both sides extending midway between the effective radius and the $r_{24}$ radius in $r$-band. The rotation velocity reached at the effective radius is almost equal to the \Hi~$V_{max}$ but the curve goes on increasing beyond and none of the models (Courteau as well as Zhao) suggests any significant change in the slope beyond that radius.

HRS 146. This galaxy is almost edge-on and we adopted the morphological values for the inclination and for the PA of the major axis when computing our \Ha~\rc. Despite some dispersion, the resulting \rc~is rather regular and symmetric, but barely reaches the effective radius in $r$-band. The rotation velocity reached there is around 70 km s$^{-1}$ (against 90 km s$^{-1}$ for the \Hi~$V_{max}$). Courteau's model suggests that the \rc~reaches a plateau beyond the optical radius (with a velocity close to the \Hi~$V_{max}$) whereas Zhao's model favors a continuous increase beyond.

HRS 148. The disc of this almost edge-on galaxy appears more extended on the southeastern side (redshifted), both on continuum images and \Ha~images (also on the \Ha~+ [NII]  image from \citealp{Boselli:2015}), maybe because of dust absorption. We adopted the morphological values for the inclination and for the PA of the major axis when computing our \Ha~\rc. The resulting \rc~is fairly regular and symmetric, with both sides extending up to the effective radius in $r$-band, where the rotation velocity is equal to the \Hi~$V_{max}$. The redshifted side enables to draw the curve up to the $r_{25}$ radius in B-band (here almost coincident with the $r_{24}$ radius in $r$-band). This side of the curve seems to reach a plateau at the effective radius but rises again shortly after, with a significant slope, although it seems to diminish when reaching the optical radius.    

HRS 149. This galaxy is almost edge-on and we adopted the morphological values for the inclination and for the PA of the major axis when computing our \Ha~\rc. The resulting \rc~is perfectly regular and symmetric, with both sides extending up to the effective radius in $r$-band, where the rotation velocity is slightly above the \Hi~$V_{max}$. The shape of the curve suggests that it goes on rising beyond the effective radius, in agreement with both models (Courteau and Zhao).

HRS 151. We find a significant difference between the morphological and kinematical values of the PA of the major axis (21$^{\circ}$ difference) when computing our \Ha~\rc. The resulting curve is quite chaotic and asymmetric, anyway both sides extend slightly beyond the effective radius in $r$-band and bend slightly there, suggesting that the curve is about to reach a plateau, with a rotation velocity around the \Hi~$V_{max}$. Both models (Courteau as well as Zhao) suggest however that the curve goes on rising much beyond, even after the optical radius.

HRS 152. This galaxy is not far from face-on, we adopted the morphological value for the inclination of its disc and find a good agreement between the morphological and kinematical values of the PA of the major axis (6$^{\circ}$ difference) when computing our \Ha~\rc. The resulting \rc~is fairly regular and symmetric, extending slightly beyond the effective radius in $r$-band (N.B. The redshifted side almost extends midway between the effective radius and the $r_{24}$ radius in $r$-band). Courteau's model suggests that the \rc~is not far from reaching a plateau beyond the optical radius (with a rotation velocity slightly above the \Hi~$V_{max}$) whereas Zhao's model favors a continuous rising in the outer parts.

HRS 153. The disc of this galaxy is peppered with many HII regions covering its whole optical extent. It is faintly inclined and we adopted the morphological value for the inclination of the disc of this galaxy while finding a fairly good agreement between the morphological and kinematical values of the PA of the major axis (10$^{\circ}$ difference) when computing our \Ha~\rc. The resulting \rc~is regular and symmetric, extending almost out to the $r_{24}$ radius in $r$-band for both sides. Both models (but Courteau better than Zhao) suggest that the \rc~reaches a plateau after the optical radius, with a rotation velocity close to the \Hi~$V_{max}$.

HRS 154. This galaxy is not far from face-on, we adopted the morphological value for the inclination of its disc and find a large difference between the morphological and kinematical values of the PA of the major axis (56$^{\circ}$ difference) when computing our \Ha~\rc. Despite this disagreement, the resulting \rc~is not so bad and, despite a strong dispersion in the central rising part (especially for the redshifted side) both sides exhibit the same trend and extend out to the $r_{24}$ radius in $r$-band, almost reaching a plateau there, with a rotation velocity close to the \Hi~$V_{max}$, as suggested by both models (Courteau and Zhao).

HRS 156. This galaxy is almost edge-on and we adopted the morphological values for the inclination and for the PA of the major axis when computing our \Ha~\rc. The resulting \rc~is fairly regular and symmetric, with both sides extending up to the effective radius in $r$-band, where the rotation velocity is almost twice the \Hi~$V_{max}$. Zhao's model suggests that the maximum of the \Ha~\rc~is reached there, whereas Courteau's model suggests that it continuous rising after the effective radius and reaches a plateau beyond the optical radius.

HRS 157. The \Ha~emission is rich and perfectly covers the whole optical disc of this Sbc type galaxy. We find a perfect agreement between the morphological and kinematical values of the inclination and a rather good agreement between the morphological and kinematical values of the PA of the major axis (7$^{\circ}$ difference). The resulting \Ha~\rc~is fairly regular and symmetric and reaches the $r_{24}$ radius in $r$-band for both sides. The rotation velocity reached there is slightly above the \Hi~$V_{max}$ and both models (Courteau and Zhao) suggest that it goes on rising beyond, although with a smaller slope for Courteau's model, favoring a plateau beyond the optical radius.

HRS159. The \Ha~emission is limited to the inner disc of this Sa type galaxy, as can be seen also on the \Ha~+ [NII]  image from \cite{Boselli:2015}. Despite the relatively high inclination of the disc, the velocity gradient of the \Ha~velocity field of this galaxy is rather faint but, most strangely, it suggests that the PA of the major axis is almost at right angle with respect to the morphological value (92$^{\circ}$ difference). All of this suggests that the behavior of the ionized gas do not reflect that of the main body of the galaxy. Anyway, we traced our \Ha~\rc~with the kinematical values suggested by our velocity field (85$^{\circ}$ for the inclination and 0$^{\circ}$ for the PA of the major axis, against 62$^{\circ}$ and 92$^{\circ}$ respectively for the morphological values). The resulting \rc~is quite chaotic but suggests a solid body rotation for both sides of the central disc of ionized gas. The curve does not reach the effective radius but its maximum rotation velocity is close to the \Hi~$V_{max}$. The strange behavior of the ionized gas in the center of this Sa galaxy leads to think that it could be the remnant of a merger \citep{Boselli:NGC4424}.

HRS 160 (KPG 338A). We find a significant difference between the morphological and kinematical values of the inclination (17$^{\circ}$) as well as for the PA of the major axis (17$^{\circ}$ difference) when computing our \Ha~\rc. Despite some chaotic behavior within the first 10 arcsec, the resulting curve is nevertheless fairly regular and symmetric, extending beyond the effective radius in the $r$-band. The maximum rotation velocity reached is however clearly below the \Hi~$V_{max}$. Zhao's model suggests that the curve goes on rising with an almost constant slope, whereas Courteau's model suggests that we almost reach a plateau. 

HRS 164. No significant \Ha~emission can be detected in this Sa type galaxy, as confirmed by the \Ha~+ [NII]  image from \cite{Boselli:2015}. We could not draw any velocity field nor \rc.

HRS 165. This galaxy is almost edge-on and we adopted the morphological values for the inclination and for the PA of the major axis when computing our \Ha~\rc. The resulting \rc~is quite chaotic but reveals a solid body rotation for both sides, extending up to the effective radius in $r$-band, where the rotation velocity reaches the \Hi~$V_{max}$. Zhao's model suggests that the \rc~goes on rising beyond, whereas Courteau's model suggests that it reaches a plateau at this radius.

HRS 168. We find a fairly good agreement between the morphological and kinematical values of the PA of the major axis (9$^{\circ}$ difference) when computing our \Ha~\rc. The resulting curve is regular and symmetric, extending beyond the $r_{24}$ radius in $r$-band for both sides. Both models (but Courteau better than Zhao) suggest that our curve reaches a plateau at the optical radius, with a rotation velocity slightly below the \Hi~$V_{max}$.

HRS 169. We adopted the morphological value for the inclination of the disc of this galaxy and find a good agreement between the morphological and kinematical values of the PA of the major axis (6$^{\circ}$ difference) when computing our \Ha~\rc. Despite some dispersion for the redshifted side, the resulting curve is fairly regular and symmetric, extending beyond the effective radius in $r$-band for both sides and almost out to the $r_{25}$ radius in B-band for the blueshifted side. The rotation velocity reaches the \Hi~$V_{max}$ at the effective radius but both models (Zhao more than Courteau) suggest that our curve goes on rising much beyond the optical radius.

HRS 171. The \Ha~emission of this Sbc galaxy is very bright but limited to the inner disc. The extension that can be seen on the northern side of the galaxy on our maps seems doubtful when comparing with the \Ha~+ [NII]  image from \cite{Boselli:2015} and we did not take it into account when tracing our \Ha~\rc~(indeed, this extension produces extreme values on the residual velocity field). We adopted the morphological value for the inclination of the disc of this galaxy and find a good agreement between the morphological and kinematical values of the PA of the major axis (6$^{\circ}$ difference) when computing our \Ha~\rc. The resulting curve is fairly regular and symmetric, it reaches the effective radius in $r$-band for both sides, with a rotation velocity close to the \Hi~$V_{max}$. Zhao's model suggests that our curve reaches a maximum between the effective radius and the $r_{24}$ radius, whereas Courteau's model suggests that it goes on rising much beyond. 

HRS 172. The \Ha~emission of this Sb galaxy is very bright but limited to the inner disc. We find a rather good agreement  between the morphological and kinematical values of the inclination (10$^{\circ}$ difference) and a perfect agreement for the PA of the major axis (1$^{\circ}$ difference) when computing our \Ha~\rc. The resulting curve is fairly regular and symmetric but does not even reach the effective radius in $r$-band. The rotation velocity reached is close to the \Hi~$V_{max}$ but both models (Courteau and Zhao) suggest that the curve goes on rising much beyond.

HRS 177. This Sa type galaxy has a surprisingly rich \Ha~emission, covering almost all of its optical disc. We find a good agreement between the morphological and kinematical values of the inclination (5$^{\circ}$ difference) and a rather good agreement for the PA of the major axis (10$^{\circ}$ difference) when computing our \Ha~\rc. The resulting curve is fairly regular and symmetric, it extends midway between the effective radius and the $r_{24}$ radius in $r$-band. The rotation velocity reached there is slightly below the \Hi~$V_{max}$. Zhao's model suggests that the \rc~goes on rising beyond the $r_{24}$ radius, whereas Courteau's model suggests that it reaches a plateau at this radius, with a rotation velocity remaining below the \Hi~$V_{max}$.

HRS 182. We find a good agreement between the morphological and kinematical values of the inclination (6$^{\circ}$ difference) and a perfect agreement for the PA of the major axis (1$^{\circ}$ difference) when computing our \Ha~\rc. The resulting curve is fairly regular and symmetric, it extends almost out to the $r_{25}$ radius in B-band (close here from the $r_{24}$ radius in B-band) at least for the blueshifted side. The rotation velocity reaches the \Hi~$V_{max}$ at the effective radius in $r$-band and both models (Zhao and Courteau) suggest that the \Ha~\rc~reaches a plateau at the optical radius, with a rotation velocity slightly above the \Hi~$V_{max}$.

HRS 184. There is but a small patch of \Ha~emission in the center of this Sa type galaxy, as confirmed by the \Ha~+ [NII]  image from \cite{Boselli:2015}. The other small spots on our maps are but noise or ghost image. We could not draw any realistic velocity field nor \rc.

HRS 185. The \Ha~emission of this Sa galaxy has a faint surface brightness and is mainly located in a central ring of about 10 arcsec radius, as can be seen also on the \Ha~+ [NII]  image from \cite{Boselli:2015}. Its disc is almost face-on and we adopted the morphological value for the inclination but find a significant difference between the morphological and kinematical values of the PA of the major axis (36$^{\circ}$ difference) when computing our H \rc. The resulting curve is poorly defined and quite chaotic, with only a few points between 5 and 20 arcsec from the center. It does not even reach the effective radius and the rotation velocity remains much below the \Hi~$V_{max}$ (about half). 

HRS 187 (KPG 343A). Although this object is a pair of galaxies, both rich in \Ha~emission, the velocity field of the smaller one (NGC4496B, located about 1 arcmin southeast from NGC4496A) seems completely melted in the velocity field of the bigger one. As a result, we reduced the data as if it were a single galaxy. We find a good agreement between the morphological and kinematical values of the inclination (8$^{\circ}$ difference) as well as for the values of the PA of the major axis (7$^{\circ}$ difference) when computing our \Ha~\rc. The resulting curve is fairly regular and symmetric, it extends almost out to the $r_{24}$ radius in $r$-band for both sides. The rotation velocity reaches the \Hi~$V_{max}$ at the effective radius in $r$-band and remains slightly above beyond, with an almost constant value. Both models (Zhao and Courteau) agree that the \Ha~\rc~reaches its maximum at the $r_{24}$ radius.

HRS 189. The \Ha~emission of this galaxy is asymmetric, with a very bright patch north of the nucleus, otherwise it is distributed rather uniformly in the central disc, as can be seen also on the \Ha~+ [NII]  image from \cite{Boselli:2015}. We find a rather good agreement between the morphological and kinematical values of the inclination (14$^{\circ}$ difference) and an excellent agreement between the values of the PA of the major axis (2$^{\circ}$ difference) when computing our \Ha~\rc. The resulting curve is quite chaotic, especially for the redshifted side (with extreme values on that side of the residual velocity field) and barely reaches the effective radius in $r$-band. The rotation velocity there is much below the \Hi~$V_{max}$. Zhao's model suggests that the curve goes on rising beyond, whereas Courteau's model favors a plateau. 

HRS 191. We find an excellent agreement between the morphological and kinematical values of the inclination (2$^{\circ}$ difference) as well as between the values of the PA of the major axis (1$^{\circ}$ difference) when computing our \Ha~\rc. The resulting curve is fairly regular and symmetric and extends midway between the effective radius and the $r_{24}$ radius in $r$-band. Both models (Zhao and Courteau) suggest that it goes on rising, until it reaches the \Hi~$V_{max}$ at the $r_{25}$ radius in B-band.
 
HRS 192. The \Ha~emission of this Sa type galaxy is limited to a few spots in the very center, as can be seen also on the \Ha~+ [NII]  image from \cite{Boselli:2015}. We nevertheless tried to draw the \Ha~\rc, using the morphological values of the inclination and PA of the major axis, but it gives nothing reliable. 

HRS 193. We adopted the morphological value for the inclination but find a significant difference between the morphological and kinematical values of the PA of the major axis (21$^{\circ}$ difference) when computing our H \rc. Anyway, the resulting curve is fairly regular and symmetric and extends a bit after the effective radius in $r$-band for both sides (slightly more for the blueshifted side). The rotation velocity reaches the \Hi~$V_{max}$ after the effective radius and both models (Zhao and Courteau) suggest that it goes on rising much beyond.

HRS 195. No significant \Ha~emission can be seen in this Sab galaxy, as confirmed by the \Ha~+ [NII]  image from \cite{Boselli:2015}. The few spots seen on our maps are probably but noise. We could not draw any realistic velocity field nor \rc.

HRS 197. This galaxy is almost edge-on and we adopted the morphological value for the inclination. We then find an excellent agreement between the morphological and kinematical values of the PA of the major axis (2$^{\circ}$ difference) when computing our \Ha~\rc. The resulting \rc~is fairly regular and symmetric and extends almost out to the effective radius in $r$-band for both sides. The rotation velocity reaches the \Hi~$V_{max}$ at the effective radius and both models (Zhao and Courteau) suggest that it goes on rising much beyond.

HRS 198. We find a good agreement between the morphological and kinematical values of the inclination (5$^{\circ}$ difference) as well as between the values of the PA of the major axis (5$^{\circ}$ difference) when computing our \Ha~\rc. The resulting curve is somewhat chaotic within the first 10 arcsec (for both sides) but becomes more regular and symmetric beyond. It extends almost midway between the effective radius and the $r_{24}$ radius in $r$-band where the rotation velocity is close to the \Hi~$V_{max}$. Zhao's model suggests that our \Ha~curve reaches its maximum there and declines afterwards, whereas Courteau's model favors a plateau slightly above the \Hi~$V_{max}$.

HRS 199. The \Ha~emission of this Sb galaxy is very bright but limited to the inner disc as can be seen also on the \Ha~+ [NII]  image from \cite{Boselli:2015}. We find an acceptable agreement between the morphological and kinematical values of the inclination (16$^{\circ}$ difference) but a quite large difference between the morphological and kinematical values of the PA of the major axis (25$^{\circ}$ difference) when computing our \Ha~\rc. Despite some dispersion, the resulting curve shows a solid body rotation for both sides but, because of the small extension of the disc of ionized gas, it is far from reaching the effective radius in $r$-band. However, the maximum velocity rotation reached is above the \Hi~$V_{max}$.

HRS 207. We find an excellent agreement between the morphological and kinematical values of the inclination (2$^{\circ}$ difference) but a quite large difference between the morphological and kinematical values of the PA of the major axis (31$^{\circ}$ difference) when computing our \Ha~\rc. The resulting curve is somewhat chaotic for the blueshifted side (especially within the 5 first arcsec) but much more regular for the redshifted side. The general shape for both sides is that of a solid body rotation, up to the effective radius in $r$-band. The blueshifted side of the curve extends a bit further, with a marked bending and a rotation velocity remaining slightly above the \Hi~$V_{max}$. Zhao's model suggests that our \Ha~curve reaches its maximum there and declines afterwards, whereas Courteau's model favors a plateau above the \Hi~$V_{max}$.

HRS 212 (KPG 346A). This galaxy has bright HII regions almost all over its optical disc. We adopted the morphological value for the inclination but find a significant difference between the morphological and kinematical values of the PA of the major axis (22$^{\circ}$ difference) when computing our \Ha~\rc. The shape of the isovelocity lines of our \Ha~velocity field suggests that the disc of this galaxy is warped, explaining such a difference. The velocity gradient is very faint in the center, so that our \rc~is almost flat for both sides, up to a radius of 10 arcsec, then the curve rises very steeply and reaches a plateau at the effective radius in $r$-band, with different velocities depending on the side. The redshifted side exhibits a higher velocity than the blueshifted side (about 70 km s$^{-1}$ against 50 km s$^{-1}$) but both remain much below the \Hi~$V_{max}$. 

HRS 225. The \Ha~emission of this Sb type galaxy is limited to a fuzzy patch on the southeastern side of the nucleus, as can be seen also on the \Ha~+ [NII]  image from \cite{Boselli:2015}. The faint northern spot seen on our \Ha~map is probably but noise. We nevertheless tried to draw the \Ha~\rc, using the morphological values of the inclination and PA of the major axis, but it gives nothing reliable.

HRS 226. We find an excellent agreement between the morphological and kinematical values of the inclination (1$^{\circ}$ difference) as well as between the morphological and kinematical values of the PA of the major axis (2$^{\circ}$ difference) when computing our \Ha~\rc. The resulting \rc~is fairly regular and symmetric and extends out to the effective radius in $r$-band for both sides. Both models (Zhao and Courteau) agree that our curve reaches a plateau at the effective radius, with a rotation velocity slightly above the \Hi~$V_{max}$.

HRS 230. We find an excellent agreement between the morphological and kinematical values of the inclination (less than 1$^{\circ}$ difference) as well as between the morphological and kinematical values of the PA of the major axis (1$^{\circ}$ difference) when computing our \Ha~\rc. The resulting \rc~is fairly regular and symmetric and extends out to the effective radius in $r$-band for both sides. Both models (but Courteau better than Zhao) agree that our curve reaches a plateau slightly after the effective radius, with a rotation velocity remaining below the \Hi~$V_{max}$.

HRS 237. Except in the very center, the \Ha~emission of this Im type galaxy has a faint surface brightness. We adopted the morphological value for the inclination when computing our \Ha~\rc. Despite some dispersion, the trend of the curve for both sides is that of a solid body rotation, albeit with a rather faint slope. Both sides extend a bit further than the effective radius in $r$-band and the rotation velocity remains below the \Hi~$V_{max}$.

HRS 238. The \Ha~emission of this Sm galaxy is rather faint and patchy. The large extension bordering the whole northeastern side of the disc has a very faint surface brightness and seems doubtful (N.B. It was not imaged by \citealp{Boselli:2015}). The fact that all this area appears with the same velocity (coded with blue color on our velocity field map) suggests that our data are polluted there by an OH night skyline insufficiently subtracted. We adopted the morphological value for the inclination but find a large difference between the morphological and kinematical values of the PA of the major axis (27$^{\circ}$ difference) when computing our \Ha~\rc. The resulting curve is rather poor, because of the faintness of the \Ha~emission, its trend is that of a solid body rotation for both sides. It reaches the effective radius in $r$-band, with a rotation velocity which is only half the \Hi~$V_{max}$.

HRS 239 (KPG 351B). This galaxy is almost edge-on and we adopted the morphological values for the inclination and for the PA of the major axis when computing our \Ha~\rc. The resulting curve is fairly regular and symmetric up to a radius of 20 arcsec, showing a solid body rotation with a steep slope. Then the slope of the blueshifted side diminishes whereas the redshifted side stops for a while before climbing again with about the same slope as the blueshifted side. Both sides of the curve extend slightly after the effective radius in $r$-band and the models (Zhao as well as Courteau) suggest that our curve reaches a plateau at the optical radius, with a rotation velocity above the \Hi~$V_{max}$.
 
HRS 249. The \Ha~emission of this Sb type galaxy is limited to a bright spot on the left of the nucleus and another (less bright) spot on its right. We could not draw any realistic velocity field nor \rc.

HRS 252. The disc of this Sd galaxy is peppered with many bright HII regions covering almost completely the southern part of its optical disc, but less numerous on the northern side (redshifted) as can be seen also on the \Ha~+ [NII]  image from \cite{Boselli:2015}. We find an excellent agreement between the morphological and kinematical values of the inclination (less than 1$^{\circ}$ difference) as well as between the morphological and kinematical values of the PA of the major axis (3$^{\circ}$ difference) when computing our \Ha~\rc. The resulting curve is fairly regular and symmetric up to the effective radius in $r$-band for both sides. The blueshifted side then enables to trace the curve up to the $r_{24}$ radius in $r$-band (found here slightly beyond the $r_{25}$ radius in B-band) where it reaches a rotation velocity higher than the \Hi~$V_{max}$. Zhao's model suggests that our curve reaches a maximum there before declining, whereas Courteau's model favors a continuous rising, although with a smaller slope.

HRS 255. The spiral pattern of this Scd galaxy is underlined by many HII regions, as can be seen on the \Ha~+ [NII]  image from \cite{Boselli:2015}. This galaxy is not far from face-on, we adopted the morphological value for the inclination of its disc and find a large difference between the morphological and kinematical values of the PA of the major axis (55$^{\circ}$ difference) when computing our \Ha~\rc. The resulting curve is quite chaotic, especially in the central part, but the general trend, for both sides, is that of a solid body rotation with a faint slope. Both sides extend beyond the $r_{24}$ radius in $r$-band, where the rotation velocity is slightly above the \Hi~$V_{max}$. The redshifted side enables to draw the \rc~almost out to the $r_{25}$ radius in B-band, keeping the same slope, but with a large uncertainty.
 
HRS 256. The \Ha~emission is bright but limited to the inner disc of this Sa type galaxy, as can be seen also on the \Ha~+ [NII]  image from \cite{Boselli:2015}. Despite the relatively high inclination of the disc, the velocity gradient of the \Ha~velocity field of this galaxy is rather faint but, most strangely, it suggests that the PA of the major axis is almost at right angle with respect to the morphological value (95$^{\circ}$ difference). All of this suggests that the behavior of the ionized gas do not reflect that of the main body of the galaxy. We adopted the morphological value for the inclination of its disc when computing our \Ha~\rc. The resulting curve is quite chaotic for the redshifted side while suggesting a solid body rotation for the blueshifted side of the central disc of ionized gas. The curve does not reach the effective radius and its maximum rotation velocity is very far from the \Hi~$V_{max}$. The strange behavior of the ionized gas in the center of this Sa galaxy leads to think that it could be the remnant of a merger (as already seen for HRS142 and HRS159). 

HRS 261. This galaxy is almost edge-on and we adopted the morphological values for the inclination and for the PA of the major axis when computing our \Ha~\rc. Except some strong dispersion around 10 arcsec radius for the redshifted side, the resulting curve is fairly regular and symmetric, rising up to the effective radius in $r$-band for both sides. The redshifted side enables to draw the curve a bit further (midway between the effective and $r_{24}$ radius in $r$-band) where it reaches the \Hi~$V_{max}$. Both models (Zhao as well as Courteau) suggest that our curve goes on rising beyond the optical radius, although with a smaller slope for Courteau's model.

HRS 264. This galaxy is almost edge-on and we adopted the morphological values for the inclination and for the PA of the major axis when computing our \Ha~\rc. The resulting curve is fairly regular and symmetric, almost reaching the effective in $r$-band for both sides. The rotation velocity at that radius is not far from the \Hi~$V_{max}$ but both models (Zhao as well as Courteau) suggest that our curve goes on rising beyond the optical radius.

HRS 267. The \Ha~emission of this almost edge-on Scd type galaxy is asymmetric, much brighter on the southwestern side (blueshifted), probably because of absorption by dust in the disc. We adopted the morphological values for the inclination and for the PA of the major axis when computing our \Ha~\rc. The resulting curve is quite chaotic, more especially for the redshifted side that exhibits a marked bump at about 10 arcsec radius before rising again. The curve do not even reach the effective radius in $r$-band and the rotation velocity remains much below the \Hi~$V_{max}$.

HRS 268. We adopted the morphological values for the inclination and for the PA of the major axis when computing our \Ha~\rc. The resulting curve is fairly regular for the blueshifted side and extends out to the $r_{25}$ radius in B-band (which is here slightly larger than the $r_{24}$ radius in $r$-band). The redshifted side is shorter and more irregular, although with the same general trend. The rotation velocity reaches the \Hi~$V_{max}$ and the blueshifted side seems to mark a plateau at this value, although both models (Zhao and Courteau) favor a continuous rising of the \rc~beyond the optical radius.

HRS 271. This galaxy is almost edge-on and we adopted the morphological values for the inclination and for the PA of the major axis when computing our \Ha~\rc. The resulting curve is quite irregular within 10 arcsec radius from the center but the general trend is the same for both sides, that of a solid body rotation with a diminishing slope when reaching the effective radius in $r$-band. The rotation velocity reached at that radius is equal to the \Hi~$V_{max}$. Zhao's model suggests that our \rc~reaches a maximum at the optical radius whereas Courteau's model favors a continuous rising beyond.  
 
HRS 273. This galaxy is almost edge-on and we adopted the morphological values for the inclination when computing the \rc. We find an excellent agreement between the morphological and kinematical values of the PA of the major axis (less than 2$^{\circ}$ difference) and, despite some dispersion in the rising part, the resulting curve is fairly regular and symmetric. The curve extends midway between the effective radius and the $r_{24}$ radius in $r$-band. The maximum rotation velocity reached is close to the \Hi~$V_{max}$ and both models (but Courteau better than Zhao) suggest that our \rc~reaches a plateau at the optical radius.

HRS 275. The optical disc of this Sd type galaxy is peppered with many HII regions covering it completely. The inclination of the disc is rather faint and we adopted the morphological value when computing our \Ha~\rc. We find a rather large difference between 
the morphological and kinematical values of the PA of the major axis (35$^{\circ}$ difference) but the resulting \rc~is quite regular and symmetric, despite some dispersion. Indeed both sides show the same trend, that of a solid body rotation, almost out to the $r_{24}$ radius in $r$-band where the rotation velocity is about to reach the \Hi~$V_{max}$. Zhao's model suggests that our curve reaches a maximum slightly after the $r_{24}$ radius, whereas Courteau's model suggests a continuous rising beyond. 

HRS 276. The \Ha~emission of this Sbc type galaxy covers fairly well the northeastern side of the optical disc but is patchy for the southwestern side (redshifted) as can be seen also on the \Ha~+ [NII]  image from \cite{Boselli:2015}. We adopted the morphological value for the inclination of the disc, when computing our \Ha~\rc, and find a rather good agreement between the morphological and kinematical values of the PA of the major axis (7$^{\circ}$ difference). The redshifted side of the curve is rather ill defined, because of the lack of velocity points, but the blueshifted side is quite regular and enables to trace the \rc~much beyond the effective radius in $r$-band, after which it seems to reach a plateau. Indeed Courteau's model suggests that our curve reaches a plateau after the effective radius, with a velocity close to the \Hi~$V_{max}$, whereas Zhao's model favors a continuous rising beyond that radius.

HRS 278. The \Ha~emission of this edge-on Sb type galaxy has a faint surface brightness and is limited to the inner disc as can be seen also on the \Ha~+ [NII]  image from \cite{Boselli:2015}. We adopted the morphological values for the inclination and for the PA of the major axis when computing our \Ha~\rc. Despite some dispersion, both sides of the curve show the same trend, that of a solid body rotation, and extend out to about $3/4$ of the effective radius in $r$-band. The rotation velocity reached for both sides is clearly above the \Hi~$V_{max}$. 

HRS 279. The optical disc of this Sd type galaxy is peppered with many HII regions covering it completely. We find a fairly good agreement between the morphological and kinematical values of the inclination (7$^{\circ}$ difference) as well as between the morphological and kinematical values of the PA of the major axis (5$^{\circ}$ difference) when computing our \Ha~\rc. Despite some dispersion within the first 20 arcsec, the resulting curve is fairly regular and symmetric, almost reaching the $r_{24}$ radius in $r$-band (at least for the redshifted side) which is coincident here with the $r_{25}$ radius in B-band. The rotation velocity reaches the \Hi~$V_{max}$ for both sides of the curve. Zhao's model suggests that our curve reaches a maximum before the optical radius and declines strongly beyond, whereas Courteau's model favors a continuous rising beyond the optical radius, although with a smaller slope.

HRS 280. The \Ha~emission of this Sb type galaxy shows a bright inner disc, with an extension of faint surface brightness on the western extremity of the disc (blueshifted side) as can be seen also on the \Ha~+ [NII]  image from \cite{Boselli:2015}. The inclination of the disc is rather strong and we adopted the morphological value when computing our \Ha~\rc. We find a rather good agreement between the morphological and kinematical values of the PA of the major axis (7$^{\circ}$ difference) and, despite the relative small number of velocity points, the resulting \rc~is quite regular and symmetric. The curve extends out to the effective radius in $r$-band for both sides, reaching there a rotation velocity equal to the \Hi~$V_{max}$. The blueshifted side enables to trace our \rc~a bit further and seems to stop rising. Indeed both models (but Courteau better than Zhao) suggest that our curve reaches a plateau slightly above the \Hi~$V_{max}$ just after the effective radius.

HRS 282. We detected no significant \Ha~emission in this BCD type galaxy (N.B. It was not imaged by \citealp{Boselli:2015}) and could not draw any reliable \Ha~velocity field nor \rc.

HRS 283. The disc of this galaxy is completely covered by \Ha~emission and we find an excellent agreement between the morphological and kinematical values of the inclination (1$^{\circ}$ difference only) as well as between the morphological and kinematical values of the PA of the major axis (2$^{\circ}$ difference only) when computing our \Ha~\rc. The resulting \rc~is fairly regular and symmetric, despite some waves on the redshifted side, it extends out to the $r_{24}$ radius in $r$-band (almost coincident here with the $r_{25}$ radius in B-band) for both sides. The rotation velocity reached there is equal to the \Hi~$V_{max}$, but both models (Zhao and Courteau) suggest that our curve goes on rising beyond the optical radius, although with a smaller slope for Courteau's model. 

HRS 284. The disc of this galaxy is strongly inclined and we adopted the morphological values for the inclination and for the PA of the major axis when computing our \rc. The resulting curve is fairly regular and symmetric, with the same trend for both sides, that of a solid body rotation. The curve extends out to about $3/4$ of the $r_{24}$ radius in $r$-band and the rotation velocity reached is slightly above the \Hi~$V_{max}$. Both models (Zhao and Courteau) suggest that our \rc~reaches a plateau shortly after. 

HRS 287. This Scd type galaxy has a thick bar and a strong spiral pattern, both underlined by bright HII regions. We adopted the morphological value for the inclination of the disc, when computing our \Ha~\rc, and find a significant difference between the morphological and kinematical values of the PA of the major axis (25$^{\circ}$ difference). Despite some distortions, the agreement between both sides of our curve is not so bad for the first 10 arcsec (about half the effective radius in $r$-band), but quite different beyond. The redshifted side then keeps an almost constant rotation velocity, equal to the \Hi~$V_{max}$, out to the $r_{24}$ radius in $r$-band (almost coincident here with the $r_{25}$ radius in B-band). The blueshifted side is not so extended and its velocity, after a small pause, goes on rising continuously beyond the effective radius. Zhao's model suggests that our \rc~reaches a maximum (close to the \Hi~$V_{max}$) between the effective radius and $r_{24}$ radius before declining very slowly, while Courteau's model suggests that it reaches a plateau with a rotation velocity close to the \Hi~$V_{max}$.    

HRS 289. This Sbc type galaxy has a nice spiral pattern, underlined by many bright HII regions. We find a reasonable agreement between the morphological and kinematical values of the inclination (10$^{\circ}$ difference) as well as between the morphological and kinematical values of the PA of the major axis (14$^{\circ}$ difference) when computing our \Ha~\rc. The resulting \rc~is fairly regular and symmetric, extending almost out to the $r_{24}$ radius in $r$-band (at least for the blueshifted side). The central part of the curve is ill defined, because of the lack of \Ha~emission there, and the velocity bump seen at about 10 arcsec from the center is likely to be explained by the strong central bulge. Both models (but Courteau better than Zhao) suggest that our curve rapidly reaches a plateau with a rotation velocity slightly above the \Hi~$V_{max}$.

HRS 290. The H$\alpha$ emission of this almost edge-on Sa type galaxy is concentrated in the inner disc. We adopted the morphological values for the inclination and for the PA of the major axis when computing our \rc. Despite the relatively small number of velocity points, the resulting curve is fairly regular and symmetric, almost reaching the effective radius in $r$-band. The redshifted side reaches a higher rotation velocity than the blueshifted side and both are clearly above the \Hi~$V_{max}$.  

HRS 291. The \Ha~emission of this Sab type galaxy is limited to a few faint spots around the nucleus. We cannot check if they are real or mere noise since it was not imaged by \cite{Boselli:2015}. We could not draw any reliable \Ha~velocity field nor \rc.

HRS 292. The \Ha~emission of this Sb type galaxy is very bright but limited to the inner disc, as can be seen also on the \Ha~+ [NII]  image from \cite{Boselli:2015}. We find an acceptable agreement between the morphological and kinematical values of the PA of the major axis (14$^{\circ}$ difference). The resulting curve is regular and symmetric, with both sides reaching the effective radius in $r$-band. The curve rises steeply up to a plateau at the \Hi~$V_{max}$ as confirmed by both models (Zhao and Courteau). 

HRS293. The whole optical disc of this Sdm galaxy is well covered by HII regions. We find a reasonable agreement between the morphological and kinematical values of the inclination (7$^{\circ}$ difference) as well as between the morphological and kinematical values of the PA of the major axis (16$^{\circ}$ difference) when computing our \Ha~\rc. Despite some waves on the redshifted side, we find a fairly regular and symmetric curve, extending out to the optical radius (the $r_{24}$ radius in $r$-band and $r_{25}$ radius in B-band are coincident here). The rotation velocity reached there is the \Hi~$V_{max}$ but both models (Zhao and Courteau) suggest that our curve goes on rising beyond the optical radius.

HRS 294. Apart from a few bright spots, the \Ha~emission of this edge-on Sb type galaxy has a rather faint surface brightness, as can be seen also on the \Ha~+ [NII]  image from \cite{Boselli:2015}. We adopted the morphological values for the inclination and for the PA of the major axis when computing our \rc. Despite the relatively small number of velocity points, the resulting curve is fairly regular and symmetric but do not even reach the effective radius in $r$-band. Both sides show a solid body rotation that could reach the \Hi~$V_{max}$ at the effective radius if extrapolated keeping the same slope.

HRS 297. The whole optical disc of this Sbc type galaxy is peppered with bright HII regions. It is strongly inclined and we adopted the morphological values for the inclination as well as for the PA of the major axis when computing our \rc. Despite some waves on both sides, the resulting curve is fairly regular and symmetric, extending almost out to the $r_{24}$ radius in $r$-band where the rotation velocity reaches the \Hi~$V_{max}$. Zhaos's model suggests that our curve goes on rising beyond the optical radius, almost with the same slope, but Courteau's model apparently fails fitting our data correctly here. 

HRS 298 (KPG 397A). The \Ha~emission of this peculiar galaxy is very bright but limited to the inner disc. The large and faint structure surrounding the northern and western side of the disc on our maps seems doubtful since nothing can be seen there on the \Ha~+ [NII]  image from \cite{Boselli:2015}. The fact that it pops out with the same velocity, color coded green on the map of our velocity field, strongly suggests that it is the remnant of an OH night skyline insufficiently subtracted (it also gives extreme values on the residual velocity field). Anyway, it was not taken into account when computing our \Ha~\rc. We adopted the morphological value for the inclination of the disc, when computing our \Ha~\rc, and find a rather good agreement between the morphological and kinematical values of the PA of the major axis (9$^{\circ}$ difference). The resulting \rc~is very regular and symmetric, extending almost out to the $r_{25}$ radius in B-band where it reaches a rotation velocity higher than the \Hi~$V_{max}$. Zhao's model suggests that our curve goes on rising beyond the optical radius but Courteau's model favors a plateau shortly after.  

HRS 300. There is no significant \Ha~emission in this Sab type galaxy, as confirmed by the \Ha~+ [NII]  image from \cite{Boselli:2015} that shows only a very faint fuzzy spot in the center. The spots on our maps are probably more noise than signal and we could not draw any reliable \Ha~velocity field nor \rc.

HRS 301. The whole optical disc of this Sc type galaxy is peppered with numerous HII regions as can be seen also on the \Ha~+ [NII]  image from \cite{Boselli:2015}. We find a good agreement between the morphological and kinematical values of the inclination (8$^{\circ}$ difference) as well as between the morphological and kinematical values of the PA of the major axis (6$^{\circ}$ difference) when computing our \Ha~\rc. The resulting \rc~is fairly regular and symmetric, except some dispersion (for both sides) around the 10 arcsec radius.  The curve extends out to the $r_{24}$ radius in $r$-band and reaches a rotation velocity higher than the \Hi~$V_{max}$. Both models (Zhao and Courteau) suggest that our curve goes on rising beyond the optical radius but with a diminishing slope.  

HRS 303. The \Ha~emission of this Sb type galaxy is very bright but limited to the inner disc. We find a rather good agreement between the morphological and kinematical values of the inclination (10$^{\circ}$ difference) and an excellent agreement  between the morphological and kinematical values of the PA of the major axis (2$^{\circ}$ difference) when computing our \Ha~\rc. The resulting curve is fairly regular and symmetric, with both sides extending a bit further than the effective radius in $r$-band. Courteau's model suggests that our \rc~reaches a plateau at the optical radius with a rotation velocity around 120 km s$^{-1}$ whereas Zhao's model favors a continuous rising curve.

HRS 304. The disc is strongly inclined and we adopted the morphological values for the inclination as well as for the PA of the major axis when computing our \rc. The resulting curve is fairly regular and symmetric, with both sides extending a bit further than the effective radius in $r$-band. The curve reaches the \Hi~$V_{max}$ there but both models (Zhao and Courteau) suggest that it goes on rising beyond that radius, although with a smaller slope for Courteau's model.

HRS 305. The \Ha~emission of this peculiar galaxy is limited to two relatively bright spots on each side of the nucleus, as can be seen also on the \Ha~+ [NII]  image from \cite{Boselli:2015}. Some velocity gradient can be seen anyway on our velocity field and we adopted the morphological values for the inclination as well as for the PA of the major axis when computing our \rc. The resulting curve is quite chaotic and limited to one third of the effective radius in $r$-band. The blueshifted side is less dispersed than the redshifted side but no more extended and its rotation velocity remains much below the \Hi~$V_{max}$. 

HRS 308. The \Ha~emission of this Sb type galaxy is limited to a faint fuzzy central patch as can be seen also on the \Ha~+ [NII]  image from \cite{Boselli:2015}. Anyway, our velocity field shows some gradient and we tried to draw the \rc. The disc is strongly inclined and we adopted the morphological values for the inclination as well as for the PA of the major axis in order to compute our \rc. The result is however quite disappointing and not reliable. 

HRS 309. The optical disc of this Sm type galaxy is peppered with many bright HII regions. We find a rather good agreement between the morphological and kinematical values of the inclination (11$^{\circ}$ difference) and an excellent agreement  between the morphological and kinematical values of the PA of the major axis (3$^{\circ}$ difference) when computing our \Ha~\rc. Despite a rather faint spatial sampling, the resulting curve is fairly regular and symmetric, with both sides extending a bit further than the $r_{24}$ radius in $r$-band. Our curve remains clearly below the \Hi~$V_{max}$ but both models (Zhao and Courteau) suggest that it goes on rising beyond the optical radius, although with a smaller slope for Courteau's model.

HRS 313. The disc of this Sbc type galaxy is peppered with many bright HII regions. The disc is strongly inclined and we adopted the morphological values for the inclination as well as for the PA of the major axis in order to compute our \rc. Despite some dispersion, the resulting curve is fairly regular and symmetric, with both sides extending out halfway between the effective radius and the $r_{24}$ radius in $r$-band. Our curve just reaches the \Hi~$V_{max}$ but both models (Zhao and Courteau) suggest that it goes on rising beyond the optical radius, although with a smaller slope for Courteau's model which is almost flat in the outer part.

HRS 314. The \Ha~emission of this Sd type galaxy is asymmetric, with brighter HII regions on the northern side (blueshifted) as can be also on the \Ha~+ [NII]  image from \cite{Boselli:2015}.  The whole optical disc is nevertheless well covered on both sides by ionized gas. We find a good agreement between the morphological and kinematical values of the inclination (6$^{\circ}$ difference) and an acceptable agreement between the morphological and kinematical values of the PA of the major axis (13$^{\circ}$ difference) when computing our \Ha~\rc. The resulting curve is quite chaotic, with a larger dispersion on the redshifted side for which the \Ha~emission is the less bright. Nevertheless, both sides of the curve exhibit the same general trend, steeply rising up to the effective radius in $r$-band where the rotation velocity reaches the \Hi~$V_{max}$. Beyond that radius, the rotation velocity remains almost constant for both sides, out to the $r_{25}$ radius in B-band for the redshifted side. Zhao's model suggests that the curve reaches a maximum (with a rotation velocity close to the \Hi~$V_{max}$) before the optical radius and declines beyond, whereas Courteau's model favors a plateau beyond the optical radius (with a rotation velocity close to the \Hi~$V_{max}$).

HRS 317. This galaxy is almost edge-on and we adopted the morphological value for the inclination. We then find an excellent agreement between the morphological and kinematical values of the PA of the major axis (less than 1$^{\circ}$ difference) when computing our \Ha~\rc. Despite the faint sampling and some dispersion, the resulting \rc~is fairly regular and symmetric and extends almost out to the $r_{25}$ radius in B-band for both sides. The rotation velocity just reaches the \Hi~$V_{max}$ but both models (Zhao and Courteau) suggest that it goes on rising beyond the optical radius.

HRS318. The faint extension seen on our maps, on the northeastern side of the disc, is doubtful since nothing can be seen there on the \Ha~+ [NII]  image from \cite{Boselli:2015}. We adopted the morphological value for the inclination and find an acceptable agreement between the morphological and kinematical values of the PA of the major axis (16$^{\circ}$ difference) when computing our \Ha~\rc. The resulting curve is fairly regular and symmetric, despite some waves in the outer part of the redshifted side. The curve extends out to the $r_{25}$ radius in B-band (almost coincident here with the $r_{24}$ radius in $r$-band). Both models (Zhao and Courteau) suggest that our curve goes on rising beyond the optical radius, athough with a lower slope for Courteau's model.

HRS 319. The spiral pattern of this Scd galaxy is underlined by bright HII regions. We find an excellent agreement between the morphological and kinematical values of the inclination (1$^{\circ}$ difference) as well as between the morphological and kinematical values of the PA of the major axis (3$^{\circ}$ difference) when computing our \Ha~\rc. However, although the rising part of our curve is fairly regular and symmetric, it becomes quite chaotic and asymmetric beyond the effective radius in $r$-band. The blueshifted side of the curve rises up to the \Hi~$V_{max}$, extending midway between the effective radius and the $r_{24}$ radius in $r$-band. The redshifted side extends out to the $r_{25}$ radius in B-band with velocities remaining clearly below the \Hi~$V_{max}$. Zhao's model suggests a declining curve beyond the effective radius, whereas Courteau's model favors a plateau after that radius.

HRS 321. The \Ha~emission of this Sb type galaxy is very bright but limited to the inner disc as can be seen also on the \Ha~+ [NII]  image from \cite{Boselli:2015}. We adopted the morphological value for the inclination and find a good agreement between the morphological and kinematical values of the PA of the major axis (4$^{\circ}$ difference) when computing our \Ha~\rc. The resulting curve is fairly regular and symmetric but extends out only midway between the effective radius in $r$-band and the $r_{25}$ radius in B-band (at least for the redshifted side). The rotation velocity remains much below the \Hi~$V_{max}$ and both models (Zhao and Courteau) suggest that it stops rising beyond the effective radius in $r$-band. Zhao's model suggests that our curve declines beyond that radius, whereas Courteau's model favors a plateau.

HRS 322. The spiral pattern of this almost face-on Sa type galaxy is underlined by HII regions tracing a ring with about 1.5 arcmin radius. Because of this peculiar distribution of the H alpha emission, our \rc~has no points in its rising part and it is difficult to optimize the kinematical parameters. We adopted the morphological value of the inclination and choose the value of the PA of the major axis by eye-estimate from the gradient of the velocity field. The resulting curve shows a plateau at about 210 km s$^{-1}$, above the \Hi~$V_{max}$.

HRS 323. The \Ha~emission of this almost edge-on Sb galaxy is asymmetric, much brighter on the northern side (blueshifted), as can be seen also on the \Ha~+ [NII]  image from \cite{Boselli:2015}. We adopted the morphological values for the inclination as well as for the PA of the major axis when computing our \rc. The resulting curve is quite chaotic in the central part, out to a radius of 10 arcsec radius, but more regular and symmetric beyond. The blueshifted side (which is the brighter) is better defined than the redshifted side but both sides have about the same slope in their outer part, extending a bit further than the effective radius in $r$-band. Both models (Zhao and Courteau) suggest that our curve goes on rising with about the same slope and reaches the \Hi~$V_{max}$ at the $r_{24}$ radius in $r$-band.

\clearpage
\section{\Ha~profiles}
\label{profiles}
\begin{figure*}
\begin{minipage}{180mm}
\begin{center}
\includegraphics[width=3.5cm]{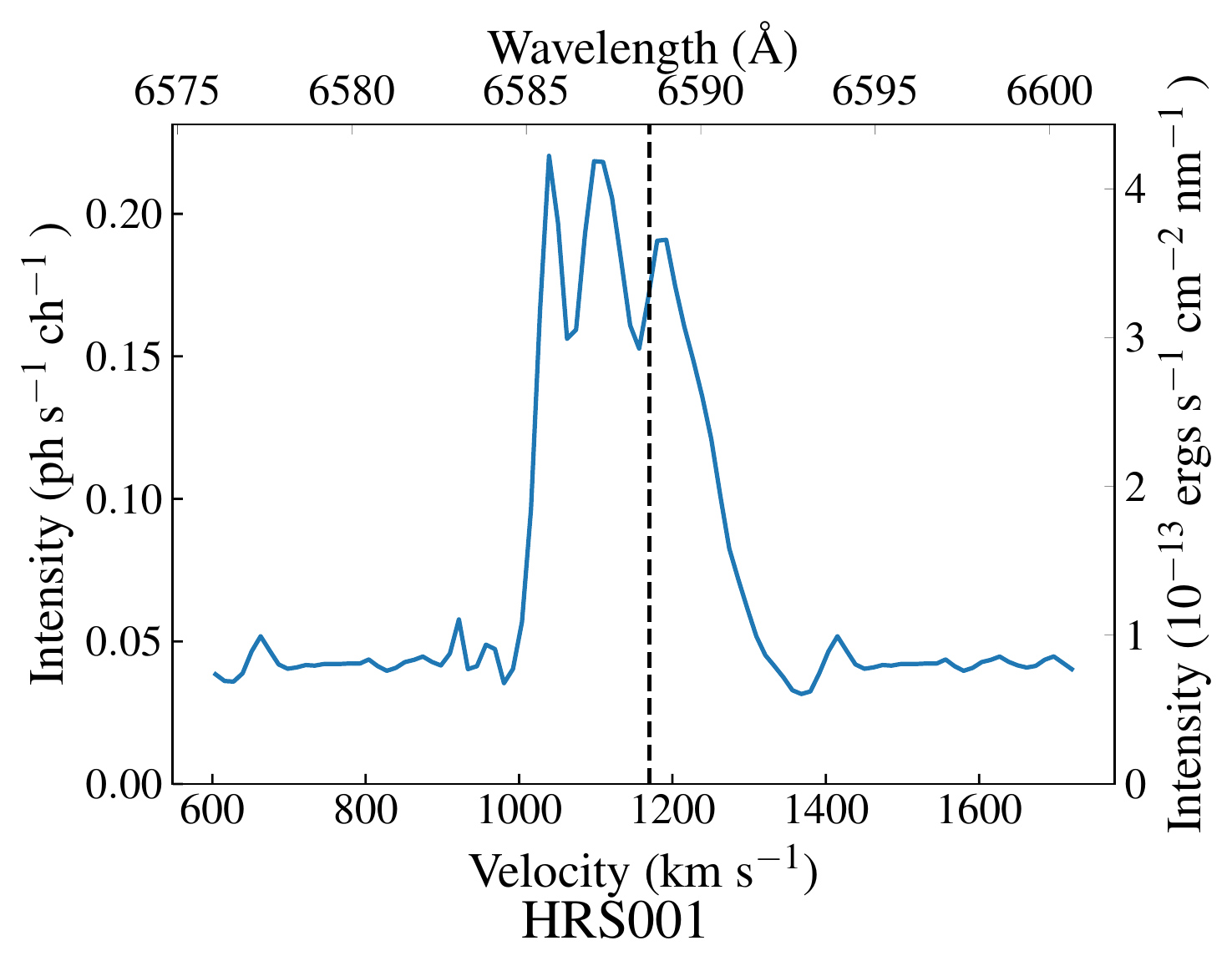}
\includegraphics[width=3.5cm]{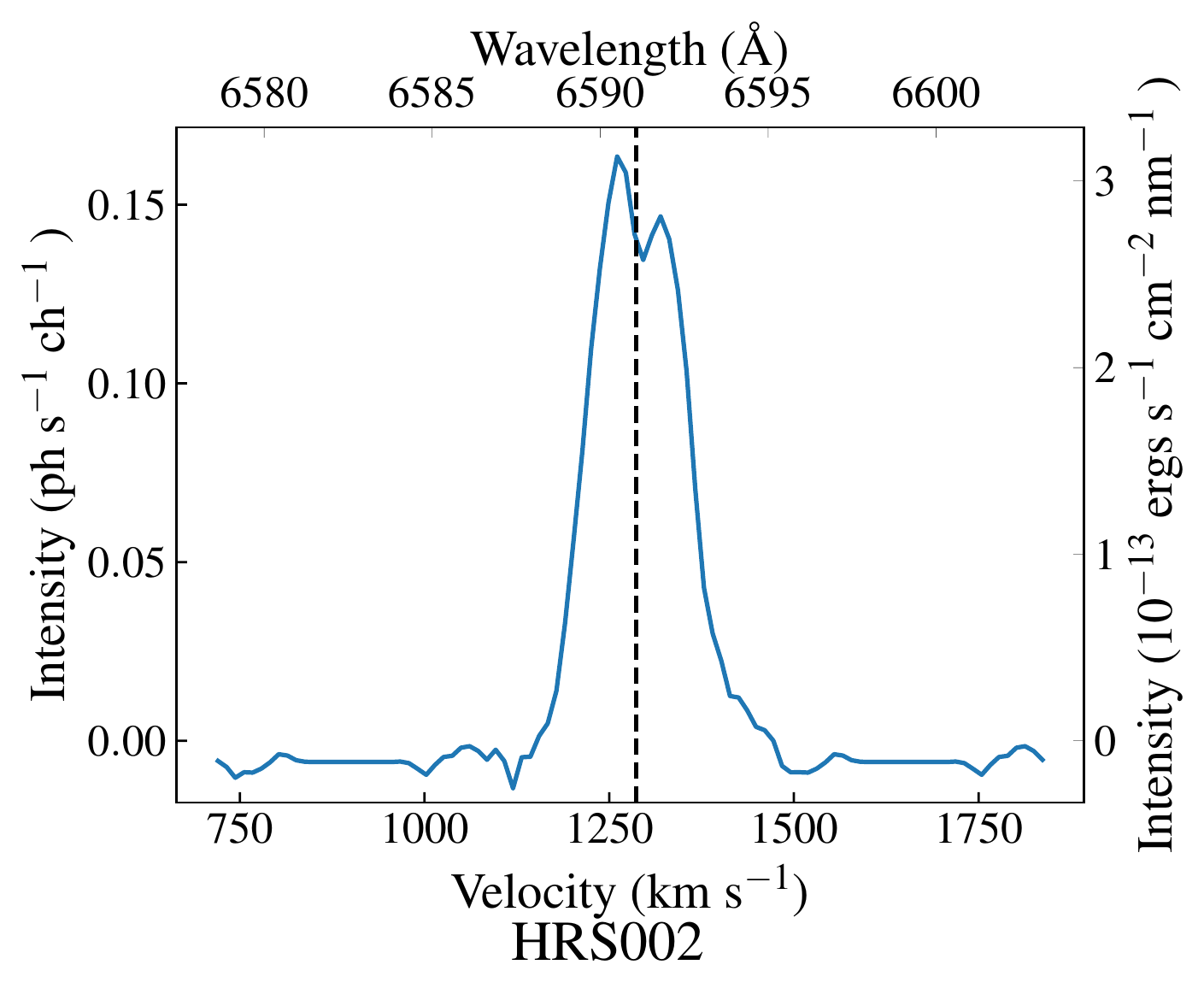}
\includegraphics[width=3.5cm]{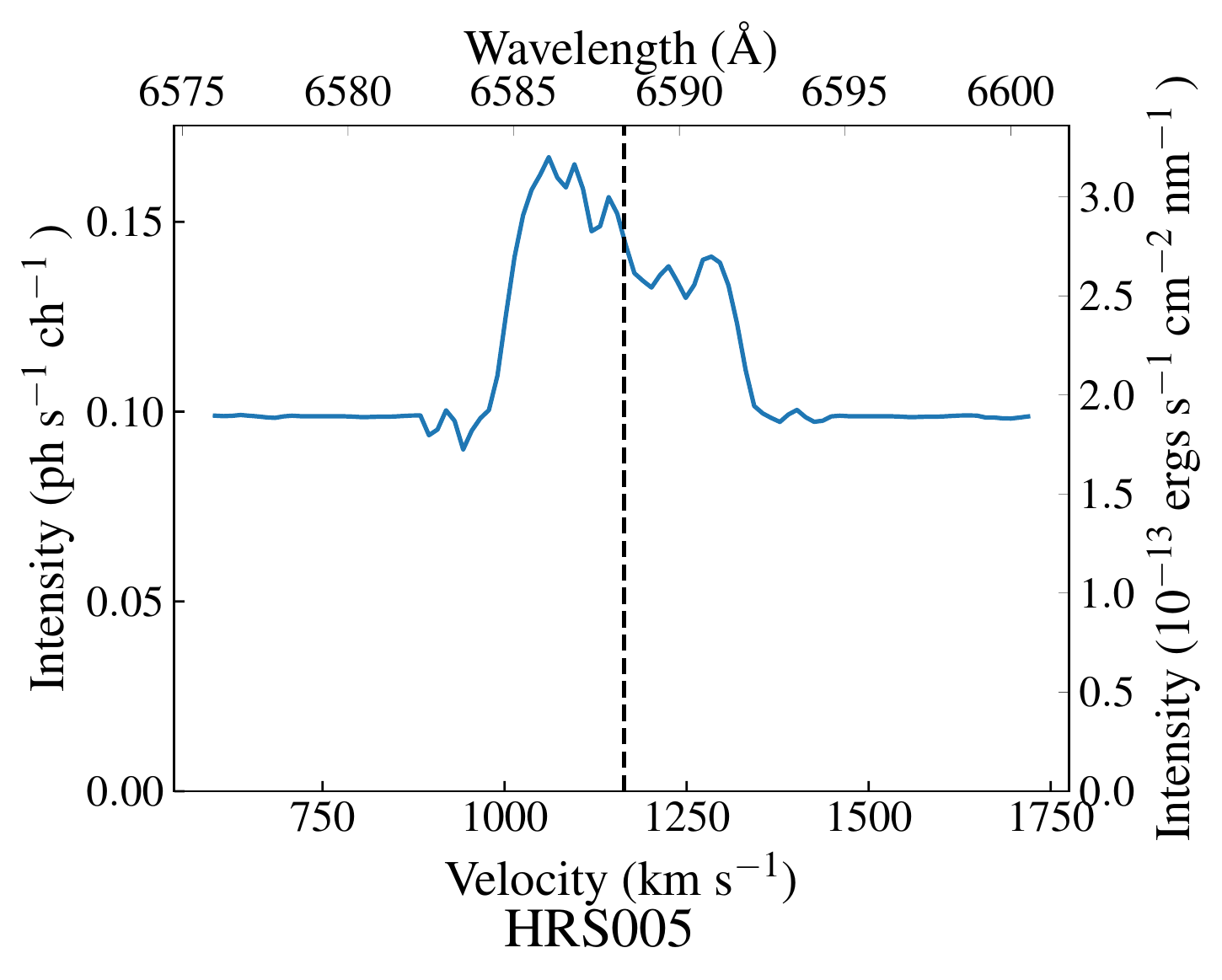}
\includegraphics[width=3.5cm]{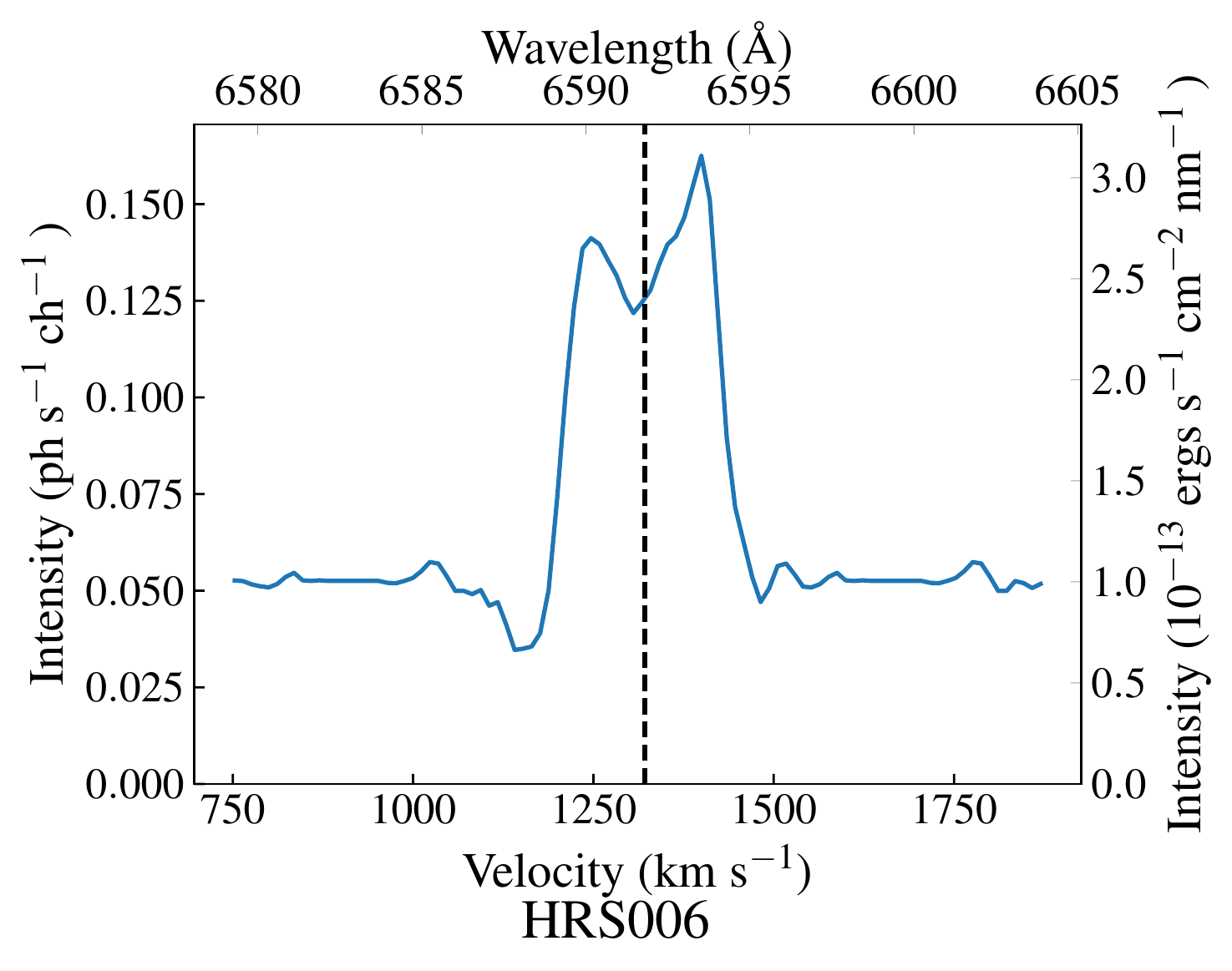}
\includegraphics[width=3.5cm]{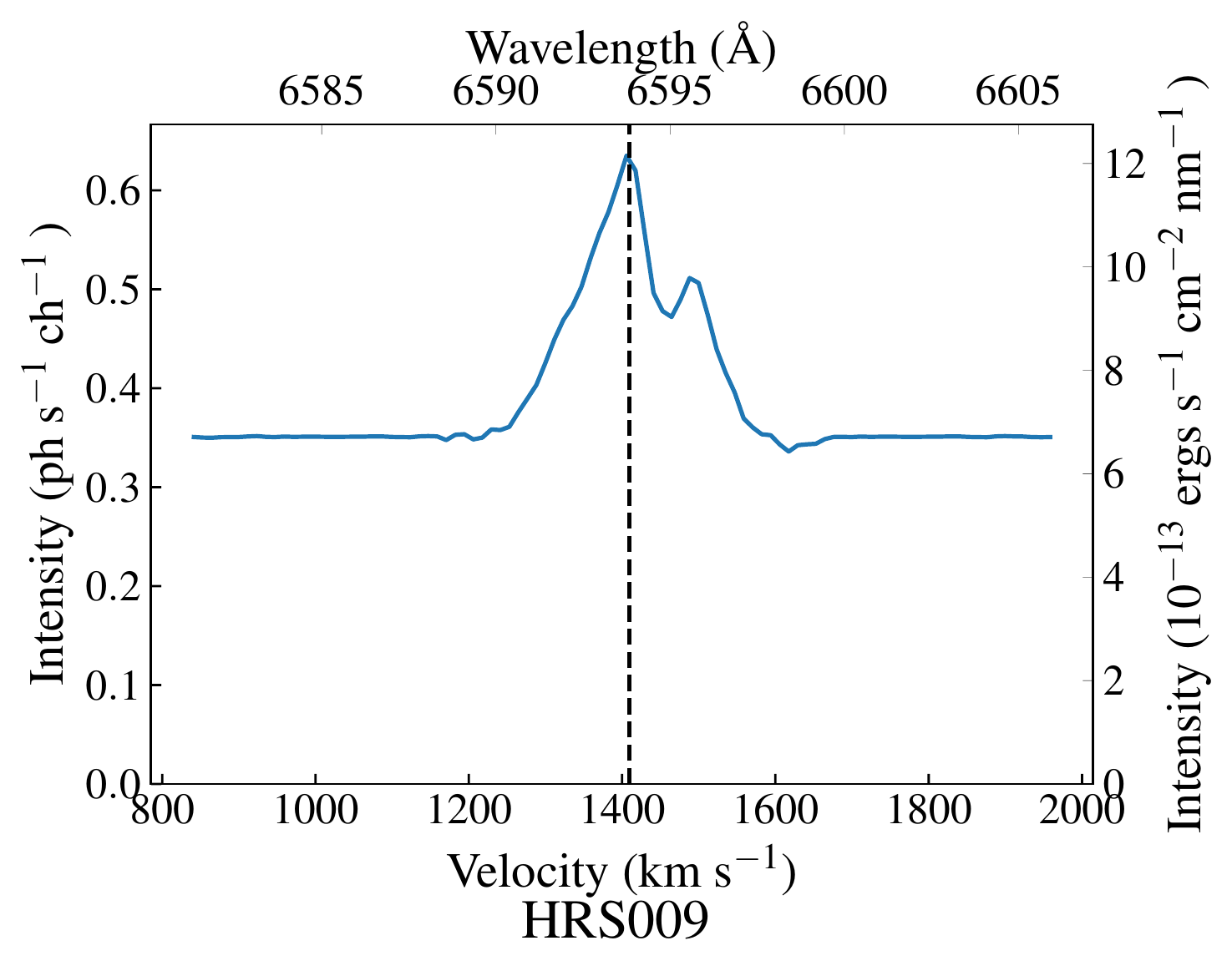}
\includegraphics[width=3.5cm]{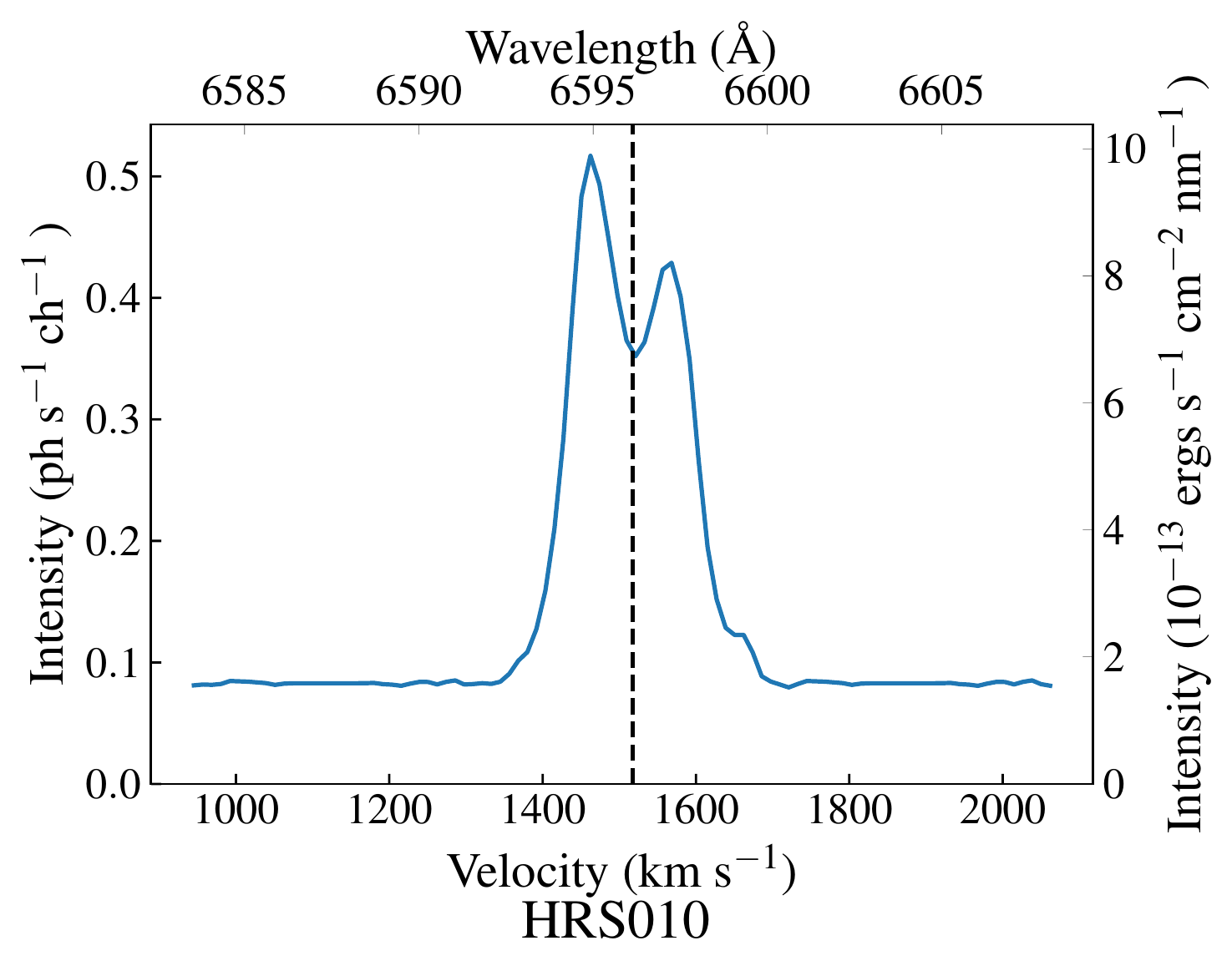}
\includegraphics[width=3.5cm]{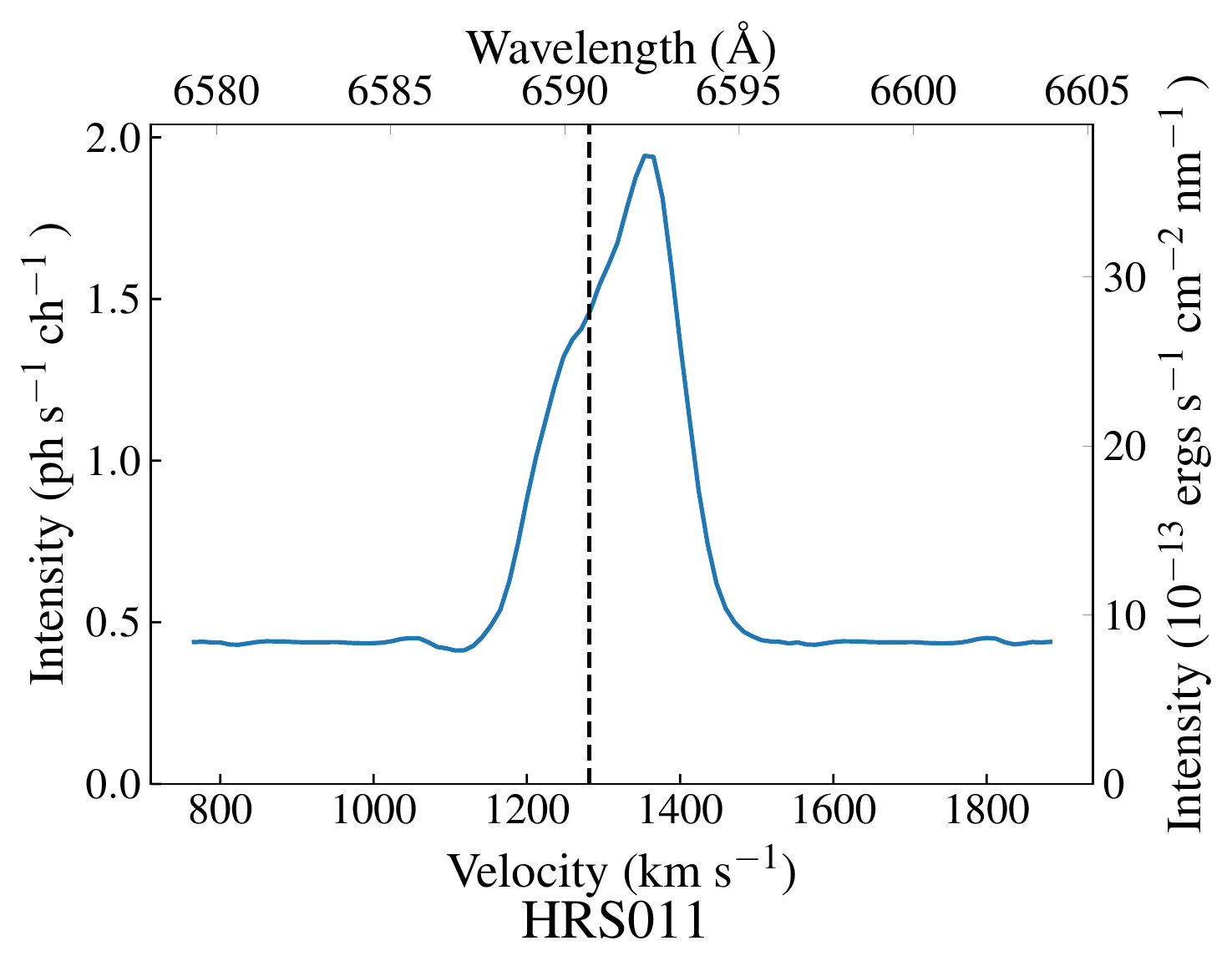}
\includegraphics[width=3.5cm]{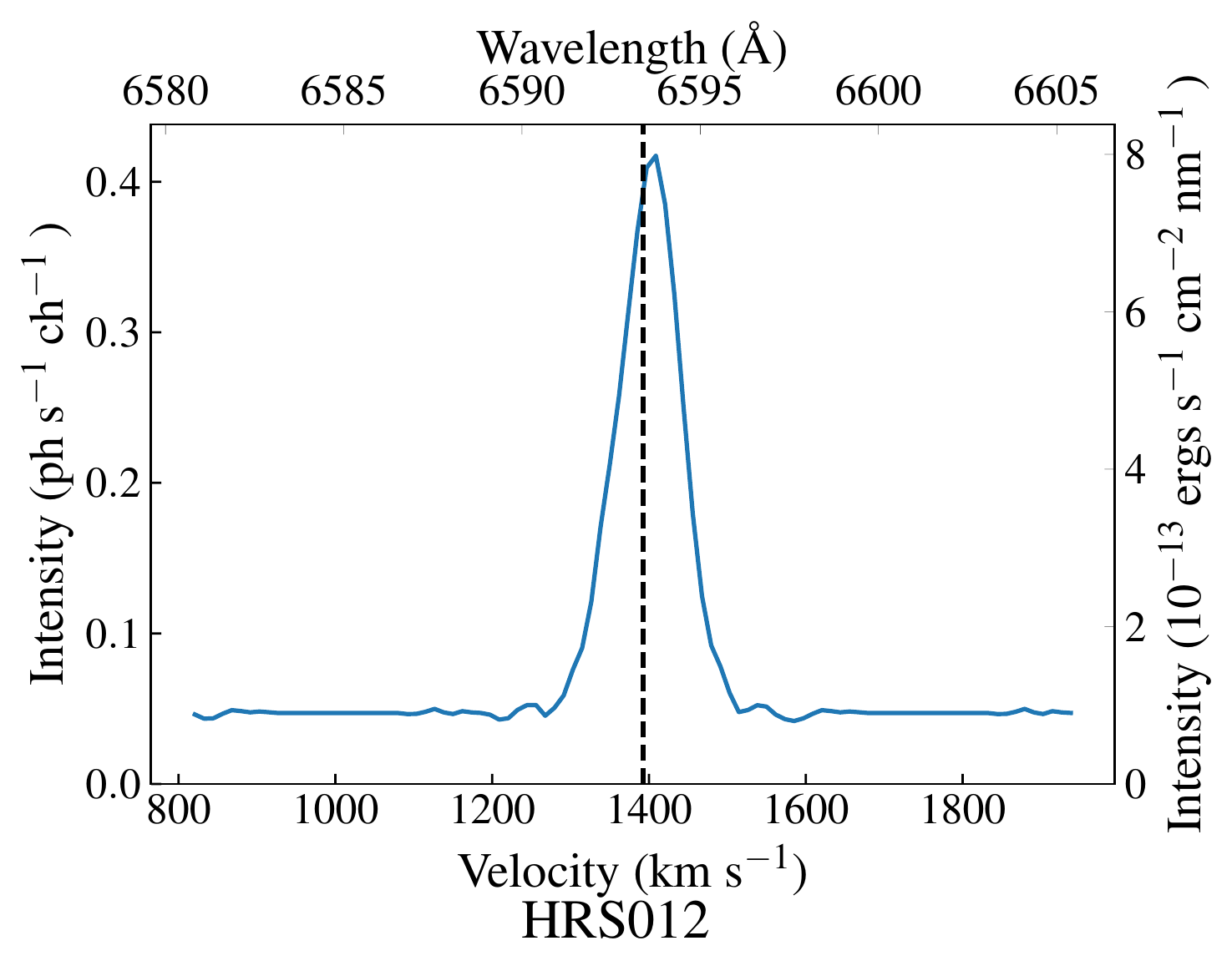}
\includegraphics[width=3.5cm]{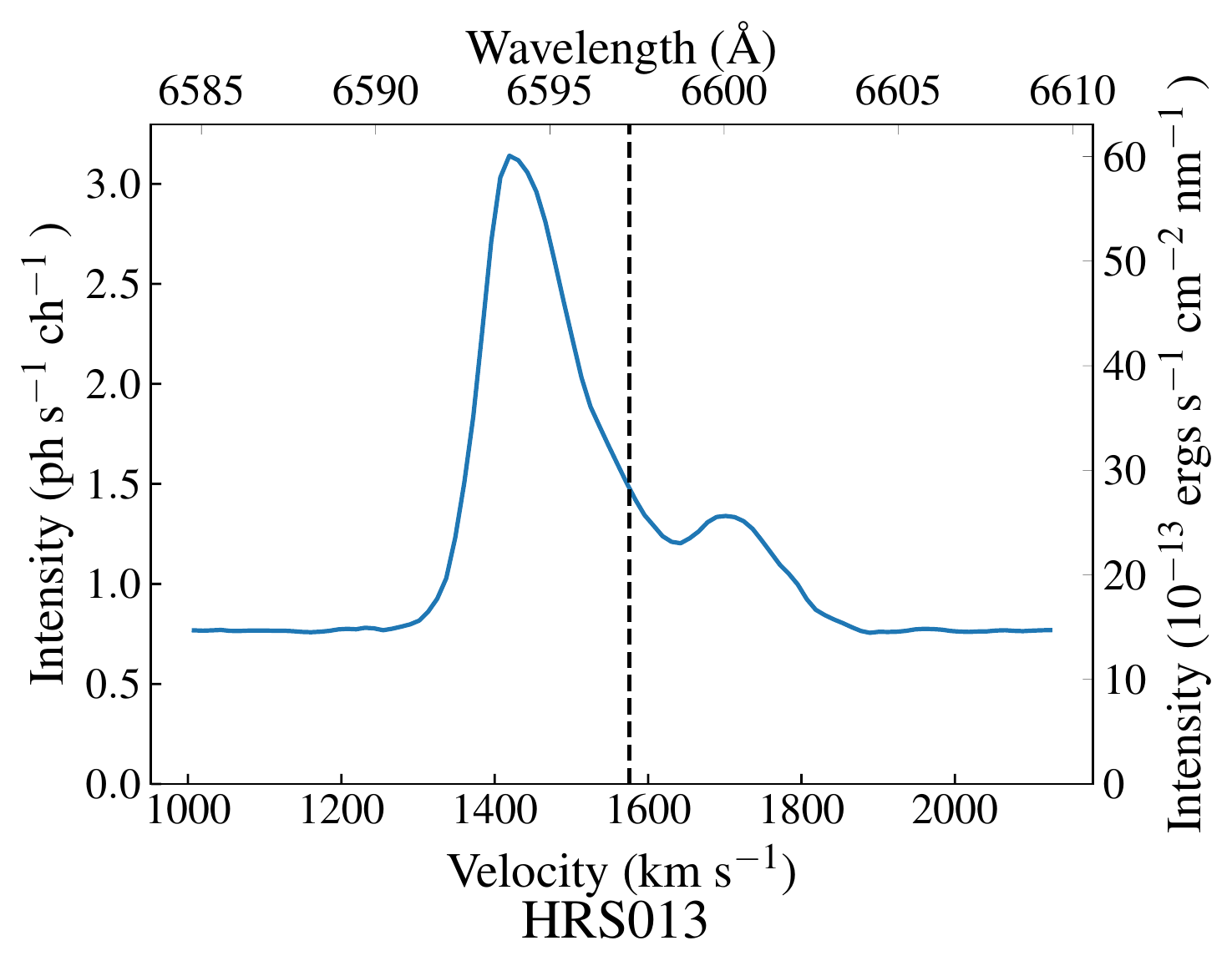}
\includegraphics[width=3.5cm]{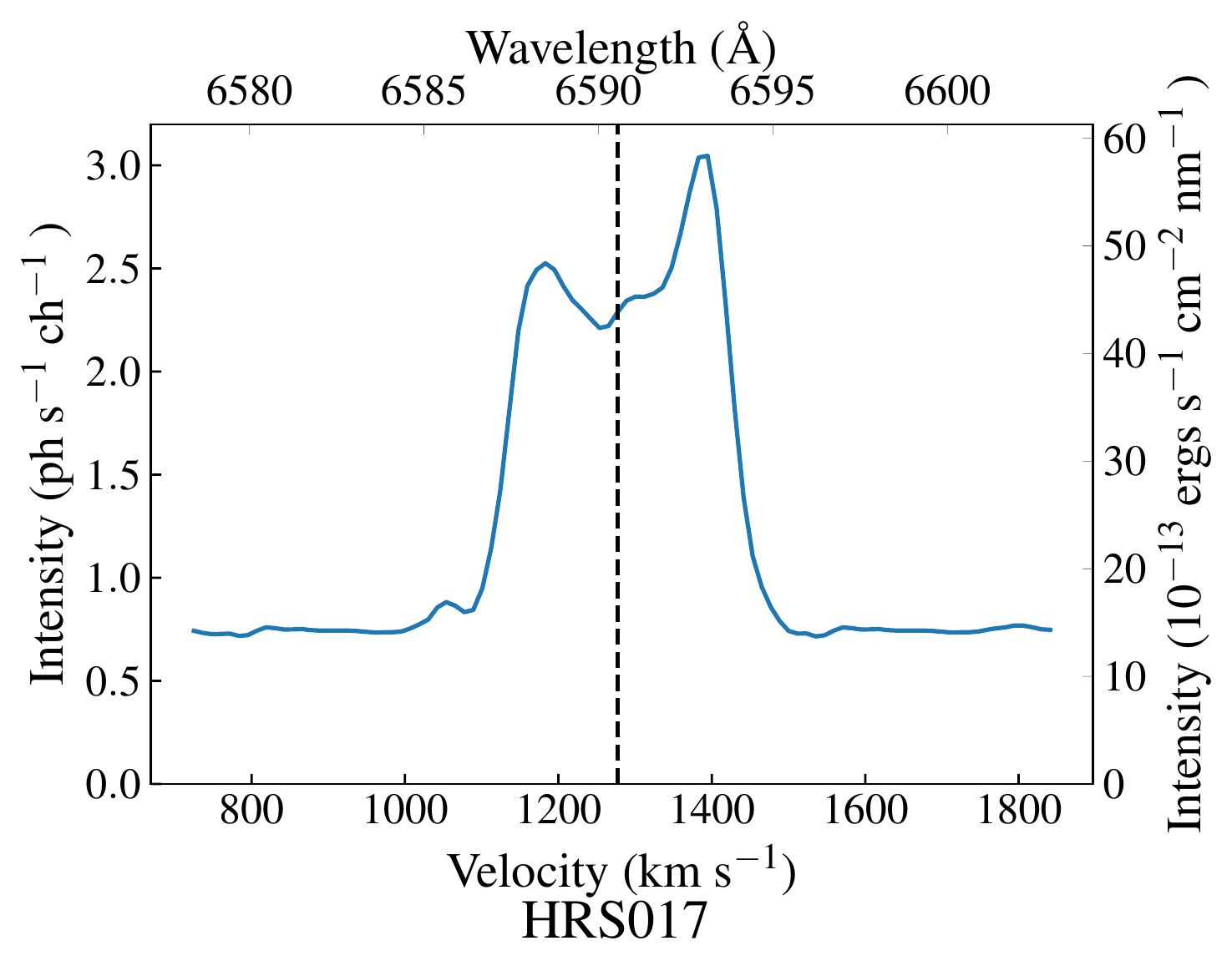}
\caption{Integrated \Ha~profiles for the 152 HRS galaxies concerning this work. The profiles are displayed over three times the free spectral range ($\sim$25$\AA$ top label, 1100 km s$^{-1}$ bottom label). The calibrated intensity in erg seg$^{-1}$ cm$^{-2}$ nm$^{-1}$ appears on the right side of the Y-axis. The black dashed vertical line represents the systemic velocity computed by the kinematical models; in the cases where a kinematical modelling was not possible, the corresponding systemic velocity taken from \cite{Boselli:2010} is plotted instead. Full version of the Figure available in the journal version.}
\label{fig:profiless}
\end{center}
\end{minipage}
\end{figure*}
\clearpage



\clearpage
\section{Individual Maps}
\label{maps}
\begin{figure*}
\begin{center}
\includegraphics[width=15cm]{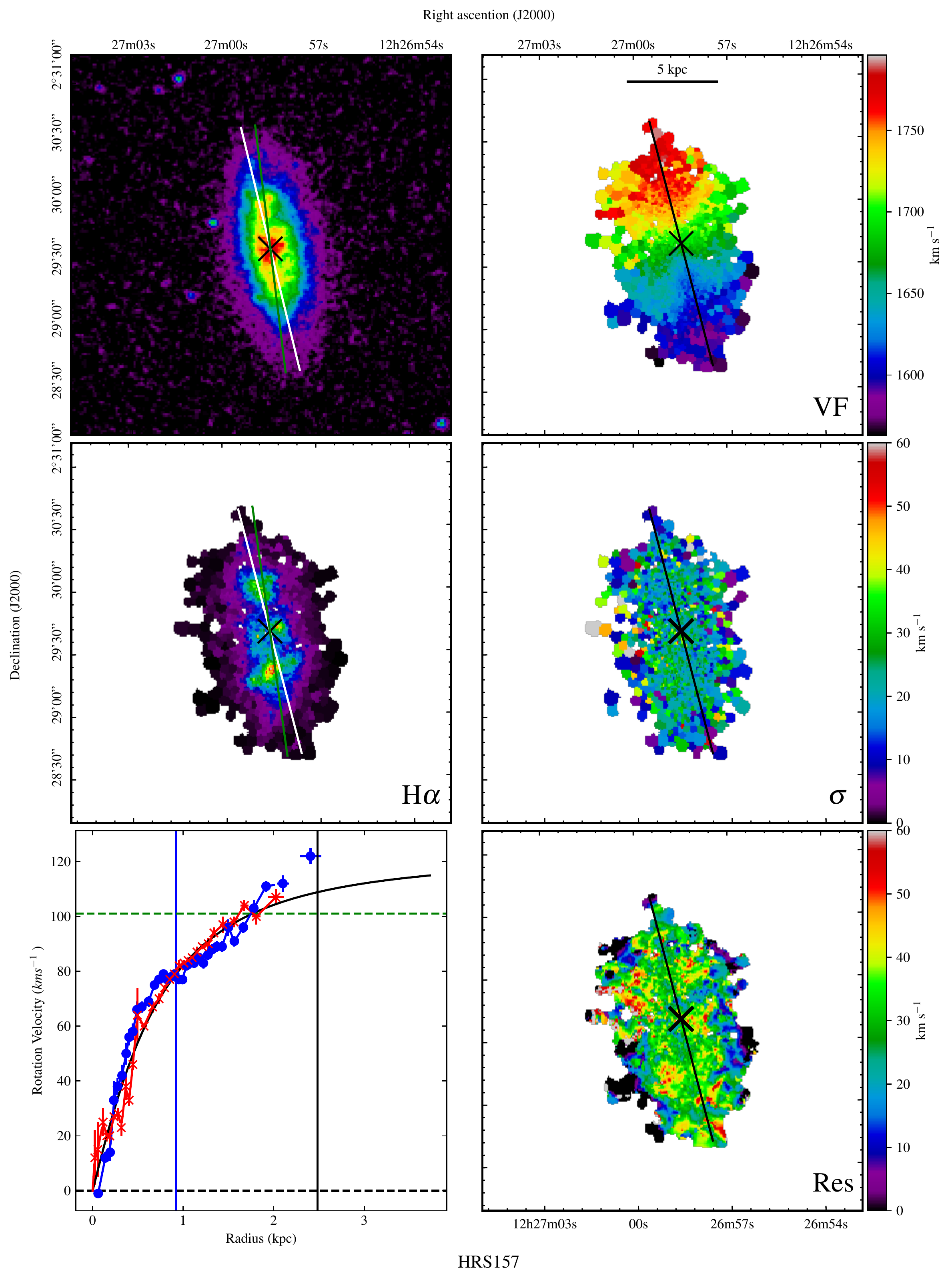}
\end{center}
\caption{Galaxy HRS 157. Top left: XDSS $R$-band image. Top right: \Ha~velocity field. Middle left: \Ha~monochromatic image. Middle right: \Ha~velocity dispersion field. Bottom left: rotation curve (if computed). Bottom right: residual velocity field (if computed).The black cross is the kinematical center. In the maps, the green line is the morphological major axis, while the black/white line is the estimated kinematical major axis, its length represents the $2 \times r_{opt}$. In the rotation curve, blue dots indicate the approaching side while red crosses the receding side, the black solid vertical line represents the $r_{opt}$, and the black dashed line represents the $r_{25}$; the blue solid line the $r_{eff}$ (if available in \citealp{Cortese:2012}) and the green horizontal line the HI $V_{max}$ (if available in \citealp{Boselli:2014}); the solid black curve is the best fit Courteau function to the rotation curve.}
\label{fig:hrs157}
\end{figure*}



\clearpage
\section{Tables}
\label{tables}
\begin{landscape}
\begin{table}
\caption{Herschel Reference Survey.}
\label{table_hrs}
\scalebox{0.78}{
\begin{tabular}{cccccccccccccccccccc}
\noalign{\medskip} \hline
 HRS  & CGCG & VCC & UGC & NGC & IC  & RA(2000)    & Dec(2000)  & D & Type  & r$_{eff}$ & D$_{24}$ & D$_{25}$ & $i$ &  IRAC1 & log $M_{star}$ & E(B-V) & Virgo D & Memb & $HI_{Def}$  \\
     &  &  &  &  &   & (h m s) & ($^{\circ}$ ' '') & Mpc &  & kpc & kpc & kpc  & mag & mag & M\Sun & & \Deg  & & \\
 (1)&(2)&(3)&(4)&(5)&(6)&(7)&(8)&(9)&(10)&(11)&(12)&(13)&(14)&(15)&(16)&(17)&(18)&(19)&(20) \\
\hline
1 &   123-035   &   -   &   -   &   -   &   -   &   10 17 39.66   &   +22 48 35.9   &16.79&   Pec   &0.45&2.32&2.44&13.899&   -   &8.72&0.0311&34.88&   Leo Cl.   &   -   \\
2 &   124-004   &   -   &5588&   -   &   -   &   10 20 57.13   &   +25 21 53.4   &18.44&   S?   &0.39&1.96&1.39&13.259&   -   &8.77&0.023&34.96&   Leo Cl.   &0.07\\
5 &   94-052   &   -   &   -   &   -   &610&   10 26 28.37   &   +20 13 41.5   &16.71&   Sc   &1.07&4.91&4.52&12.59&12.253&9.62&0.0203&32.06&   Leo Cl.   &0.66\\
6 &   154-016   &   -   &5662&   -   &   -   &   10 27 01.16   &   +28 38 21.9   &18.89&   SB(s)b   &1.80&6.32&9.09&13.545&14.206&8.94&0.0267&34.96&   Leo Cl.   &0.62\\
9 &   154-026   &   -   &5731&3277&   -   &   10 32 55.45   &   +28 30 42.2   &20.21&   SA(r)ab;HII   &1.40&8.20&5.73&11.037&11.46&10.08&0.0266&33.6&   Leo Cl.   &0.49\\
10 &   183-028   &   -   &5738&   -   &   -   &   10 34 29.82   &   +35 15 24.4   &21.66&   S?   &0.53&2.77&2.87&13.47&   -   &8.71&0.0271&36.99&   Leo Cl.   &0.12\\
11 &   124-038   &   -   &5742&3287&   -   &   10 34 47.31   &   +21 38 54.0   &18.93&   SB(s)d   &1.56&6.75&5.75&11.85&12.282&9.45&0.023&30.45&   Leo Cl.   &0.24\\
12 &   124-041   &   -   &   -   &   -   &   -   &   10 35 42.07   &   +26 07 33.7   &19.89&   cI   &0.43&2.17&1.71&14.088&   -   &8.47&0.023&31.89&   Leo Cl.   &0.06\\
13 &   183-030   &   -   &5753&3294&   -   &   10 36 16.25   &   +37 19 28.9   &22.47&   SA(s)c   &2.45&11.86&11.6&10.71&10.829&10.17&0.0195&37.97&   Leo Cl.   &0.25\\
17 &   95-019   &   -   &5887&3370&   -   &   10 47 04.05   &   +17 16 25.3   &18.3&   SA(s)c   &1.22&6.95&8.41&11.57&11.747&9.46&0.0308&26.39&   Leo Cl.   &-0.09\\
\hline
\end{tabular}
}
\\ The general properties of the 152 HRS galaxies observed at OHP with GHASP instrument. (1) HRS galaxy number. (2)(3)(4)(5)(6) Name of the galaxy in the CGCG/VCC/UGC/NGC/IC catalogs when available. (7)(8) Coordinates (in J2000) of the photometric center of the galaxy (\cite{Boselli:2010}). (9) Distance to the galaxy in Mpc from \cite{Boselli:2015}. (10) Morphological type according to \cite{Boselli:2010} . (11)(12) Effective radius (r$_eff$) and D$_{24}$ in $r$-band when available in \cite{Cortese:2012} converted to kpc. (13) D$_{25}$ in B-band from \cite{Boselli:2010} converted to kpc. (14)(15) Apparent asymptotic magnitude in the $i$-band and in the IRAC1 $3.6$$\mu$m from \cite{Cortese:2012} and \cite{Sheth:2010} respectively and if available. (16) Stellar mass from \cite{Cortese:2012} and \cite{Boselli:2015}, determined using the prescription given in \cite{Zibetti:2009} based on the $i$-band luminosity and $g - i$ mass-to-light ratio. For those objects without SDSS $g$ and $i$-band data (11 objects, marked with "a" in Table \ref{table_hrs}), the stellar masses were calculated as in \cite{Boselli2009} based on the $H$-band luminosity and $B - H$ mass-to-light ratio. (17) Color index from \cite{Cortese:2012}. (18)(19)(20) Distance from Virgo cluster (Virgo A or Virgo B) in degrees, name of the cloud where the galaxy is member, and HI-deficiency parameter from \cite{Boselli:2015}. Full version of the table available in the journal version.
\end{table}
\end{landscape}

\begin{table}
\caption{Log of the observations.}
\label{tablelog}
\begin{tabular}{ccccc}
\noalign{\medskip} \hline
  HRS   &   $\lambda_{c}$ & date & exp & Filter \\
    &   & & time & tilt\\
            & \AA & dd/mm/yy & s &\Deg \\
 (1)&(2)&(3)&(4)&(5)\\
\hline
1  &  6595  &  24/02/17  &  7360  &  2.5\\ 
2  &  6595  &  23/02/17  &  7360  &  2.5\\ 
5  &  6595  &  21/02/17  &  8960  &  2.5\\ 
6  &  6595  &  19/02/17  &  7360  &  0\\ 
9  &  6595  &  01/03/16  &  9720  &  0\\ 
10  &  6595  &  20/02/17  &  5760  &  0\\ 
11  &  6595  &  04/02/17  &  7380  &  0\\ 
12  &  6595  &  25/02/17  &  7200  &  0\\ 
13  &  6605  &  07/02/16  &  7380  &  5\\ 
17  &  6595  &  04/02/16  &  7020  &  0\\ 
18  &  6605  &  02/03/16  &  7530  &  5\\ 
19  &  6605  &  10/02/16  &  7380  &  5\\ 
21  &  6585  &  22/02/17  &  7040  &  0\\ 
23  &  6595  &  18/02/17  &  7360  &  0\\ 
25  &  6595  &  04/02/16  &  7020  &  0\\ 
26  &  6605  &  28/02/17  &  6340  &  3\\ 
27  &  6605  &  26/02/17  &  5760  &  3\\ 
28  &  6595  &  03/02/16  &  10440  &  0\\ 
32  &  6595  &  18/04/17  &  7039  &  5\\ 
35  &  6595  &  08/02/18  &  5440  &  0\\ 
37  &  6595  &  05/03/16  &  6000  &  0\\ 
38  &  6595  &  25/02/17  &  6400  &  0\\ 
39  &  6605  &  15/03/08  &  8160*  &  4\\ 
40  &  6605  &  06/03/16  &  7380  &  5\\ 
44  &  6595  &  22/05/17  &  6080  &  4\\ 
47  &  6595  &  05/02/16  &  7380  &  0\\ 
48  &  6595  &  02/02/16  &  10794  &  4\\ 
50  &  6595  &  15/02/18  &  6400  &  1\\ 
52  &  6595  &  20/05/17  &  6720  &  4\\ 
56  &  6605  &  08/03/16  &  7380  &  3\\ 
59  &  6605  &  24/02/17  &  7040  &  2.5\\ 
60  &  6585  &  11/02/16  &  7380  &  0\\ 
61  &  6595  &  19/04/17  &  6340  &  4\\ 
62  &  6605  &  19/02/17  &  7360  &  4\\ 
63  &  6595  &  05/02/16  &  10744  &  0\\ 
64  &  6595  &  02/05/17  &  7040  &  2.5\\ 
65  &  6595  &  16/02/16  &  6000  &  0\\ 
66  &  6595  &  07/02/16  &  7380  &  0\\ 
67  &  6595  &  12/03/16  &  7380  &  0\\ 
68  &  6595  &  15/02/18  &  7040  &  1\\ 
70  &  6605  &  27/04/17  &  6400  &  5\\ 
72  &  6605  &  10/03/16  &  8100  &  5\\ 
74  &  6585  &  14/02/16  &  7020  &  0\\ 
76  &  6585  &  07/03/16  &  7620  &  0\\ 
77  &  6595  &  10/02/16  &  6060  &  0\\ 
78  &  6595  &  11/03/16  &  7380  &  2.5\\ 
79  &  6595  &  19/03/2018  &  6720  &  0\\ 
80  &  6585  &  15/02/16  &  7080  &  0\\ 
82  &  6585  &  10/02-12/02/2018  &  8320  &  0\\ 
\hline
\end{tabular}
\\Log of the Fabry-Perot observations carried out so far at OHP and presented in this work. (1) HRS galaxy number. (2) Central wavelength of the interference filter used (FWHM = $15 \AA$). (3) Date of the observations. (4) Total exposure time in second; observations scanned with 48 instead of 32 channels are marked with a star. (5) Filter tiltination when the case.
\end{table}

\begin{table}
\begin{tabular}{ccccc}
\noalign{\medskip} \hline
   HRS   &   $\lambda_{c}$ & date & exp & Filter \\
    &   & & time & tilt\\
            & \AA & dd/mm/yy & s &\Deg \\
 (1)&(2)&(3)&(4)&(5)\\
\hline

83  &  6595  &  22/03/18  &  6400  &  4\\ 
84  &  6615  &  28/05/17  &  6720  &  0\\ 
86  &  6595  &  03/02/16  &  7800  &  0\\ 
88  &  6595  &  07/02/16  &  6741  &  0\\ 
92  &  6615  &  20/03/18  &  6400  &  0\\ 
95  &  6575  &  22/04/17  &  6400  &  0\\ 
98  &  6575  &  21/03/2018  &  6400  &  4\\ 
104  &  6585  &  15/02-22/02-22/03/18  &  6320  &  0\\ 
106  &  6625  &  28/04/17  &  6720  &  1.5\\ 
107  &  6605  &  08/02/18  &  7040  &  2.3\\ 
112  &  6615  &  24/04/17  &  7380  &  0\\ 
117  &  6585  &  17/03/18  &  5760*  &  0\\ 
118  &  6595  &  10/02/16  &  7380  &  1.7\\ 
121  &  6595  &  23/04/17  &  6400  &  4\\ 
132  &  6585  &  06/03/16  &  7380  &  5\\ 
133  &  6595  &  18/03/18  &  7040  &  4\\ 
134  &  6595  &  23/04/17  &  6720  &  4\\ 
136  &  6585  &  02/05/17  &  8000  &  5\\ 
139  &  6595  &  21/05/17  &  6080  &  5\\ 
141  &  6585  &  12/02/16  &  8616  &  0\\ 
142  &  6605  &  28/02/17  &  5900  &  3\\ 
143  &  6595  &  20/03/2018  &  6080  &  0\\ 
145  &  6595  &  24/04/17  &  7380  &  4\\ 
146  &  6595  &  24/05/17  &  6400  &  4\\ 
148  &  6565  &  25/02/17  &  6000  &  3\\ 
149  &  6575  &  13-15/03/18  &  17280*  &  4\\ 
151  &  6565  &  28/04/17  &  6720  &  0\\ 
152  &  6615  &  19/04/17  &  8639  &  0\\ 
153  &  6595  &  18/02/17  &  7360  &  0\\ 
154  &  6595  &  03/02/16  &  7800  &  0\\ 
156  &  6565  &  21/02/17  &  7040  &  5\\ 
157  &  6605  &  19/02/17  &  5760  &  4\\ 
159  &  6575  &  21/04/17  &  14400  &  0\\ 
160  &  6595  &  14/02/16  &  7380  &  0\\ 
164  &  6585  &  24/02/17  &  6080  &  4\\ 
165  &  6615  &  20/03/18  &  4160  &  0\\ 
168  &  6595  &  22/03/18  &  6400  &  4\\ 
169  &  6605  &  12/02/18  &  7380  &  3\\ 
171  &  6585  &  22/03/18  &  5760  &  0\\ 
172  &  6605  &  08/03/16  &  7380  &  3\\ 
177  &  6615  &  26/05/17  &  5760  &  0\\ 
182  &  6615  &  25/05/17  &  6400  &  0\\ 
184  &  6575  &  13/03/16  &  9180  &  0\\ 
185  &  6605  &  19/03/18  &  6400  &  0\\ 
187  &  6605  &  15/02/16  &  7080  &  1\\ 
189  &  6615  &  22/03/18  &  5300  &  4\\ 
191  &  6605  &  23/05/17  &  5760  &  5\\ 
192  &  6585  &  24/05/17  &  6400  &  5\\ 
193  &  6565  &  26/05/17  &  4800  &  4\\ 
195  &  6585  &  31/05/17  &  3000  &  0\\ 
197  &  6615  &  13/03/18  &  7680*  &  0\\ 
198  &  6595  &  22/02/17  &  7040  &  5\\ 
\hline
\end{tabular}
\end{table}

\begin{table*}
\begin{tabular}{ccccc}
\noalign{\medskip} \hline
  HRS   &   $\lambda_{c}$ & date & exp & Filter \\
    &   & & time & tilt\\
            & \AA & dd/mm/yy & s &\Deg \\
 (1)&(2)&(3)&(4)&(5)\\
\hline

199  &  6625  &  28/05/17  &  5120  &  5\\ 
207  &  6595  &  04/02/16  &  5420  &  0\\ 
212  &  6595  &  12/03/18  &  7680*  &  0\\ 
225  &  6625  &  08/02/18  &  7040  &  0\\ 
226  &  6615  &  27/05/17  &  6400  &  0\\ 
230  &  6585  &  20/05/17  &  6400  &  5\\ 
237  &  6585  &  01/05/17  &  7040  &  3.5\\ 
238  &  6595  &  23/05/17  &  7040  &  4\\ 
239  &  6565  &  28/04/17  &  5440  &  0\\ 
249  &  6585  &  30/04/17  &  7680  &  3.5\\ 
252  &  6605  &  13/02/18  &  7040  &  3\\ 
255  &  6585  &  16/02/16  &  7380  &  0\\ 
256  &  6595  &  28/02/17  &  3520  &  5\\ 
261  &  6595  &  20/04/17  &  3520  &  4\\ 
264  &  6595  &  15/02/18  &  7040  &  1\\ 
267  &  6595  &  19/03/18  &  6080  &  4\\ 
268  &  6605  &  19/03/18  &  6400  &  0\\ 
271  &  6595  &  24/02/17  &  6080  &  2.5\\ 
273  &  6595  &  21/03/18  &  6700  &  4\\ 
275  &  6605  &  11/03/16  &  7380  &  5\\ 
276  &  6625  &  12/03/16  &  7380  &  0\\ 
278  &  6595  &  22/05/17  &  5760  &  4\\ 
279  &  6595  &  20/02/17  &  7360  &  0\\ 
280  &  6625  &  11/02/18  &  7040  &  0\\ 
282  &  6625  &  08/02-11/02/18  &  6080  &  0\\ 
283  &  6585  &  22/02/17  &  3420  &  0\\ 
284  &  6595  &  22/04/17  &  6400  &  4\\ 
287  &  6595  &  22/02/17  &  6710  &  2.5\\ 
289  &  6605  &  25/02/17  &  7040  &  4\\ 
290  &  6595  &  18/04/17  &  5439  &  5\\ 
291  &  6595  &  20/04/17  &  7040  &  4\\ 
292  &  6595  &  19/04/17  &  5439  &  4\\ 
293  &  6585  &  21/02/17  &  7040  &  0\\ 
294  &  6605  &  22/04/17  &  6720  &  4\\ 
297  &  6595  &  12/03-15/03/18  &  14480*  &  2\\ 
298  &  6595  &  27/02/17  &  6400  &  2\\ 
300  &  6595  &  18/02/17  &  8640  &  0\\ 
301  &  6595  &  10/02/16  &  6480  &  0\\ 
303  &  6605  &  27/02/2017  &  6600  &  3\\ 
304  &  6595  &  25/02/17  &  6760  &  0\\ 
305  &  6595  &  20/04/17  &  7320  &  4\\ 
308  &  6595  &  21/05/17  &  5760  &  4\\ 
309  &  6595  &  10/03/16  &  6231  &  1.5\\ 
313  &  6595  &  26/02/17  &  7040  &  0\\ 
314  &  6595  &  12/03/16  &  8346  &  0\\ 
317  &  6605  &  18/04/17  &  6420  &  2.5\\ 
318  &  6595  &  28/02/17  &  6720  &  0\\ 
319  &  6595  &  15/02/16  &  6420  &  0\\ 
321  &  6605  &  22/04/17  &  6080  &  4\\ 
322  &  6595  &  16/02/16  &  7080  &  0\\ 
323  &  6605  &  21/04/17  &  6079  &  4\\

\hline
\end{tabular}
\end{table*}

\begin{landscape}
\begin{table}
\caption{Galaxy parameters.}
\label{kinparam}
\scalebox{0.98}{
\begin{tabular}{cccccccccccccccc}
\noalign{\medskip} \hline
 HRS & $\epsilon$ & $\chi^2_{red}$& $i_{morph}$ & $i_{kin}$ & V$_{sys}$ & V$_{sys_{FP}}$ & PA$_{morph}$ & PA$_{kin}$ & V$_{max}$ & V$_{max}$ flag &  $\overline{res}$  & $\sigma_{res}$ & f (\Ha) & r$_{RC}$/r$_{eff}$ & N$_{beams}$\\
       & & model &\Deg & \Deg & km s$^{-1}$ & km s$^{-1}$ & \Deg & \Deg & km s$^{-1}$ & & km s$^{-1}$ & km s$^{-1}$ & 10$^{-13}$ erg s$^{-1}$ cm$^{-2}$ & & \\
 (1)&(2)&(3)&(4)&(5)&(6)&(7)&(8)&(9)&(10)&(11)&(12)&(13)&(14)&(15)&(16)\\
\hline
1  &  0.60  &  2.8  &  67  &  67*  &  1175  &  1170 $\pm$ 1  &  175  &  177 $\pm$ 2  &  103*  &  2-B  &  0.17  &  10.04  &  1.58 & 2.58 & 450\\ 
2  &  0.28  &  1.32  &  45  &  28 $\pm$ 9 &  1291  &  1287 $\pm$ 1  &  35  &  39 $\pm$ 1  &  101*  &  3-B  &  0.22  &  7.05  &  1.22 & 2.50 & 159 \\ 
5  &  0.76  &  2.25  &  77  &  77*  &  1170  &  1164 $\pm$ 1  &  29  &  30$\pm$3  &  145*  &  4-B  &  -0.10  &  9.50  &  0.10  & 2.30 & 536\\ 
6  &  0.85  &  4.01  &  85  &  85*  &  1322  &  1321$\pm$ 3  &  330  &  330*  &  130*  &  4-B  &  -0.55  &  12.00  &  1.00  & 1.75 & 1013\\ 
9  &  0.28  &  4.75  &  45  &  54 $\pm$ 9  &  1415  &  1409 $\pm$ 3  &  216  &  198 $\pm$ 2  &  159*  &  4-B  &  0.76  &  10.68  &  0.25  & 2.93 & 3054\\ 
10  &  0.30  &  0.97  &  47  &  47*  &  1516  &  1518 $\pm$ 1 &  32  &  30 $\pm$ 2  &  100*  &  2-A  &  -0.22  &  6.37  &  2.57  & 2.60 & 467\\ 
11  &  0.53  &  0.93  &  63  &  61 $\pm$ 6  &  1325  &  1282 $\pm$ 1  &  20  &  19 $\pm$ 1  &  136  &  1-B  &  0.03  &  6.59  &  9.58  & 2.61 & 1843\\ 
12  &  0.34  &  0.59  &  49  &  58 $\pm$ 16  &  1392  &  1393 $\pm$ 1  &  30  &  51 $\pm$ 3  &  50  &  2-B  &  0.35  &  4.48  &  1.82  & 2.16& 250\\ 
13  &  0.54  &  2.74  &  63  &  59$\pm$ 3  &  1573  &  1576 $\pm$ 2  &  116  &  119 $\pm$ 1  &  213*  &  1-A  &  1.00  &  11.54  &  14.01  & 2.54& 5815\\ 
17  &  0.47  &  1.31  &  58  &  55 $\pm$ 2  &  1281  &  1277 $\pm$ 1  &  323  &  325 $\pm$ 1  &  160  &  1-A  &  0.75  &  7.91  &  19.7  & 2.41& 5062\\ 
\hline
\end{tabular}
}
\\Kinematical parameters computed from the models for the 152 HRS galaxies concerning this work. (1)  HRS galaxy number. (2) Elipticity from \cite{Cortese:2012}. (3) $\chi^2_{red}$ of the kinematical model. (4) Morphological inclination calculated with the elipticity value as in \cite{Masters:2010}. (5) Inclination calculated by our kinematical model (6) Systemic velocity from \cite{Boselli:2010}. (7) Systemic velocity computed by our kinematical models (8) Morphological PA taken from \cite{Cortese:2012}; if necessary, the value of the PA has been transformed in order to be able to compare with the kinematical PA easily. (9) PA calculated by our kinematical models (The case of galaxy HRS 322 is very explained in Appendix \ref{notes}). (10) V$_{max}$ derived from the Courteau function fitted to the \rc, marked with a star when extrapolated. (11) Flag of the \rc. (12) Mean residual velocity dispersion of the residual velocity fields. (13) Mean velocity dispersion of the residual velocity fields. (14) Integrated flux deduced from our comparison to \cite{Boselli:2014} data. (15) $r_{RC}$ normalised to $r_{eff}$. (16) Number of resolution elements per galaxy spatial coverage. Full version of the table available in the journal version.
\end{table}
\end{landscape}

\begin{table}
\caption{HRS 157 \RC}
\label{rclabel_hrs157}
\begin{tabular}{cccccccc}
\noalign{\medskip} \hline

           r   &  $\sigma_r$   &             r   &    $\sigma_r$  &          v   & $\sigma_v$  &            N bins  &              side \\
            kpc  &   kpc  &             "  &    "  &         \kms  & \kms  &              &               \\
            (1)&(2)&(3)&(4)&(5)&(6)&(7)&(8)\\
\hline
            0.05  &            0.00  &             0.6  &             0.0  &              12  &              10  &               1  &               r \\
            0.12  &            0.00  &             1.4  &             0.0  &              15  &              10  &               1  &               r \\
            0.13  &            0.01  &             1.5  &             0.2  &              -1  &               0  &               2  &               a \\
            0.23  &            0.02  &             2.8  &             0.2  &              25  &               5  &               2  &               r \\
            0.28  &            0.02  &             3.4  &             0.3  &              12  &               2  &               4  &               a \\
            0.31  &            0.01  &             3.8  &             0.1  &              20  &               2  &               3  &               r \\
            0.38  &            0.02  &             4.6  &             0.3  &              20  &               3  &               5  &               r \\
            0.39  &            0.02  &             4.7  &             0.3  &              14  &               3  &               2  &               a \\
            0.47  &            0.02  &             5.7  &             0.2  &              33  &               7  &               4  &               a \\
            0.47  &            0.02  &             5.7  &             0.3  &              27  &               1  &               3  &               r \\
            0.56  &            0.03  &             6.7  &             0.3  &              38  &               3  &               6  &               a \\
            0.57  &            0.02  &             6.9  &             0.2  &              28  &               2  &               7  &               r \\
            0.63  &            0.02  &             7.7  &             0.2  &              23  &               3  &               2  &               r \\
            0.65  &            0.02  &             7.9  &             0.3  &              42  &               4  &               7  &               a \\
            0.73  &            0.02  &             8.9  &             0.2  &              38  &               5  &               4  &               r \\
            0.74  &            0.02  &             9.0  &             0.3  &              50  &               4  &               5  &               a \\
            0.81  &            0.02  &             9.8  &             0.2  &              56  &               3  &               7  &               a \\
            0.81  &            0.00  &             9.8  &             0.0  &              33  &               2  &               2  &               r \\
            0.89  &            0.02  &            10.8  &             0.2  &              46  &               2  &              11  &               r \\
            0.89  &            0.02  &            10.8  &             0.2  &              58  &               3  &               9  &               a \\
            0.98  &            0.02  &            11.9  &             0.3  &              66  &               3  &              11  &               a \\
            0.99  &            0.03  &            12.0  &             0.4  &              64  &              10  &               3  &               r \\
            1.08  &            0.05  &            13.2  &             0.6  &              67  &               2  &              23  &               a \\
            1.13  &            0.07  &            13.8  &             0.8  &              60  &               1  &              23  &               r \\
            1.24  &            0.05  &            15.0  &             0.6  &              69  &               2  &              23  &               a \\
            1.33  &            0.04  &            16.1  &             0.5  &              67  &               2  &              23  &               r \\
            1.37  &            0.03  &            16.6  &             0.4  &              75  &               1  &              23  &               a \\
            1.47  &            0.04  &            17.8  &             0.5  &              70  &               2  &              23  &               r \\
            1.47  &            0.03  &            17.8  &             0.3  &              77  &               1  &              23  &               a \\
            1.56  &            0.03  &            19.0  &             0.3  &              79  &               1  &              23  &               a \\
            1.59  &            0.03  &            19.3  &             0.4  &              74  &               1  &              23  &               r \\
            1.65  &            0.02  &            20.0  &             0.3  &              77  &               1  &              23  &               a \\
            1.71  &            0.03  &            20.7  &             0.4  &              77  &               1  &              23  &               r \\
            1.74  &            0.02  &            21.1  &             0.3  &              78  &               2  &              23  &               a \\
            1.80  &            0.03  &            21.9  &             0.4  &              79  &               1  &              23  &               r \\
            1.82  &            0.02  &            22.0  &             0.3  &              79  &               1  &              23  &               a \\
            1.91  &            0.03  &            23.1  &             0.4  &              77  &               1  &              23  &               a \\
            1.92  &            0.03  &            23.2  &             0.4  &              82  &               2  &              23  &               r \\
            1.99  &            0.02  &            24.1  &             0.2  &              77  &               1  &              23  &               a \\
            2.02  &            0.03  &            24.5  &             0.3  &              83  &               1  &              23  &               r \\
            2.07  &            0.02  &            25.1  &             0.3  &              82  &               1  &              23  &               a \\
            2.11  &            0.02  &            25.6  &             0.3  &              84  &               1  &              23  &               r \\
            2.16  &            0.03  &            26.1  &             0.3  &              83  &               2  &              23  &               a \\
            2.20  &            0.03  &            26.7  &             0.4  &              85  &               1  &              23  &               r \\
            2.24  &            0.04  &            27.2  &             0.4  &              83  &               1  &              23  &               a \\
            2.31  &            0.03  &            28.0  &             0.3  &              87  &               2  &              23  &               r \\
            2.35  &            0.03  &            28.5  &             0.4  &              85  &               1  &              23  &               a \\
            2.42  &            0.04  &            29.4  &             0.5  &              89  &               1  &              23  &               r \\
            2.45  &            0.03  &            29.7  &             0.4  &              83  &               2  &              23  &               a \\
            2.55  &            0.03  &            30.9  &             0.3  &              86  &               2  &              23  &               a \\
            2.55  &            0.03  &            31.0  &             0.4  &              90  &               2  &              23  &               r \\
\hline
\end{tabular}
\\Numerical table of the \rc~for the galaxy HRS 157. (1)(3) Galactic radius. (2), (4) Dispersion around the galactic radius (if zero, it is because only one bin was used). (5) Rotation velocity. (6) Dispersion on the rotation velocity. (7) Number of velocity bins. (8) Receding -- r -- or approaching -- a -- side.
\end{table}

\begin{table}
\begin{tabular}{cccccccc}
\noalign{\medskip} \hline
           r   &  $\sigma_r$   &             r   &    $\sigma_r$  &          v   & $\sigma_v$  &            N bins  &              side \\
            kpc  &   kpc  &             "  &    "  &         \kms  & \kms  &              &               \\
            (1)&(2)&(3)&(4)&(5)&(6)&(7)&(8)\\
\hline
            2.65  &            0.03  &            32.1  &             0.3  &              88  &               2  &              23  &               a \\
            2.68  &            0.04  &            32.5  &             0.5  &              94  &               2  &              23  &               r \\
            2.74  &            0.03  &            33.3  &             0.4  &              89  &               2  &              23  &               a \\
            2.86  &            0.03  &            34.7  &             0.4  &              89  &               2  &              23  &               a \\
            2.88  &            0.07  &            34.9  &             0.8  &              97  &               3  &              23  &               r \\
            3.00  &            0.04  &            36.4  &             0.5  &              96  &               3  &              23  &               a \\
            3.12  &            0.08  &            37.9  &             0.9  &              98  &               2  &              23  &               r \\
            3.13  &            0.04  &            38.0  &             0.5  &              91  &               2  &              23  &               a \\
            3.33  &            0.07  &            40.4  &             0.8  &              96  &               2  &              23  &               a \\
            3.35  &            0.09  &            40.7  &             1.1  &             104  &               2  &              23  &               r \\
            3.56  &            0.07  &            43.2  &             0.9  &             103  &               3  &              23  &               a \\
            3.62  &            0.09  &            43.9  &             1.1  &             100  &               3  &              23  &               r \\
            3.83  &            0.09  &            46.5  &             1.1  &             111  &               2  &              23  &               a \\
            4.04  &            0.18  &            49.1  &             2.2  &             107  &               3  &              20  &               r \\
            4.20  &            0.14  &            51.0  &             1.6  &             112  &               3  &              23  &               a \\
            4.82  &            0.24  &            58.4  &             2.9  &             122  &               3  &              16  &               a \\
            2.55  &            0.03  &            31.0  &             0.4  &              90  &               2  &              23  &               r \\
            2.65  &            0.03  &            32.1  &             0.3  &              88  &               2  &              23  &               a \\
            2.68  &            0.04  &            32.5  &             0.5  &              94  &               2  &              23  &               r \\
            2.74  &            0.03  &            33.3  &             0.4  &              89  &               2  &              23  &               a \\
            2.86  &            0.03  &            34.7  &             0.4  &              89  &               2  &              23  &               a \\
            2.88  &            0.07  &            34.9  &             0.8  &              97  &               3  &              23  &               r \\
            3.00  &            0.04  &            36.4  &             0.5  &              96  &               3  &              23  &               a \\
            3.12  &            0.08  &            37.9  &             0.9  &              98  &               2  &              23  &               r \\
            3.13  &            0.04  &            38.0  &             0.5  &              91  &               2  &              23  &               a \\
            3.33  &            0.07  &            40.4  &             0.8  &              96  &               2  &              23  &               a \\
            3.35  &            0.09  &            40.7  &             1.1  &             104  &               2  &              23  &               r \\
            3.56  &            0.07  &            43.2  &             0.9  &             103  &               3  &              23  &               a \\
            3.62  &            0.09  &            43.9  &             1.1  &             100  &               3  &              23  &               r \\
            3.83  &            0.09  &            46.5  &             1.1  &             111  &               2  &              23  &               a \\
            4.04  &            0.18  &            49.1  &             2.2  &             107  &               3  &              20  &               r \\
            4.20  &            0.14  &            51.0  &             1.6  &             112  &               3  &              23  &               a \\
            4.82  &            0.24  &            58.4  &             2.9  &             122  &               3  &              16  &               a \\
\hline
\end{tabular}
\end{table}



\end{document}